\documentclass[fleqn,usenatbib]{mnras}
\usepackage[T1]{fontenc}
\usepackage{ae,aecompl}
\usepackage{graphicx}
\usepackage{amsmath}
\usepackage{amssymb}
\usepackage{txfonts}

\title[Appreciating mergers]
{Appreciating mergers for understanding the non-linear 
$M_{\rm bh}$--$M_{\rm *,spheroid}$ 
and $M_{\rm bh}$--$M_{\rm *,galaxy}$ relations, updated herein,
and the implications for the (reduced) role of AGN feedback}

\author[Graham \& Sahu]
{
Alister W.\ Graham$^1$\thanks{E-mail: AGraham@swin.edu.au},
Nandini Sahu$^{1,2}$
\\
$^1$ Centre for Astrophysics and Supercomputing, Swinburne University of
Technology, Hawthorn, VIC 3122, Australia\\ 
$^2$ OzGrav-Swinburne, Centre for Astrophysics and Supercomputing, Swinburne
University of Technology, Hawthorn, VIC 3122, Australia
}

\date{Accepted XXX. Received YYY; in original form ZZZ}
\pubyear{2022}

\begin{document}
\label{firstpage}
\pagerange{\pageref{firstpage}--\pageref{lastpage}}
\maketitle

\begin{abstract}

We present revised (black hole mass)-(spheroid stellar mass) and 
(black hole mass)-(galaxy stellar mass) scaling relations based on 
colour-dependent (stellar mass)-to-light ratios.  Our 3.6~$\mu$m luminosities 
were obtained from multicomponent decompositions, 
which accounted for bulges, discs, bars, ansae, rings, nuclear components, etc. 
The lenticular galaxy bulges (not associated with recent mergers) follow a 
steep $M_{\rm{bh}}\propto~M_{\rm{*,bulge}}^{1.53\pm0.15}$ 
relation, offset by roughly an order of magnitude in black hole mass from the 
$M_{\rm{bh}}\propto~M_{\rm{*,ellip}}^{1.64\pm0.17}$ 
relation defined by the elliptical (E) galaxies which, in Darwinian terms,  
are shown to have evolved by punctuated equilibrium rather than gradualism. 
We use the spheroid (i.e., bulge and elliptical) size-mass relation to reveal 
how disc-galaxy mergers explain this offset and 
the dramatically lower $M_{\rm{bh}}/M_{\rm{*,sph}}$ ratios in the elliptical
galaxies.  
The deceptive near-linear $M_{\rm{bh}}$--$M_{\rm{*,sph}}$ `red sequence', followed by
neither the bulge population nor the elliptical galaxies, is shown to be 
an artefact of sample selection,  
combining bulges and elliptical galaxies from disparate
$M_{\rm{bh}}$--$M_{\rm{*,sph}}$ sequences.
Moreover, both small bulges with `undermassive' black holes and big
lenticular galaxies (including relic `red nuggets') with
`overmassive' black holes --- relative to the near-linear 
$M_{\rm{bh}}$--$M_{\rm{*,sph}}$ sequence --- are no longer viewed as outliers. 
We confirm a steep $M_{\rm{bh}}\propto~M_{\rm{*,bulge}}^{2.25\pm0.39}$
relation for spiral galaxies and 
discuss numerous implications of this work, including how 
mergers, rather than (only) feedback from active galactic nuclei, have shaped the
high-mass end of the galaxy mass function.  We also explain why there may be no
useful $M_{\rm{bh}}$--$M_{\rm{*,sph}}$--$R_{\rm{e,sph}}$ plane due to
$M_{\rm{*,sph}}$ scaling nearly linearly with $R_{\rm{e,sph}}$.

\end{abstract}


\section{Introduction}

The linear $M_{\rm bh}$--$M_{\rm *,sph}$ relation \citep{1988ApJ...324..701D,
  1989IAUS..134..217D}, see also \citet{1992ASPC...31..417Y}, sometimes
referred to as the Magorrian relation \citep{1998AJ....115.2285M}, has
repeatably been heralded as a critical ingredient to understanding the coevolution
of galaxies and their central massive black holes.  Black hole feedback is
said to regulate the gas and thereby control the star formation \citep[e.g.,][and
  references therein]{1998A&A...331L...1S, 2007MNRAS.380..877S,
  2008ApJ...676...33D, 2010MNRAS.402.1536S,
  2013MNRAS.432.3401G} and thus establish/explain the $M_{\rm bh}$--$M_{\rm
  *,sph}$ relation.\footnote{\citet{2005SSRv..116..523F}, \citet{2006ccha.book.....L} and
  \cite{2016ASSL..418..263G} review the discovery
  of black holes and their scaling relations.}  Despite early evidence for a non-linear $M_{\rm bh}$--$M_{\rm
  *,sph}$ relation \citep[e.g.,][]{1998ApJ...505L..83L, 2001ApJ...553..677L,
  1999ApJ...519L..39W, 2000MNRAS.317..488S}, there has been a tendency to
cling to the simplicity of the original trend.  However, along with
increases in sample size and improvements in galaxy decomposition
--- which have led to both a better understanding of galaxies and a better
measurement of their spheroidal component\footnote{We use the term `spheroid' to denote
  bulges and (pure) elliptical galaxies.} ---, has come an ever-refined
insight into the $M_{\rm bh}$--$M_{\rm *,sph}$ diagram through the detection
of (galaxy morphology)-dependent substructure and departures from the near-linear
relation.

Clues that something was amiss with the notion of a near-linear $M_{\rm
  bh}$--$M_{\rm *,sph}$ relation were presented in
\citet{2012ApJ...746..113G}, which reported a steeper near-quadratic relation
for spheroids with a \citet{1963BAAA....6...41S} light profile\footnote{A
  review of the \citet{1963BAAA....6...41S} model can be found in
  \citet{2005PASA...22..118G}.} and a near-linear relation for spheroids with
a core-S\'ersic light profile\footnote{\cite{1966ApJ...143.1002K} and
  \citet{1972IAUS...44...87K} noted that such galaxies have shallow inner
  light profiles notably flatter than expected from their outer $R^{1/4}$-like
  light profile.} \citep{2003AJ....125.2951G}.  This work built on a key
tip-off in the final paragraph of
\citet{2007MNRAS.379..711G}\footnote{\citet{2012ApJ...746..113G} noted that 
the final exponent in the second last
  sentence of the Appendix of \citet{2007MNRAS.379..711G} should have read
  1/0.5 rather than 0.5 to give $M_{\rm bh} \propto L^{1/0.5}$.}  and was
later expressed as a (cool gas)-rich versus (cool gas)-poor galaxy sequence in
\citet{2013ApJ...764..151G} and \citet{2013ApJ...768...76S}.
\citet{2015ApJ...798...54G} revealed that the near-quadratic relation also
appeared to encompass active galactic nuclei (AGNs) with virial
masses\footnote{Virial masses were derived using a virial factor $f=2.8$
  \citep{2011MNRAS.412.2211G}.}  as low as $2\times10^5$ M$_{\odot}$.  With
improved data, \citet{2016ApJ...817...21S} found that the distributions in the
$M_{\rm bh}$--$M_{\rm *,sph}$ diagram were better described by a `blue
sequence' for late-type galaxies (LTGs) --- which are all S\'ersic galaxies
--- and a `red sequence' for early-type galaxies (ETGs), which can be S\'ersic
galaxies or core-S\'ersic galaxies.  This red versus blue sequence was later
emphasised by others, including \citet{2016ApJ...831..134V} and
\citet{2020ApJ...898...83D}.


Doubling the sample size of spiral galaxies used by
\citet{2016ApJ...817...21S}, \citet{2019ApJ...873...85D} could better
constrain the $M_{\rm bh}$--$M_{\rm *,sph}$ relation for the LTGs, finding a
slope of 2.17$\pm$0.32 to $2.44^{+0.35}_{-0.31}$ depending on the regression
analysis used.  Doubling the sample size of ETGs used by
\citet{2016ApJ...817...21S}, \citet{2019ApJ...876..155S} measured a slope of
1.27$\pm$0.07 for the ETGs but crucially explained why this was misleading.
\citet{2019ApJ...876..155S}, and \citet{2016ApJS..222...10S}, knew which ETGs
were (pure) elliptical galaxies and which were lenticular or
ellicular\footnote{Ellicular galaxy is the name given by
  \citet{2016ApJ...831..132G} to the ES galaxy type introduced by
  \citet{1966ApJ...146...28L}.  They have intermediate-scale discs which do
  not dominate the light at large radii, in contrast to the familiar
  lenticular (S0) galaxies whose large-scale discs do
  \citep{2019MNRAS.487.4995G}.}  galaxies.  Separating the ETGs into those
with and without discs, \citet{2019ApJ...876..155S} revealed that they
followed separate $M_{\rm bh}$--$M_{\rm *,sph}$ relations with similar slopes
($\approx$1.9$\pm$0.2, based on $M_*/L_{3.6} = 0.6$) but offset by an order of
magnitude in $M_{\rm bh}$.  Therefore, as \citet{2019ApJ...876..155S}
explained, published slopes for the near-linear $M_{\rm bh}$--$M_{\rm *,sph}$
relation, i.e., the `red-sequence',  
are dependent on the sample's arbitrary number of ETGs with and without discs.  
See, for example, 
\citet[][with a slope of 0.93$\pm$0.10]{2007MNRAS.379..711G}, 
\citet[][with a slope of 1.16$\pm$0.08]{2013ARA&A..51..511K}, 
\citet[][with a slope of 0.846$\pm$0.064]{2016ApJ...818...47S}, and 
\citet[][with a slope of 1.04$\pm$0.10]{2016ApJ...817...21S}. 
The slope is not a measure of physical 
importance --- as has been thought and reported for over a quarter of a century
regarding galaxy/black hole coevolution --- but rather a reflection of 
the sample selection. This revelation has been shown to impact black hole correlations
involving not just the spheroid's stellar mass but also the spheroid's size 
\citep{2020ApJ...903...97S} and the spheroid's range of density 
measures \citep{2022ApJ...927...67S}. 

This new knowledge is important because it rewrites our understanding of the
interplay between spheroids and their central massive black holes.  This
realisation was refined by performing multicomponent decompositions of the
galaxy light, with recourse to kinematic information and accounting 
for distinct physical entities such as bars, rings, bulges, and discs 
detected in the images and the Fourier harmonic analysis of the isophotes
\citep{2015ApJ...810..120C}.  Here, 
with updated data, we offer a likely explanation for the offset 
between the relations followed by elliptical and ellicular/lenticular galaxies. 
We also raise some of the ensuing
implications.  In particular, we more clearly elucidate the origin and 
`red herring' nature of the near-linear $M_{\rm bh}$--$M_{\rm *,sph}$ relation
in regard to understanding the (limited caretaker) role for AGN feedback in elliptical
galaxies.

We previously used a simple conversion of starlight-to-mass in
our (galaxy morphology)-dependent scaling diagrams: specifically, 
$M_*/L_{*,3.6\, \mu m} = 0.6$ \citep{2014ApJ...788..144M}.\footnote{This was based on a
  \citet{2003PASP..115..763C} `initial mass function'.}\footnote{In 
practice, while $M_*/L_{obs,3.6}=0.6$ was used for the ETGs
  \citep{2019ApJ...876..155S}, a lower value of $M_*/L_{\rm obs,3.6}=0.453$
  was applied to the LTGs \citep{2019ApJ...873...85D} because the observed
  luminosity at 3.6~$\mu$m includes both starlight and the glow of warm dust.
  This reduced ratio encapsulated the mean ratio $L_{*,3.6}/L_{\rm obs.3.6}
  \approx 0.75$ for LTGs \citep{2015ApJS..219....5Q}.}  Such an approach
meshes well with the notion that many compact `red nuggets' at redshifts
$z \sim 2.5 \pm1$ (both massive and not so massive) have become the bulges of
some of today's lenticular and spiral galaxies \citep{2015ApJ...804...32G,
  2016MNRAS.457.1916D, 2017ApJ...840...68G, 2022MNRAS.514.3410H}.  Such an origin for
these bulges would make them old, as \citet{1996AJ....111.2238P} 
and \citet{2009MNRAS.395...28M} reported, and therefore require a high mass-to-light
ratio.  However, not every bulge needs to be old. Here we explore 
colour-dependent $M_*/L_{obs,3.6}$ ratios for a sample of $\sim$100 galaxies 
pre-observed with the Spitzer Space Telescope and close
enough to resolve their bulges ($R_{\rm e} \gtrsim 2\arcsec$).  That is,
we allow for departures from the assumption that all the bulges have the same
$M_*/L \equiv \Upsilon_*$ ratio.  Here, we use a $B-V$ colour-dependent
mass-to-light ratio prescription to derive the stellar masses.
Appendix~\ref{Apdx1} offers an alternative optical-NIR prescription for the
$\Upsilon_*$ ratio based on the $V-$[3.6] colour. It provides an analysis
less sensitive to star formation (given that star formation may be more
reflective of the disc than the spheroid).
%

In Section~\ref{Sec_DaA}, we summarise the salient features of our galaxy
sample and describe the prescription for deriving their colour-dependent
$M_*/L_{obs,3.6}$ ratios.  We have also updated a few black hole masses, some
spheroid luminosities, and many galaxy distances, slightly impacting the black
hole masses and absolute magnitudes.  We provide a data table of final values
with sufficient information to trace the origin of the data readily.  In
Section~\ref{Sec_Results}, we present the $M_{\rm bh}$--$M_{\rm *,sph}$ and
$M_{\rm bh}$--$M_{\rm *,gal}$ diagrams and relations as a function of galaxy
morphology (E, ES/S0, and S).

Section~\ref{Sec_b2e} presents the size-mass relation for our sample of
spheroids and uses this to reveal how dry mergers, and the transition from bulges
to E galaxies, naturally produce the offset $M_{\rm bh}$--$M_{\rm
  *,sph}$ and $M_{\rm bh}$--$R_{\rm e,sph}$ relations for E galaxies
relative to the bulges in ES/S0 galaxies and also the offset between ES/S0 and E galaxies in the
$M_{\rm bh}$--$M_{\rm *,gal}$ diagram.  Section~\ref{Sec_Disc_2} explains the
apparent overmassive and underermassive black holes (in bulges) relative to the
original near-linear relation, with Section~\ref{Sec_Disc_3} presenting the
location of relic `red nuggets' at the top of the bulge sequence.  The
stripped S0 galaxy M32 --- the prototype for the `compact elliptical' 
galaxy class --- is discussed in Section~\ref{Sec_Disc_4}.  
Section~\ref{Sec_Disc_5} identifies and discusses what may be the primary
bivariate black hole relation in the $M_{\rm bh}$--$M_{\rm *,sph}$ diagram.
Section~\ref{Sec_Disc_6} then discusses the galaxy stellar mass function and
the (moot) role of AGN feedback in shaping it instead of potentially just
maintaining it.  Finally, several other implications are briefly mentioned in
Section~\ref{Sec_imp}.

It is important to note that the authors have been mindful of using the strict
interpretation of morphological terms in this paper.  An 
elliptical (E) galaxy has no substantial disc other than perhaps a small
nuclear disc, whereas ellicular (ES) and lenticular (S0) galaxies have an
intermediate-scale and a large-scale disc, respectively.  The expression
`early-type galaxy' (ETG) is used to generically refer to the E, ES, and S0
galaxies without a spiral pattern, while the expression `late-type galaxy'
(LTG) refers to spiral (S) and irregular (Irr) galaxies.  This notation is confined to
high-surface brightness galaxies that define the galaxy classification grid
seen in \cite{2019MNRAS.487.4995G} and built on the
Aitken-Jeans-Lundmark-Hubble galaxy sequence discussed there.
The term `spheroid' refers to both an elliptical galaxy and the bulge of a
disc galaxy, while the term bulge refers to the bulges of S, ES, and S0
galaxies but not E galaxies. The only (mild)
confusion\footnote{\citet{1966ApJ...143..192Z} pointed out that these
  elliptical-like galaxies are notably more compact than the more commonly
  known `elliptical galaxies' - many pf which turned out to be lenticular
  galaxies with large-scale discs.}  to this nomenclature is that we will
sometimes refer to relic `red nuggets' --- unevolved spheroidal-shaped
galaxies from $z\sim 2.5\pm1$ which have not acquired a large-scale disc of
stars by today --- as belonging to the bulge sequence.  Why we do this will
become apparent as one reads on.

\section{Data and Analysis}\label{Sec_DaA}

\subsection{The sample}\label{Sec_Sample}

\citet{2019ApJ...873...85D} and \citet{2019ApJ...876..155S} provide galaxy
decompositions for LTGs and ETGs with directly measured black hole masses
obtained from the literature. The galaxy decomposition process involved the
extraction of a nested set of one-dimensional profiles, including the surface
brightness profile, the ellipticity profile, the position angle profile, and
an array of profiles quantifying the amplitude of Fourier Harmonic terms used
to describe the isophotal deviations from perfect ellipses
\citep{2015ApJ...810..120C}.  These one-dimensional profiles enable accurate
two-dimensional reconstruction of the galaxy without stochastic irregularities
due to, for example, star formation or undigested neighbours.  Such
irregularities remain in the `residual image', obtained by subtracting the
smooth reconstruction from the original image, where they can more readily be
studied without the (often overwhelming) glow of the host galaxy
\citep[e.g.,][]{2021ApJ...923..146G}.  The surface brightness profile of the
geometric-mean axis\footnote{The geometric-mean axis, also know as the
  `equivalent axis' $R_{\rm eq}$, is given by the geometric-mean of the major
  (a) and minor (b) axis.  These $R_{\rm eq}=\sqrt{ab}$ radii are {\it
    equivalent} to a circularised version of a galaxy's quasi-elliptical
  isophotes.\label{footReq}} is then recreated by optimally fitting a suite of
galaxy components.  One of the advantages of this approach is that it is not
limited to models in which galaxy components may have fixed ellipticity and
position angles, as with directly fitting the two-dimensional image.  For
instance, a single-component triaxial bulge with a radially-varying
ellipticity and position angle profile might get broken into two or more
components when attempting to model it in two dimensions.

The bulk of the sample was previously imaged with the Spitzer Space Telescope at
a wavelength of 3.6~$\mu$m.  The galaxies were `disassembled' to reveal their
components and better establish the luminosity of their spheroidal component.  
Their samples were supplemented by using optical and near-IR $K_s$-band images when
the Spitzer data were either unavailable or when better spatial resolution was
required to probe the bulge component.  To keep things simple, and minimise
the introduction of possible biases, here we avoid potential offsets arising
from the use of a range of filters and thus adopted stellar mass-to-light
ratios.  We do this by solely using the galaxy sample whose structural
composition was studied at 3.6~$\mu$m.  This sample consists of 73
ETGs\footnote{This sample of 73 ETGs is comprised of 40 from
  \citet{2016ApJS..222...10S}, of which three (NGC: 821; 1399; and 3377) are
  remodelled in \cite{Graham:Sahu:22}, plus 33 from
  \citet{2019ApJ...876..155S}, of which two (NGC~2787 and NGC~5419) are
  remodelled in \cite{Graham:Sahu:22}.}  
plus 31 LTGs\footnote{\citet[][their
    Table~3]{2019ApJ...873...85D} contains 28 galaxies with Spitzer data,
  including NGC~4395 and NGC 6926 which are bulgeless, and including NGC~224
  taken from \citet{2016ApJS..222...10S}.  Two of these 28 (NGC~1320 and
  NGC~4699) are remodelled in \cite{Graham:Sahu:22}.  
A further three galaxies (NGC~2273, NGC~4945 and
  UGC~3789) from \citet{2016ApJS..222...10S} are included, taking the tally to
  31.}, coming from the larger sample of 84 ETGs \citep{2019ApJ...876..155S}
and 43 LTGs \citep{2019ApJ...873...85D}.  The smaller fraction of LTGs with
useful Spitzer data is a consequence of the need to resolve the bulge
component of the galaxy. As such, more LTGs than ETGs had previously required
Hubble Space Telescope (HST) data.

In passing, it is noted that the
(peanut shell)-shaped structures associated with buckled bars
\citep{1981A&A....96..164C, 2005MNRAS.358.1477A, 2016MNRAS.459.1276C} ---
sometimes referred to as `pseudobulges' --- were either modelled as a `barlens'
\citep[e.g.,][their Figure~3]{2011MNRAS.418.1452L, 2019ApJ...876..155S} 
%
%
or effectively folded into the Ferrers bar 
component during the galaxy decomposition, which can be seen for every galaxy
in \citet{2016ApJS..222...10S}, \citet{2019ApJ...873...85D} and
\citet{2019ApJ...876..155S}.  As noted above, we revisited the decomposition of 
seven of these (73+31=) 104 galaxies in \citet{Graham:Sahu:22}, and we use the
results here.

The distances, shown in Table~\ref{Table-data}, are
regarded as luminosity distances. 
%
%
As such, the small correction for cosmological
surface brightness dimming, 2.5$\log(1+z)^4$ is (implicitly) applied when we
convert the 3.6~$\mu$m apparent magnitudes, $m$ --- given in the above four
papers --- into absolute magnitudes, $\mathfrak{M}$,
using the expression $m-\mathfrak{M}=25+5\log\,D_L$, where $D_L$ is the luminosity
distance in Mpc.  No Galactic extinction correction has been applied to the
3.6~$\mu$m data because any excess emission from dust in the Milky Way glowing
at 3.6~$\mu$m would have effectively been removed during the sky-subtraction
procedure \citep[see Section~2.2.1 in ][]{2019ApJ...876..155S}.  Finally, no
K-correction or evolutionary corrections were applied given the small
redshifts involved, typically $z \approx$ 0.01--0.02.

The spheroid and galaxy absolute magnitudes were expressed in units of solar
luminosity using $\mathfrak{M}_{\odot,3.6} = 6.02$ (AB mag), equal to 3.26
(Vega mag).  These were then converted into stellar masses using the
prescription described in the following subsection.  These masses appear in
Table~\ref{Table-data}, along with the references to where one can see each
galaxy's decomposition.  These references are also the source for the sizes of
the spheroids, quantified using the effective half-light radius, $R_{\rm
  e,sph}$, measured along the `equivalent axis', $R_{\rm eq}$ (see footnote~\ref{footReq}).  
The masses for the black holes, updated according to the new
galaxy distances, are also provided in Table~\ref{Table-data}.  Unless an
update is indicated, the nearly 100 references for these black hole masses can
be traced through \citet{2020ApJ...903...97S}.

\begin{figure*}
\begin{center}
\includegraphics[trim=0.0cm 0cm 0.0cm 0cm, height=0.35\textwidth,
  angle=0]{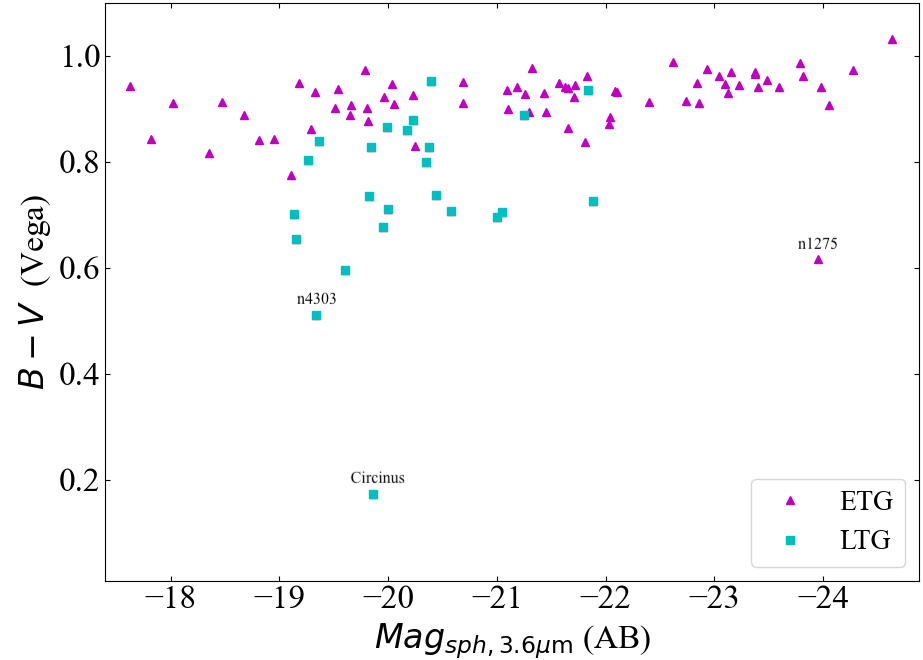} 
\includegraphics[trim=0.0cm 0cm 0.0cm 0cm, height=0.35\textwidth,
  angle=0]{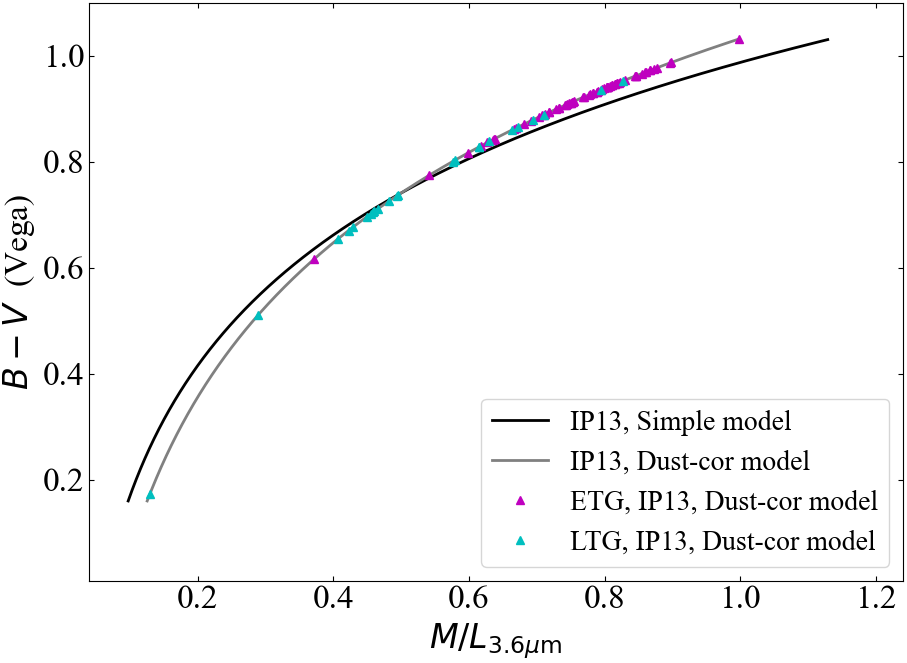}
\caption{Left-hand panel: (Galactic extinction)-corrected $B-V$ galaxy colour (Vega mag)
  versus the 3.6~$\mu$m absolute magnitude $\mathfrak{M}_{3.6}$ (AB mag) of
  the spheroid.  Both NGC~1275 and Circinus are close to the Galactic plane.
  Right-hand panel: The thick curve (with the triangles) shows the same $B-V$ colour versus the logarithm
  of the stellar mass-to-light ratio at 3.6~$\mu$m obtained from
  Equation~\ref{Eq_MonL_IP13}, adapted from the dusty galaxy model of
  \citet[][their Table~6]{2013MNRAS.430.2715I}. For reference, we also show
  with a thin curve their simple galaxy model (their Table~3, our
    Equation~\ref{Eq_MonL_simple}).  Both models have
  been converted here to a \citet{2002Sci...295...82K} IMF, see
  Section~\ref{Sec_IMF} for details.  (Circinus falls below the blue-end cut-off
  of $B-V=0.5$ for the applicable range of these mass-to-light ratio
  equations.) 
}
\label{Fig_colour_IP13} 
\end{center}
\end{figure*}

Following the exclusion of mergers by \citet{2013ARA&A..51..511K}, 
we exclude from the upcoming Bayesian linear regressions, but not the plots, 
one LTG plus four ETGs 
previously identified by others as somewhat unrelaxed mergers (NGC~2960 plus NGC~1194, NGC~1316,
NGC~5018 and NGC~5128).  We additionally exclude the stripped
galaxy NGC~4342 \citep{2014MNRAS.439.2420B} and the dwarf galaxy NGC~404
\citep[Mirach's Ghost: ][]{2017ApJ...836..237N}\footnote{NGC~404 lies 
within seven arcmin of the second magnitude star Mirach.}. NGC~404 is the only
galaxy in our sample with 
$M_{\rm bh} < 10^6$ M$_\odot$, thereby making it vulnerable to potentially biasing the
analyses due to the weight it may have in torquing the regression lines. 
The ETG with the suspiciously\footnote{NGC~1275 resides 13
  degrees from the Galactic plane and has $\sim$0.6 mag of Galactic extinction
  in the $B$-band.} blue colour of $\sim$0.6 in Figure~\ref{Fig_colour_IP13} is NGC~1275, although we left
this galaxy in the sample as its inclusion/exclusion had no appreciable impact.  
We also included the rather blue LTG NGC~4303 but needed to exclude 
NGC~5055 due to its uncertain black hole mass \citep{2004A&A...420..147B,
    2021MNRAS.500.1933S} and Circinus, an unrelaxed S galaxy known to be
undergoing considerable starbursts in addition to hosting an AGN. The 
$B-V=0.174$ (Vega) colour of Circinus is less than 0.5 and well outside of the applicability
range of the $M/L$ equations we are about to introduce. 
The nine galaxies excluded from the linear regression are marked with a dagger
symbol ($\dagger$) in Table~\ref{Table-data}, which includes all 104 galaxies
initially considered here.

\begin{table*}
\centering
\caption{Black hole, spheroid and galaxy masses, and spheroid sizes}\label{Table-data}
\begin{tabular}{llllclccccl} 
\hline
Galaxy   &  Type    & Dist.\ (Mpc) &  kpc/$\arcsec$ & $R_{\rm e,sph,equiv}$ (kpc) &  $\log(M_{\rm bh}/M_\odot)$   &    $B-V$   &  $V-$[3.6]  & 
$\log(M_{\rm *,sph}/M_\odot)$    &    $\log(M_{\rm *,gal}/M_\odot)$  &  Ref. \\
 (1)  &   (2)   &  (3)   &  (4)  &  (5)   &  (6)  &  (7)  & (8) & (9) & (10) & (11)  \\
\hline 
Circinus$^{\dagger}$ &  S       & 4.21$\pm$0.76 (T08)         & 0.020  & 0.46  & 6.25$\pm$0.11     & 0.174  &  2.41  & 9.46$\pm$0.29   &  9.97$\pm$0.22 & DGC19\\
IC 1459  &  E       & 28.1$\pm$3.6 (T01$^{\prime}$) & 0.134  & 7.68  & 9.38$\pm$0.20     & 0.966  &  3.84  & 11.69$\pm$0.17  & 11.69$\pm$0.17 & SG16\\
IC 2560  &  S       & 32.9$\pm$5.0 (NED$^i$)      & 0.157  & 0.62  & 6.52$\pm$0.11     & [0.70] & [3.17] & 9.66$\pm$0.23   & 10.69$\pm$0.18 & DGC19\\
IC 4296$^*$  &  E   & 46.9$\pm$3.7 (M00$^{\prime}$) & 0.222  & 9.12  & 9.10$\pm$0.09     & 0.954  &  3.36  & 11.72$\pm$0.14  & 11.74$\pm$0.14 & SGD19\\
NGC 0224 &  S       & 0.75$\pm$0.03 (R12)         & 0.004  & 0.69  & 8.15$\pm$0.16     & 0.865  &  3.58  & 10.23$\pm$0.15  & 11.01$\pm$0.13 & SG16\\
NGC 0253 &  S       & 3.47$\pm$0.11 (R11)         & 0.017  & 0.47  & 7.00$\pm$0.30     & [0.70] & [3.19] & 9.76$\pm$0.25   & 10.71$\pm$0.13 & DGC19\\
NGC 0404$^{\dagger}$ &  ES/S0   & 3.06$\pm$0.37 (K02)         & 0.015  & 0.058 & 5.74$\pm$0.1 [4f] & 0.889  &  2.77  & 8.03$\pm$0.50   &  9.19$\pm$0.16 & SGD19\\
NGC 0524 &  ES/S0   & 27.7$\pm$1.1 (J21)          & 0.132  & 1.10  & 9.00$\pm$0.10     & 0.977  & 3.13   & 10.88$\pm$0.15  & 11.38$\pm$0.13 & SGD19\\
NGC 0821 &  E       & 23.2$\pm$1.8 (T01$^{\prime}$) & 0.111  & 3.45  & 7.59$\pm$0.17     & 0.893  & 2.90   & 10.84$\pm$0.15  & 10.90$\pm$0.14 & GS22\\
NGC 1023 &  ES/S0   & 11.0$\pm$0.8 (T01$^{\prime}$) & 0.053  & 0.39  & 7.62$\pm$0.05     & 0.946  & 3.15   & 10.33$\pm$0.16  & 10.89$\pm$0.14 & SG16\\
NGC 1097 &  S       & 24.8$\pm$0.6 (R14)          & 0.119  & 1.35  & 8.38$\pm$0.04     & 0.726  & 3.52   & 10.84$\pm$0.25  & 11.42$\pm$0.13 & DGC19\\
NGC 1194$^{\dagger}$ &  ES/S0   & 54.1$\pm$3.8 (NED)          & 0.256  & 0.91  & 7.82$\pm$0.04     & 0.893  & 3.61   & 10.78$\pm$0.16  & 11.01$\pm$0.14 & SGD19\\
NGC 1275 &  E       & 69.0$\pm$9.0 (NED$^i$)      & 0.324  & 17.4  & 8.88$\pm$0.21     & 0.616  & 4.04   & 11.56$\pm$0.18  & 11.60$\pm$0.17 & SGD19\\
NGC 1300 &  S       & 20.7$\pm$1.5 (NED)          & 0.100  & 0.74  & 7.86$\pm$0.14     & 0.654  & 2.86   & 9.68$\pm$0.20   & 10.55$\pm$0.14 & DGC19\\
NGC 1316$^{\dagger}$ &  ES/S0   & 17.8$\pm$5.9 (NED$^i$)      & 0.086  & 1.37  & 8.16$\pm$0.29     & 0.871  & 3.60   & 11.05$\pm$0.35  & 11.69$\pm$0.31 & SG16\\
NGC 1320 &  S       & 36.8$\pm$2.6 (NED)          & 0.175  & 0.20  & 6.77$\pm$0.22     & 0.828  & 3.50   & 10.13$\pm$0.16  & 10.70$\pm$0.14 & GS22 \\
NGC 1332 &  ES/S0   & 22.0$\pm$1.8 (T01$^{\prime}$) & 0.105  & 1.89  & 9.15$\pm$0.07     & 0.932  & 3.41   & 11.15$\pm$0.15  & 11.17$\pm$0.14 & SG16\\
NGC 1374 &  ES/S0   & 19.0$\pm$1.1 (T01$^{\prime}$) & 0.091  & 1.07  & 8.76$\pm$0.05     & 0.908  & 3.19   & 10.30$\pm$0.16  & 10.59$\pm$0.13 & SGD19\\
NGC 1398 &  S       & 24.8$\pm$4.5 (T08)          & 0.119  & 1.24  & 8.03$\pm$0.11     & 0.888  & 3.36   & 10.76$\pm$0.29  & 11.44$\pm$0.20 & DGC19\\
NGC 1399 &  E       & 19.2$\pm$1.4 (T01$^{\prime}$) & 0.092  & 5.71  & 8.67$\pm$0.06     & 0.949  & 3.74   & 11.46$\pm$0.16  & 11.46$\pm$0.16 & GS22\\
NGC 1407 &  E       & 27.8$\pm$3.3 (T01$^{\prime}$) & 0.133  & 6.29  & 9.65$\pm$0.08     & 0.969  & 3.31   & 11.60$\pm$0.17  & 11.66$\pm$0.16 & SGD19\\
NGC 1600 &  E       & 71.7$\pm$2.7 (J21)          & 0.336  & 16.7  & 10.25$\pm$0.04    & 0.972  & 3.56   & 12.06$\pm$0.13  & 12.06$\pm$0.13 & SGD19\\
NGC 2273 &  S       & 30.3$\pm$4.0 (NED$^i$)      & 0.145  & 0.28  & 6.95$\pm$0.06     & 0.828  &  3.41  & 10.35$\pm$0.22  & 10.83$\pm$0.17 & SG16\\
NGC 2549 &  ES/S0   & 12.2$\pm$1.6 (T01$^{\prime}$) & 0.059  & 0.18  & 7.15$\pm$0.60     & 0.913  & 3.19   & 9.67$\pm$0.19   & 10.21$\pm$0.17 & SG16\\
NGC 2778 &  ES/S0   & 22.1$\pm$3.1 (T01$^{\prime}$) & 0.106  & 0.23  & 7.18$\pm$0.35     & 0.911  & 3.06   & 9.49$\pm$0.23   & 10.15$\pm$0.17 & SG16\\
NGC 2787 &  ES/S0   &  7.2$\pm$1.2 (T01$^{\prime}$) & 0.035  & 0.14  & 7.59$\pm$0.09     & 0.942  & 3.34   & 9.37$\pm$0.24   & 10.10$\pm$0.19 & GS22\\
NGC 2960$^{\dagger}$ &  S       & 73.0$\pm$5.1 (NED)          & 0.342  & 0.75  & 7.07$\pm$0.05     & [0.70] & [3.34] & 10.44$\pm$0.16  & 10.86$\pm$0.14 & DGC19\\
NGC 2974 &  S       & 21.5$\pm$2.5 (T08)          & 0.103  & 0.67  & 8.23$\pm$0.07     & 0.952  & 3.45   & 10.48$\pm$0.22  & 10.98$\pm$0.16 & DGC19\\
NGC 3031 &  S       &  3.48$\pm$0.13 (K12)        & 0.017  & 0.73  & 7.83$\pm$0.09     & 0.879  & 3.21   & 10.34$\pm$0.25  & 10.83$\pm$0.13 & DGC19\\
NGC 3079 &  S       & 16.5$\pm$2.9 (T08)          & 0.079  & 0.34  & 6.38$\pm$0.12     & 0.670  & 4.03   & 9.88$\pm$0.29   & 10.64$\pm$0.20 & DGC19\\
NGC 3091 &  E       & 58.6$\pm$10.9 (T07)         & 0.276  & 14.1  & 9.62$\pm$0.08     & 0.962  & 3.74   & 11.86$\pm$0.20  & 11.86$\pm$0.20 & SG16\\
NGC 3115 &  ES/S0   &  9.3$\pm$0.4 (T01$^{\prime}$) & 0.045  & 1.55  & 8.94$\pm$0.25     & 0.929  & 3.30   & 10.87$\pm$0.14  & 10.95$\pm$0.13 & SG16\\
NGC 3227 &  S       & 25.7$\pm$3.2 (K15)          & 0.123  & 1.03  & 7.97$\pm$0.14     & 0.800  & 3.19   & 10.31$\pm$0.22  & 11.07$\pm$0.16 & DGC19\\
NGC 3245 &  ES/S0   & 20.1$\pm$1.9 (T01$^{\prime}$) & 0.097  & 0.23  & 8.30$\pm$0.12     & 0.888  & 3.06   & 10.12$\pm$0.17  & 10.70$\pm$0.15 & SG16\\
NGC 3368 &  S       & 10.8$\pm$1.5 (NED$^i$)      & 0.052  & 0.25  & 6.89$\pm$0.11     & 0.838  & 3.33   &  9.95$\pm$0.17  & 10.83$\pm$0.17 & DGC19\\
NGC 3377 &  E       & 10.8$\pm$0.4 (T01$^{\prime}$) & 0.052  & 2.30  & 7.89$\pm$0.03     & 0.830  & 3.29   & 10.30$\pm$0.14  & 10.36$\pm$0.13 & GS22\\
NGC 3379 &  E       & 10.9$\pm$1.6 (K15)          & 0.053  & 2.70  & 8.62$\pm$0.13     & 0.938  & 3.44   & 10.97$\pm$0.20  & 10.97$\pm$0.20 & SG16\\
NGC 3384 &  ES/S0   & 11.2$\pm$0.7 (T01$^{\prime}$) & 0.054  & 0.30  & 7.23$\pm$0.05     & 0.907  & 3.33   & 10.14$\pm$0.16  & 10.67$\pm$0.14 & SG16\\
NGC 3414 &  E       & 24.3$\pm$3.7 (T01$^{\prime}$) & 0.117  & 2.98  & 8.38$\pm$0.09     & 0.948  & 3.39   & 10.95$\pm$0.19  & 10.98$\pm$0.18 & SG16\\
NGC 3489 &  ES/S0   & 11.6$\pm$0.8 (T01$^{\prime}$) & 0.056  & 0.095 & 6.76$\pm$0.07     & 0.816  & 2.98   & 9.53$\pm$0.20   & 10.30$\pm$0.14 & SG16\\
NGC 3585 &  E       & 19.3$\pm$1.6 (T01$^{\prime}$) & 0.093  & 8.03  & 8.49$\pm$0.13     & 0.914  & 3.84   & 11.38$\pm$0.15  & 11.39$\pm$0.14 & SG16\\
NGC 3607 &  E       & 25.0$\pm$3.2 (K15)          & 0.120  & 7.86  & 8.16$\pm$0.18     & 0.911  & 3.54   & 11.43$\pm$0.17  & 11.46$\pm$0.17 & SG16\\
NGC 3608 &  E       & 22.1$\pm$1.4 (T01$^{\prime}$) & 0.106  & 4.60  & 8.30$\pm$0.17     & 0.922  & 3.45   & 10.98$\pm$0.14  & 10.98$\pm$0.14 & SG16\\
NGC 3627 &  S       & 10.4$\pm$1.8 (K15)          & 0.050  & 0.20  & 6.94$\pm$0.09     & 0.701  & 3.24   & 9.72$\pm$0.21   & 10.76$\pm$0.20 & DGC19\\
NGC 3665 &  ES/S0   & 34.7$\pm$2.4 (T07)          & 0.166  & 2.12  & 8.76$\pm$0.10     & 0.933  & 3.54   & 11.14$\pm$0.16  & 11.39$\pm$0.14 & SGD19\\
NGC 3842 &  E       & 87.5$\pm$4.1 (J21)          & 0.407  & 30.0  & 9.94$\pm$0.13     & 0.941  & 3.84   & 11.91$\pm$0.14  & 11.93$\pm$0.13 & SG16\\
NGC 3923 &  E       & 22.1$\pm$2.9 (T01$^{\prime}$) & 0.106  & 8.35  & 9.47$\pm$0.13     & 0.929  & 3.74   & 11.55$\pm$0.17  & 11.55$\pm$0.17 & SGD19\\
NGC 3998 &  ES/S0   & 13.6$\pm$1.2 (T01$^{\prime}$) & 0.065  & 0.31  & 8.33$\pm$0.43[4a] & 0.936  & 3.47   & 10.12$\pm$0.26  & 10.61$\pm$0.15 & SG16\\
NGC 4026 &  ES/S0   & 13.2$\pm$1.7 (T01$^{\prime}$) & 0.063  & 0.15  & 8.26$\pm$0.12     & 0.901  & 3.27   & 10.19$\pm$0.22  & 10.44$\pm$0.17 & SGD19\\
NGC 4151 &  S       & 19.0$\pm$2.5 (H14)          & 0.094  & 0.56  & 7.69$\pm$0.37     & 0.706  & 3.44   & 10.28$\pm$0.22  & 10.62$\pm$0.17 & DGC19\\
NGC 4258 &  S       &  7.6$\pm$0.17 (H13)         & 0.037  & 0.98  & 7.60$\pm$0.01     & 0.676  & 3.37   & 10.02$\pm$0.19  & 10.70$\pm$0.13 & DGC19\\
NGC 4261 &  E       & 30.4$\pm$2.7 (T01$^{\prime}$) & 0.145  & 6.86  & 9.20$\pm$0.09[4b] & 0.975  & 3.69   & 11.52$\pm$0.16  & 11.54$\pm$0.15 & SG16\\
NGC 4291 &  E       & 25.2$\pm$3.7 (T01$^{\prime}$) & 0.121  & 1.86  & 8.51$\pm$0.37     & 0.927  & 3.38   & 10.80$\pm$0.18  & 10.80$\pm$0.18 & SG16\\
NGC 4303 &  S       & 19.3$\pm$0.6 (R14)          & 0.093  & 0.20  & 6.78$\pm$0.17     & 0.510  & 2.90   & 9.60$\pm$0.25   & 10.66$\pm$0.13 & DGC19\\
NGC 4339 &  ES/S0   & 15.8$\pm$1.3 (T01$^{\prime}$) & 0.076  & 0.49  & 7.62$\pm$0.33     & 0.887  & 2.96   & 9.73$\pm$0.21   & 10.22$\pm$0.14 & SGD19\\
NGC 4342$^{\dagger}$ &  ES/S0   & 22.8$\pm$0.8 (B09$^{\prime}$) & 0.110  & 0.52  & 8.65$\pm$0.18     & 0.932  & 3.52   & 10.04$\pm$0.15  & 10.36$\pm$0.15 & SGD19\\
NGC 4350 &  ES/S0   & 17.0$\pm$0.5 (B15)          & 0.082  & 1.59  & 8.87$\pm$0.41     & 0.926  & 3.43   & 10.39$\pm$0.25  & 10.66$\pm$0.13 & SGD19\\
NGC 4371 &  ES/S0   & 16.4$\pm$3.4 (T07)          & 0.079  & 0.70  & 6.84$\pm$0.12     & 0.948  & 3.34   & 9.99$\pm$0.30   & 10.70$\pm$0.22 & SGD19\\
NGC 4374 &  E       & 17.7$\pm$0.9 (T01$^{\prime}$) & 0.085  & 11.0 & 8.95$\pm$0.05     & 0.944  & 3.75  & 11.61$\pm$0.15   & 11.61$\pm$0.15 & SG16\\
NGC 4388 &  S       & 18.0$\pm$3.6 (S14)          & 0.087  & 1.24 & 6.90$\pm$0.10     & 0.711  & 3.34  & 10.07$\pm$0.30  & 10.45$\pm$0.21 & DGC19\\
NGC 4395 &  S       & 4.56$\pm$0.17 (S18)         & 0.022  &  ...  & 5.62$\pm$0.17     & 0.445  &  2.96  &   ...           &  9.15$\pm$0.13 & DGC19 \\
NGC 4429 &  ES/S0   & 16.6$\pm$0.8 (C08)          & 0.080  & 0.90 & 8.18$\pm$0.08     & 0.950  & 3.40  & 10.60$\pm$0.20  & 11.04$\pm$0.13 & SGD19\\
\hline
\end{tabular}
\end{table*}

\setcounter{table}{0}

\begin{table*}
\centering
\caption{Continued}
\begin{tabular}{llllclccccl}
\hline
Galaxy   &  Type    & Dist.\ (Mpc) &  kpc/$\arcsec$ &  $R_{\rm e,sph,equiv}$
(kpc) &  $\log(M_{\rm bh}/M_\odot)$   &    $B-V$   &  $V-$[3.6] &
$\log(M_{\rm *,sph}/M_\odot)$    &    $\log(M_{\rm *,gal}/M_\odot)$  & Ref. \\
   (1)   &   (2)    &  (3)                        & (4)    & (5)  &     (6)           & (7)    &  (8) &     (9)         & (10)           & (11)\\
\hline 
NGC 4434 &  ES/S0   & 22.3$\pm$0.7 (B09$^{\prime}$) & 0.107  & 0.57 & 7.85$\pm$0.17     & 0.861  & 3.07  & 9.95$\pm$0.20   & 10.22$\pm$0.13 & SGD19\\
NGC 4459 &  ES/S0   & 15.5$\pm$1.6 (T01$^{\prime}$) & 0.075  & 0.98 & 7.82$\pm$0.10     & 0.910  & 3.28  & 10.56$\pm$0.21  & 10.77$\pm$0.15 & SG16\\
NGC 4472 &  E       & 15.7$\pm$0.7 (T01$^{\prime}$) & 0.075  & 10.2 & 9.36$\pm$0.04     & 0.940  & 3.72  & 11.75$\pm$0.13  & 11.75$\pm$0.13 & SG16\\
NGC 4473 &  E       & 15.1$\pm$0.9 (T01$^{\prime}$) & 0.073  & 2.69 & 8.07$\pm$0.36     & 0.935  & 3.30  & 10.75$\pm$0.13  & 10.83$\pm$0.13 & SG16\\
NGC 4486 &  E       & 16.8$\pm$0.8 (EHT)          & 0.081  & 7.06 & 9.81$\pm$0.06[4c] & 0.940  & 3.60  & 11.67$\pm$0.15  & 11.67$\pm$0.15 & SG16\\
NGC 4501 &  S       & 17.0$\pm$0.5 (B15)          & 0.082  & 1.67 & 7.31$\pm$0.08     & 0.696  & 3.53  & 10.46$\pm$0.25  & 11.02$\pm$0.13 & DGC19\\
NGC 4526 &  ES/S0   & 16.3$\pm$1.5 (T01$^{\prime}$) & 0.078  & 1.16 & 8.65$\pm$0.04     & 0.940  & 3.38  & 10.79$\pm$0.26  & 11.13$\pm$0.15 & SGD19\\
NGC 4552 &  E       & 14.8$\pm$1.0 (T01$^{\prime}$) & 0.071  & 5.08 & 8.67$\pm$0.05     & 0.944  & 3.44  & 11.01$\pm$0.16  & 11.07$\pm$0.16 & SGD19\\
NGC 4564 &  ES/S0   & 14.4$\pm$1.1 (T01$^{\prime}$) & 0.069  & 0.41 & 7.77$\pm$0.06     & 0.901  & 3.20  & 10.08$\pm$0.16  & 10.35$\pm$0.14 & SG16\\
NGC 4578 &  ES/S0   & 16.2$\pm$0.5 (B09$^{\prime}$) & 0.078  & 0.49 & 7.28$\pm$0.35     & 0.842  & 3.25  & 9.79$\pm$0.15   & 10.24$\pm$0.13 & SGD19\\
NGC 4594 &  S       &  9.55$\pm$0.44 (Mc16)       & 0.046  & 1.90 & 8.81$\pm$0.03     & 0.935  & 3.11  & 11.04$\pm$0.25  & 11.26$\pm$0.13 & DGC19\\
NGC 4596 &  ES/S0   & 17.0$\pm$0.5 (B15)          & 0.082  & 0.74 & 7.90$\pm$0.20     & 0.921  & 3.37  & 10.28$\pm$0.20  & 10.86$\pm$0.13 & SG16\\
NGC 4621 &  E       & 17.6$\pm$1.6 (T01$^{\prime}$) & 0.084  & 7.64 & 8.59$\pm$0.06     & 0.912  & 3.56  & 11.24$\pm$0.16  & 11.28$\pm$0.15 & SG16\\
NGC 4649 &  E       & 16.2$\pm$1.1 (T01$^{\prime}$) & 0.078  & 6.29 & 9.66$\pm$0.10     & 0.946  & 3.58  & 11.56$\pm$0.14  & 11.56$\pm$0.14 & SGD19\\
NGC 4697 &  E       & 11.3$\pm$0.7 (T01$^{\prime}$) & 0.054  & 12.2 & 8.26$\pm$0.04     & 0.884  & 3.81  & 11.07$\pm$0.15  & 11.12$\pm$0.14 & SG16\\
NGC 4699 &  S       & 20.4$\pm$3.8 (K13)          & 0.098  & 0.23 & 8.27$\pm$0.09     & 0.860  & 3.24  & 10.30$\pm$0.25  & 11.27$\pm$0.20 & GS22 \\
NGC 4736 &  S       &  5.0$\pm$0.4 (T01$^{\prime}$) & 0.024  & 0.23 & 6.83$\pm$0.11     & 0.735  & 3.48  & 10.03$\pm$0.21  & 10.51$\pm$0.14 & DGC19\\
NGC 4742 &  ES/S0   & 14.9$\pm$1.1 (T01$^{\prime}$) & 0.072  & 0.25 & 7.13$\pm$0.18     & 0.774  & 2.91  &  9.78$\pm$0.16  & 10.07$\pm$0.14 & SGD19\\
NGC 4762 &  ES/S0   & 17.0$\pm$0.5 (B15)          & 0.082  & 0.18 & 7.24$\pm$0.14     & 0.841  & 3.35  & 9.74$\pm$0.15   & 10.83$\pm$0.13 & SGD19\\
NGC 4826 &  S       &  7.2$\pm$0.7 (T01$^{\prime}$) & 0.035  & 0.42 & 6.18$\pm$0.12     & 0.803  & 3.31  & 9.88$\pm$0.21   & 10.74$\pm$0.15 & DGC19\\
NGC 4889 &  E       & 96.3$\pm$6.7 (NED)          & 0.446  & 27.1 & 10.3$\pm$0.44     & 1.031  & 3.93  & 12.26$\pm$0.14  & 12.26$\pm$0.14 & SG16\\
NGC 4945 &  S       &  3.56$\pm$0.20 (Mo16)       & 0.017  & 0.16  & 6.13$\pm$0.30     & [0.70] & [3.09] &  9.29$\pm$0.20  & 10.42$\pm$0.13 & SG16\\ 
NGC 5018$^{\dagger}$ &  ES/S0   & 38.4$\pm$2.7 (NED)          & 0.183  & 1.13 & 8.00$\pm$0.08     & 0.836  & 3.10  & 10.93$\pm$0.16  & 11.31$\pm$0.14 & SGD19\\
NGC 5055$^{\dagger}$ &  S       &  8.87$\pm$0.39 (M17)        & 0.043  & 1.87 & 8.94$\pm$0.10     & 0.704  & 3.41  & 10.49$\pm$0.25  & 10.81$\pm$0.13 & DGC19\\
NGC 5077 &  E       & 39.8$\pm$7.4 (T07)          & 0.189  & 4.35 & 8.85$\pm$0.23     & 0.987  & 3.60  & 11.41$\pm$0.20  & 11.41$\pm$0.20 & SG16\\
NGC 5128$^{\dagger}$ &  ES/S0   &  3.76$\pm$0.05 (K07)        & 0.018  & 1.09 & 7.65$\pm$0.12     & 0.899  & 3.60  & 10.71$\pm$0.25  & 11.14$\pm$0.12 & SG16\\
NGC 5252 &  ES/S0   & 104.0$\pm$7.3 (NED)         & 0.480  & 0.71 & 9.03$\pm$0.40     & [0.90] & [3.42] & 10.97$\pm$0.27 & 11.50$\pm$0.15 & SGD19\\
NGC 5419 &  E       & 57.0$\pm$4.0 (NED)          & 0.269  & 10.8 & 9.87$\pm$0.14     & 0.986  & 3.42  & 11.87$\pm$0.16  & 11.87$\pm$0.14 & GS22\\
NGC 5576 &  E       & 24.5$\pm$1.6 (T01$^{\prime}$) & 0.118  & 5.82 & 8.19$\pm$0.10     & 0.863  & 3.35  & 10.90$\pm$0.16  & 10.90$\pm$0.16 & SG16\\
NGC 5813 &  ES/S0   & 31.0$\pm$2.6 (T01$^{\prime}$) & 0.148  & 2.10 & 8.83$\pm$0.06     & 0.940  & 3.17  & 10.96$\pm$0.16  & 11.34$\pm$0.14 & SGD19\\
NGC 5845 &  ES/S0   & 25.0$\pm$2.4 (T01$^{\prime}$) & 0.120  & 0.63 & 8.41$\pm$0.22     & 0.972  & 3.38  & 10.26$\pm$0.21  & 10.46$\pm$0.15 & SGD19\\
NGC 5846 &  E       & 24.0$\pm$2.2 (T01$^{\prime}$) & 0.115  & 9.59 & 9.04$\pm$0.06     & 0.961  & 3.79  & 11.55$\pm$0.15  & 11.55$\pm$0.15 & SG16\\
NGC 6251 &  E       & 104.6$\pm$7.3 (NED)         & 0.483  & 14.5 & 8.77$\pm$0.16     & [0.90] & [3.62] & 11.87$\pm$0.15 & 11.87$\pm$0.15 & SG16\\
NGC 6861 &  ES/S0   & 27.0$\pm$4.0 (T01$^{\prime}$) & 0.129  & 2.60 & 9.30$\pm$0.08     & 0.962  & 3.61  & 11.07$\pm$0.19  & 11.15$\pm$0.18 & SGD19\\
NGC 6926 &  S       & 85.6$\pm$6.0 (NED)          & 0.399  &  ...  & 7.68$\pm$0.50 [4d] & 0.586 &  3.02  &   ...           &  11.13$\pm$0.14 & DGC19 \\
NGC 7052 &  E       & 61.9$\pm$2.6 (J21)          & 0.291  & 5.83 & 9.35$\pm$0.05[4e] & [0.90] & [3.53] & 11.46$\pm$0.13 & 11.46$\pm$0.13 & SGD19\\
NGC 7332 &  ES/S0   & 22.2$\pm$2.0 (T01$^{\prime}$) & 0.106  & 0.26 & 7.06$\pm$0.20     & 0.877  & 3.40  & 10.17$\pm$0.17  & 10.79$\pm$0.15 & SGD19\\
NGC 7457 &  ES/S0   & 12.7$\pm$1.2 (T01$^{\prime}$) & 0.061  & 0.40 & 6.96$\pm$0.30     & 0.843  & 3.07  & 9.34$\pm$0.17   & 10.12$\pm$0.15 & SGD19\\
NGC 7582 &  S       & 22.2$\pm$4.0 (N11)          & 0.106  & 0.48 & 7.72$\pm$0.12     & 0.737  & 3.60  & 10.28$\pm$0.29  & 10.90$\pm$0.20 & DGC19\\
NGC 7619 &  E       & 46.6$\pm$1.7 (J21)          & 0.221  & 12.8 & 9.36$\pm$0.09     & 0.968  & 3.69  & 11.69$\pm$0.14  & 11.71$\pm$0.13 & SG16\\
NGC 7768 &  E       & 108.2$\pm$7.6 (NED)         & 0.499  & 21.0 & 9.09$\pm$0.15     & 0.906  & 3.83  & 11.90$\pm$0.16    & 11.90$\pm$0.16 & SG16\\
UGC 3789 &  S       & 50.7$\pm$5.2 (R13)          & 0.240  & 0.58  & 7.07$\pm$0.05     & [0.70] & [3.27] & 10.11$\pm$0.26  & 10.68$\pm$0.15 & SG16\\
\hline
\end{tabular}

Column~1: Galaxy name ($^{\dagger}$ Excluded from the Bayesian linear regression. $^*$ IC~4296 = Abell 3565-BCG). 
Column~2: Broad galaxy type from \citet[][their Table~A1]{2020ApJ...903...97S}.  
Column~3: Adopted `luminosity distance' in Mpc: 
T01$^{\prime}$ = \citet{2001ApJ...546..681T} with a 0.083 mag reduction to their distance moduli (see Section~\ref{Sec_IMF}); 
T08 = \citet{2008ApJ...676..184T, 2009AJ....138..323T};  
Y12 = \citet{2012PASJ...64..103Y}; 
M00$^{\prime}$ = \citet{2000A&A...361...68M} with a 0.083 mag reduction to their distance moduli; 
R12 = \citet{2012ApJ...745..156R}; 
R11 = \citet{2011ApJS..195...18R}; 
K02 = \citet{2002A&A...389..812K}; 
R14 = \citet{2014AJ....148..107R}; 
R13 = \citet{2013ApJ...767..154R}; 
NED$^i$ = median redshift-independent distance from NED (and the
standard deviation associated with the {\it mean} redshift-independent distance); 
NED = (Virgo + GA + Shapley)-corrected Hubble flow distance based on
$H_0=73$ km s$^{-1}$ Mpc$^{-1}$;  
T07 = \citet{2007A&A...465...71T} mean value via NED; 
K12 = \citet{2012ApJ...747...15K}; 
K15 = \citet{2015AstBu..70....1K}; 
H14 = \citet{2014Natur.515..528H}; 
H13 = \citet{2013ApJ...775...13H}; 
B09$^{\prime}$ = \citet{2009ApJ...694..556B} with a 0.023 mag reduction to
  their distance moduli (see Section~\ref{Sec_IMF}), and using the distance for NGC~4365 in the
  case of NGC~4342  \citep{2014MNRAS.439.2420B}; 
S14 = \citet{2014MNRAS.444..527S}; 
S18 = \citet{2018ApJS..235...23S}; 
C08 = \citet{2008ApJ...683...78C}; 
EHT = \citet{2019ApJ...875L...1E}; 
Mc16 = \citet{2016AJ....152..144M}; 
Mo16 \ \citet{2016MNRAS.457.1419M}; 
B15 = \citep{2015A&A...579A.102B}; 
K13 = \citet{2013MNRAS.429.2677K}; 
J21 = \citet{2021ApJS..255...21J}; 
M17 = \citet{2017AJ....154...51M}; 
K07 = \citet{2007AJ....133..504K}; 
N11 = \citet{2011A&A...532A.104N}. 
Column~4: Scale size conversion based on the `angular-size distance' (not shown) and using 
$H_0=73$ km s$^{-1}$ Mpc$^{-1}$, $\Omega_m=0.3$ and $\Omega_{\Lambda}=0.7$. 
Column~5: Effective half-light radius of the spheroid, derived from the
geometric-mean axis, 
equivalent to a circularised version of the quasi-elliptical isophotes. 
Column~6: Black hole masses from the compilation in
\citet[][and the reference-chain therein]{2020ApJ...903...97S}, after
adjusting to the distances in Column~3. 
Exceptions are: 
4a = \citet{2018MNRAS.473.2930D}; 
4b = \citet{2021ApJ...908...19B}; 
4c = \citet{2019ApJ...875L...6E}; 
4d = \citet[][upper limit]{2018ApJ...854..124Z}; 
4e = \citet{2021MNRAS.503.5984S}; and 
4f = \citet{2020MNRAS.496.4061D}. 
Columns~7 and 8: (Galactic extinction)-corrected $B-V$ and $V-$[3.6] colours, obtained from
NED. The former are used in Equation~\ref{Eq_MonL_IP13} 
and the latter in the Appendix equation~\ref{Eq-Schom-21}, 
to derive the 3.6~$\mu$m
stellar mass-to-light ratios for calculating the stellar-masses.
Values in [brackets] are assumed rather than measured. 
Column~9: Spheroid stellar mass based on the 
Spitzer apparent magnitude reported by either: 
SG16 \citep{2016ApJ...817...21S}; 
DGC19 \citep{2019ApJ...873...85D}; 
SGD19 \citep{2019ApJ...887...10S}; or 
GS22 \citep{Graham:Sahu:22}.  
The revised distances, colour-dependent $M_*/L$ ratios, and updated magnitudes for
seven systems presented in SG22, results in the updates, shown here, for the stellar 
masses reported in \citet{2020ApJ...903...97S}. 
Column~10: Updated galaxy stellar mass. 
Column~11: Reference displaying the multicomponent decomposition used to obtain
both the spheroid magnitude and size, and the galaxy magnitude. 

\end{table*}

\subsection{Stellar mass-to-light ratios}\label{Sec_IP13} 

As illustrated by, for example, \citet[][their Figure~1]{2014AJ....148...77M},
\citet[][their Section~7]{2017ApJS..233...13Z}, and 
%
%
\citet[][their Figure~4]{2019ApJ...876..155S}, the colour-dependent
mass-to-light ratio prescriptions from different papers do not agree with each
other.  Even after correcting for the different assumptions in the stellar
population models, the equations from different papers do not agree
\citep[][their Figure~6]{2014AJ....148...77M}.  \citet[][their
  Figure~10]{2019MNRAS.483.1496S} present half a dozen ($B-V$)-dependent
relations for the mass-to-light ratio.  The relation from
\citet{2013MNRAS.430.2715I} sits in the middle 
and is therefore adopted here as a middle ground.
In the Appendix, we additionally show the result of adopting the latest relation from
\citet{2022arXiv220202290S}. 

We have taken the $B_T$ and $V_T$ total galaxy magnitudes from
\citet{1991rc3..book.....D}, as listed in the NASA/IPAC Extragalactic Database
(NED)\footnote{\url{http://nedwww.ipac.caltech.edu}}, and then corrected these for
Galactic extinction based on the near-infrared maps of
\citet{2011ApJ...737..103S}, again, as provided by NED.  These Johnson-Cousins $B$
and $V$ magnitudes are Vega magnitudes and benefit from having been (i)
derived from wide field-of-view imaging from which the sky-background was
readily available, and (ii) taken in both the Northern and Southern
hemisphere, thereby capturing most of our sample.  
For three ETGs (NGC~6251, NGC~5252, and NGC~7052), and
three LTGs (IC~2560, NGC~253, and NGC~2960), 
either the $B$- or the $V$-band magnitude was not available.  
For these three ETGs we assigned a $B-V$ colour of 0.9, and for these three
LTGs we assigned a $B-V$ colour of 0.7. 
In passing, we recognise that
spheroid colours, rather than galaxy colours, would be advantageous.  However,
this would encompass considerable additional work, requiring multicomponent
decomposition of two optical bands.  Moreover, while both LTGs and ETGs can 
have colour gradients --- i.e., varying colour with radius, which can be due 
to the bulge-to-disc transition --- our sample is dominated by early-type spirals
(Sa--Sb) and early-type galaxies (E-S0) for which the bulge and disc colours within
individual galaxies may not be too dissimilar \citep{1996AJ....111.2238P}.

The (Galactic extinction)-corrected $B_T-V_T$ galaxy colour, hereafter $B-V$,
is shown in Figure~\ref{Fig_colour_IP13} and provided in
Table~\ref{Table-data}.  It was used to obtain the stellar
mass-to-light ratio at 3.6~$\mu$m as follows.  We started with the ($B-V$)-dependent
expression for the $K$-band $M_*/L_K$ ratio taken from Table~6 of
\citet{2013MNRAS.430.2715I}.  It is based on realistic dusty models, designed
for ``samples that include a range of morphologies, intrinsic colours and
random inclinations''.  It is such that
\begin{equation}
\label{EqIP13_tab6}
\log(M_*/L_K) = 0.866 (B-V) - 0.926,
\end{equation}
which is reportedly based on the \citet{1998ASPC..134..483K} `initial mass
function' (IMF)\footnote{The IMF is the histogram of stellar birth masses.}, 
and good for $0.5 < B-V < 1.1$. 
\citet{2013MNRAS.430.2715I} report a $\pm$0.1~dex (25\%) uncertainty on these $M_*/L_K$ ratios.
As they noted, the combination of 
dust attenuation (dimming the optical magnitudes) and reddening (of the $B-V$ colour) 
somewhat cancel to provide $M_*/L_K$ ratios
that are consistent with those derived from their simpler (dust free) galaxy model. 
This partial nulling behaviour was noted by \citet{2003ApJS..149..289B} 
and can be seen in \citet[][their Fig.~13]{2007MNRAS.379.1022D}, 
when assuming the dust models of \citet{2000A&A...362..138P}. 

Here, we convert Eq.~\ref{EqIP13_tab6} into a 3.6~$\mu$m 
equation using the following relation taken from the start of Section
5.6 in \citet[][see their Fig.~7]{2019MNRAS.483.1496S}: 
\begin{equation}
m_K - m_{3.6} = 0.54 - 0.42(B-V). \nonumber
\end{equation} 
Taking 2.5 times the logarithm of 
$(M_*/L_K)L_K = (M_*/L_{3.6})L_{3.6}$ 
and substituting in this $m_K - m_{3.6}$ colour term, one obtains 
%
% ...
%
\begin{equation}
\log(M_*/L_{3.6}) = 1.034(B-V) - 1.142. 
\label{Eq-ML36}
\end{equation}
While the use of individual $m_K - m_{3.6}$ colours rather than the above
mean ($B-V$)-dependent relation might seem preferable, in practice it can become 
problematic due to the different method used to 
determine the total $K$-band and Spitzer magnitudes \citep[e.g.,][their
  Fig.~2]{2013ApJ...768...76S}.

As noted above, \citet{2013MNRAS.430.2715I} initially derived a
colour-dependent $M_*/L$ relation
for a less complicated galaxy model based on composite stellar populations. 
This may be more applicable for the ETGs, and is such that 
$\log(M_*/L_K) = 1.055 (B-V) - 1.066$, for $0.2 < B-V < 1.0$. 
Morphing this in the same manner as done to Equation~\ref{EqIP13_tab6} gives the relation 
\begin{equation} 
\log(M_*/L_{3.6}) = 1.223(B-V) - 1.282.
\label{Eq_ippy1}
\end{equation} 
In the following subsection, we adjust these expressions (equations~\ref{Eq-ML36}
and \ref{Eq_ippy1}) to align them with the \citet{2002Sci...295...82K} IMF.

\pagebreak

\subsubsection{Consideration of the IMF}\label{Sec_IMF}

The stellar mass-to-light ratios from the above, and all, stellar population
models are dependent upon the assumed IMF \citep[see 
][]{2013pss5.book..115K}.  To convert from the \citet{1998ASPC..134..483K}
IMF\footnote{The concluding section in \citet{1998ASPC..134..483K} notes an
  IMF slope $\alpha_1=1.3$, and $\alpha_2=2.2$ from their previous sections,
  along with $\alpha_3=2.7$ for representing the IMF of the Galactic field.}
(spanning 0.1--100 M$_\odot$) --- which was used by
\citet{2013MNRAS.430.2715I} and is inherent in the previous equations --- to
an alternative IMF, the logarithm of the $M_*/L$ ratio needs to be adjusted by
a (near) constant factor \citep[e.g.,][]{2014ARA&A..52..415M}.  This is shown
by 
%
%
\citet[][their Figure~12]{2006MNRAS.372.1149F}, where it can be seen that 
switching to the Salpeter IMF (spanning 0.1--100 M$_\odot$)\footnote{The mass limits
  within which the IMF used by \citet{2013MNRAS.430.2715I} was integrated are stated in
  \citet{2004MNRAS.347..691P}.} requires adding 0.225 dex to the above
equations. 
%
%
%
Conversion to other assumed IMFs can be done following the offsets provided
by, for example, \citet[][their p.306]{2003ApJS..149..289B} or 
\citet[][their Table~2]{2010MNRAS.404.2087B}. 
In this paper, we adopt the \citet{2002Sci...295...82K} IMF, detailed further
in \citet{2013pss5.book..115K},  and have therefore
added 0.225 dex and subtracted 0.30 dex from Eq.~\ref{Eq-ML36} 
to give 
\begin{equation} 
\log(M_*/L_{3.6}) = 1.034(B-V) - 1.067.
\label{Eq_MonL_IP13} 
\end{equation}
Following \citet{2013MNRAS.430.2715I}, 
we assign a 25 percent uncertainty to these $M_*/L_{3.6}$ ratios. 

In the same way, Equation~\ref{Eq_ippy1} becomes 
\begin{equation}
\log(M_*/L_{3.6}) = 1.223(B-V) - 1.207.
\label{Eq_MonL_simple}
\end{equation}
We have included this additional relation (Equation~\ref{Eq_MonL_simple}), 
shown in the right-hand panel of 
Figure~\ref{Fig_colour_IP13}, simply to demonstrate that it yields similar
$M_*/L_{3.6} \equiv \Upsilon_{*,3.6}$ ratios to those from
Equation~\ref{Eq_MonL_IP13}.  

We proceed using Equation~\ref{Eq_MonL_IP13} to
derive the spheroid and galaxy stellar masses for all.
In Table~\ref{Table-data}, we list the spheroid and galaxy
stellar-masses, and the black hole masses taken from the compilation in 
\citet{2020ApJ...903...97S}, unless indicated otherwise. 
The luminosity distances are  also  tabulated here.  Distances 
from \citet{2001ApJ...546..681T} have been reduced by 
$\sim$4 percent due to a 0.083 mag reduction in their distance moduli.
This arose from a 0.06 mag reduction after a recalibration by  \citep[][their
  Section 4.6]{2002MNRAS.330..443B}\footnote{We opt not to use the attempted
  refinement offered by Equation~A1 and Figure~7 in
  \citet{2010ApJ...724..657B}.} 
plus a 0.023 mag reduction due to a reduced 
distance modulus for the Large Magellanic Cloud \citep{2019Natur.567..200P}
involved in the initial calibration.   The black hole masses depend linearly
on the angular distance to the host galaxies, and these masses have been updated here to reflect this. 

Following \citet[][their Eq.~9]{2019ApJ...876..155S}, the quoted uncertainties on the stellar 
masses include three uncertainties added in quadrature.  These relate to 
the distance (see Table~\ref{Table-data}), 
the $M_*/L$ ratio (a 25 percent uncertainty is suggested by \citet{2013MNRAS.430.2715I}), 
and the apparent magnitude. 
Here, we use a 0.15 mag uncertainty for the 
galaxy magnitude, and thus also for the spheroidal component of pure elliptical
galaxies.  This primarily captures uncertainty in the extrapolation of the
light profile  to  large radii \citep[][their Figure~1]{2005PASA...22..118G}
and this value also falls in the
middle of the $-$0.11 to +0.18 range reported by \citet[][their 
  Section~4.2.4]{2016ApJS..222...10S}. 
For those galaxies with two or more components, we assign uncertainties 
reflecting the complexity of the decomposition and thus the 
accuracy of the spheroid magnitude. These uncertainties were at  elevated
levels  in
\citet{2016ApJS..222...10S} and, in turn, \citet{2019ApJ...876..155S} because 
they were based on the published range of spheroid magnitudes from 
decompositions that, in retrospect, were clearly in error due to,
for example, missed discs or bars.  Having narrowed in on a better suite of
components for each galaxy, the typical uncertainty on the spheroid magnitude is 
reduced.  We adopt the following grading schema for the uncertainties on the
magnitudes: 
Grade 0 (0.15 mag: single-component galaxy); 
Grade 1 (0.2 mag: ES galaxies and those with only minor inner components); 
Grade 1.5 (0.25 mag: typically a clean bulge+disc fit, or if several
arcseconds of inner data were excluded, or if intracluster light (ICL) is present); 
Grade 2 (0.40 mag: usually a bar+bulge+disc fit); 
Grade 3 (0.55  mag: typically many components present). 
The forthcoming regressions were, however, tested and found to be stable (at
the 1$\sigma$ level) to a broad range of uncertainties.

In Appendix~\ref{Apdx1}, we repeat the forthcoming analysis of
Section~\ref{Sec_Results} using an alternative optical-infrared colour-dependent
prescription for the mass-to-light ratio, taken from
\citet{2022arXiv220202290S}.  This additional analysis supports one of our primary
conclusions: that violent, disc-destroying, mergers of (red) bulge$+$disc galaxies
\citep[e.g.,][]{2006ApJ...636L..81N} produce an offset population of
elliptical galaxies in the $M_{\rm bh}$--$M_{\rm *,sph}$ diagram.  Neither the
initial (bulge) nor the secondary (elliptical galaxy) relations have a
near-linear slope in this diagram.

\begin{figure}
\begin{center}
\includegraphics[trim=0.0cm 0cm 0.0cm 0cm, width=1.0\columnwidth, angle=0]{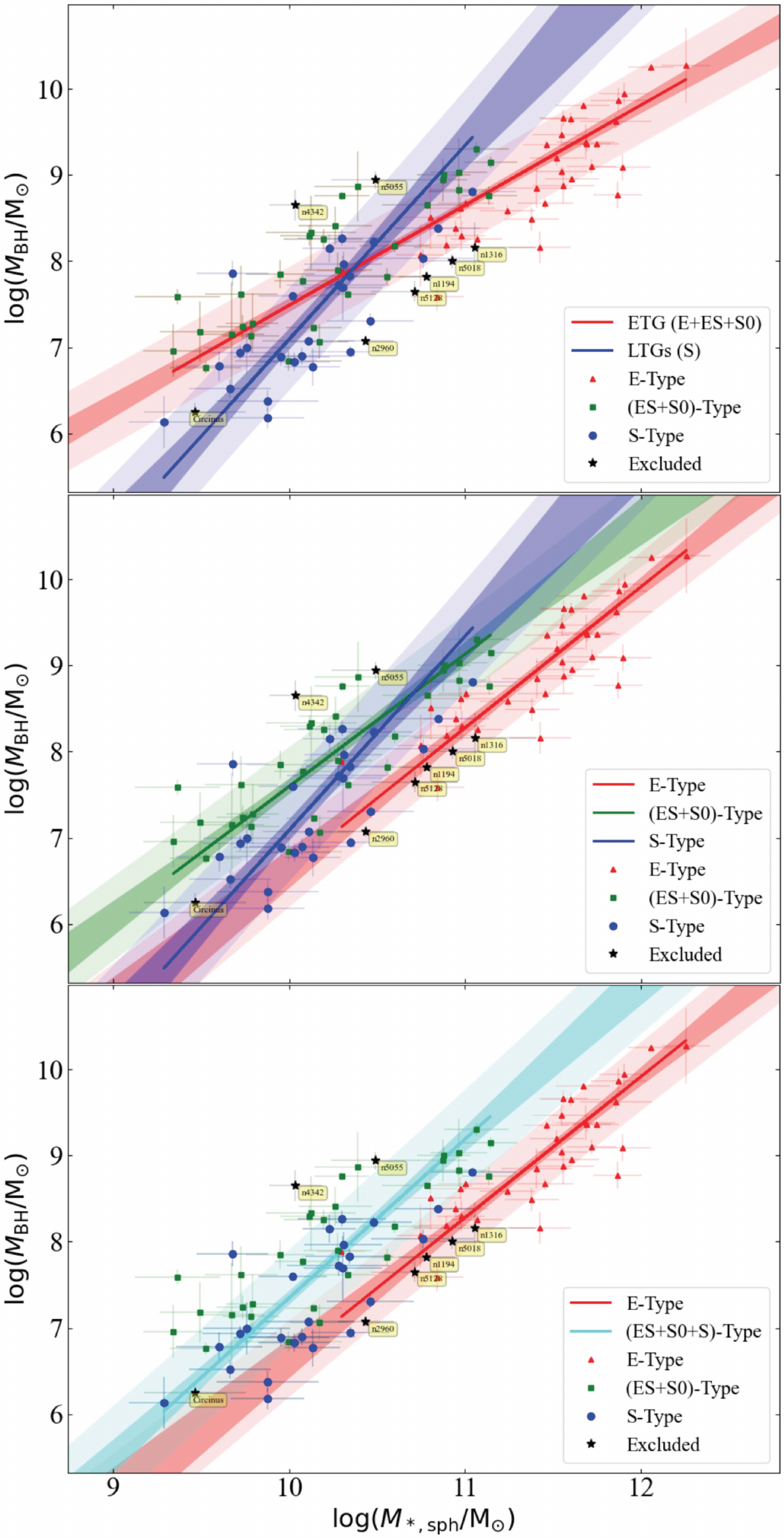}
\caption{Black hole mass versus spheroid stellar mass
  (obtained via Equation~\ref{Eq_MonL_IP13}) for different galaxy morphologies (E,
  ES/S0 and S).  As noted in the inset legends, each panel sampled the
  galaxies differently.  The darker shading reveals the 1$\sigma$ uncertainty
  on each relations' slope and intercept 
 --- as determined by  `confband.py' from SciPy \citep{2020SciPy-NMeth} ---, 
  while the lighter shading shows the
  root mean square (rms) scatter.  All quantities are shown in
  Table~\ref{Table-IP13}. The middle panel reveals that the single relation for ETGs, 
  shown in the upper panel, overlooks a key division between ETGs with and
  without discs.  Similarly, the single relation for disc galaxies, shown in the
  lower panel, overlooks the division between disc galaxies with and
  without a spiral pattern and thus the varying abundance of cold gas and star formation.
  While one may use Table~\ref{Table-data} to identify every galaxy shown here, for some
  galaxies mentioned in the text, we have 
  added small labels which can be seen by zooming in. These are 
  small to avoid overly detracting from the underlying patterns.  
}
\label{Fig_M_Msph_IP13}
\end{center}
\end{figure}

\begin{figure}
\begin{center}
\includegraphics[trim=0.0cm 0cm 0.0cm 0cm, width=\columnwidth, angle=0]{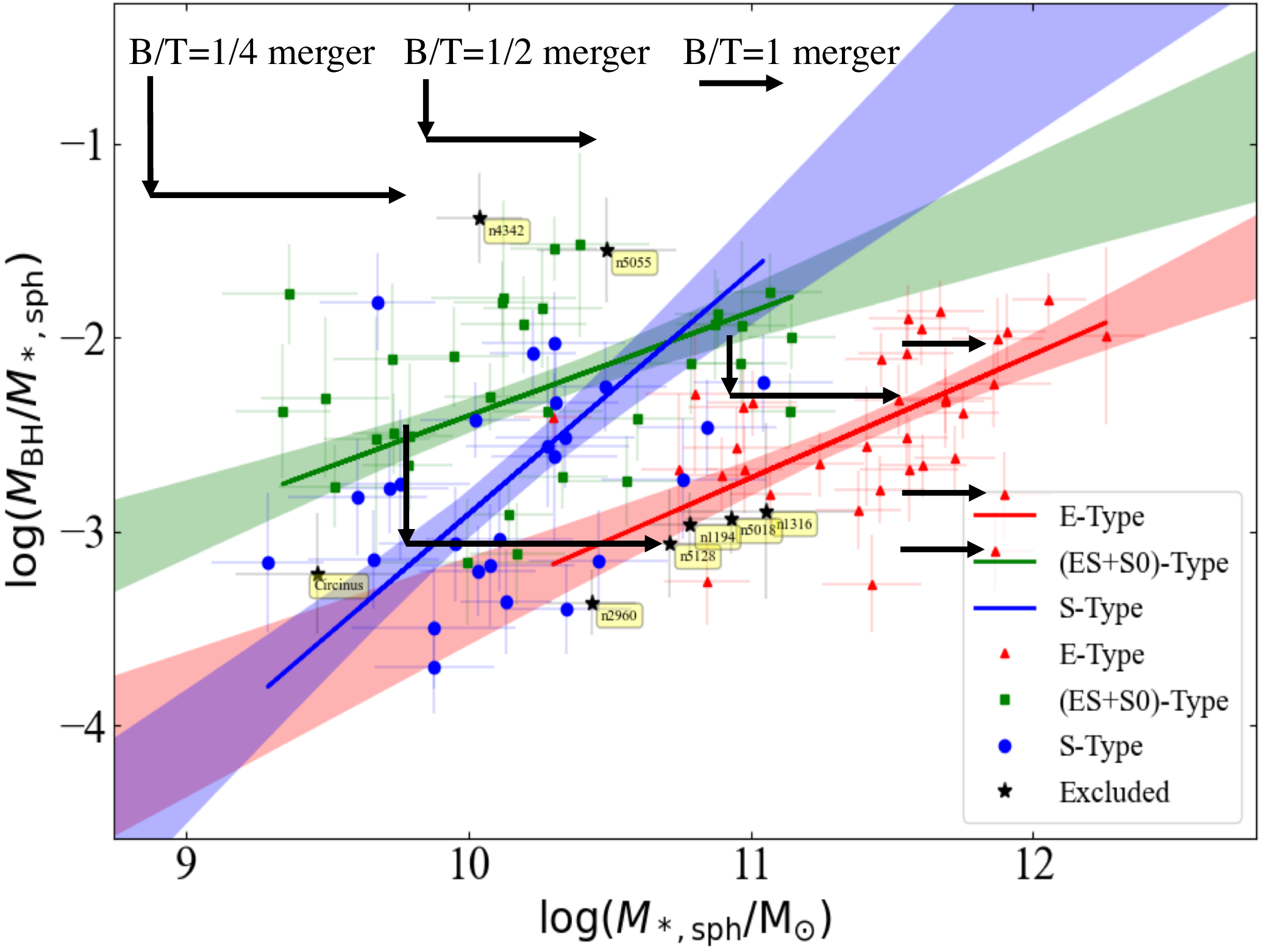} 
\caption{$M_{\rm bh}/M_{\rm *,sph}$ mass ratio versus $M_{\rm *,sph}$.  We
  have mapped the lines, and their 1$\sigma$ uncertainty, from the middle
  panel of Figure~\ref{Fig_M_Msph_IP13} For a given spheroid mass, the mean
  $M_{\rm bh}/M_{\rm *,sph}$ ratio is different by roughly an order of
  magnitude for ETGs with and without discs.  As shown by the arrows, 
  equal-mass mergers of S0 galaxies (illustrated here with two different bulge-to-total stellar
  mass ratios, $B/T=1/4$ and $1/2$) can shift systems towards the lower right. For example,
  an E galaxy built from an equal-mass merger of two S0 galaxies with
  $B/T=1/2$ will enact both a downward shift of $\sim$0.3 dex ($=\log2$) and a
  rightward shift of $\sim$0.6 dex ($=\log4$).  Given the slope of
  $0.53\pm0.15$ for the distribution of ES/S0 galaxies in this diagram,
  equal-mass mergers between S0 galaxies with $B/T=1/2$ (or $=1/4$) will
  create a new relation with a vertical offset of $\sim$0.6 (or $\sim$0.8)
  dex.  On the other hand, equal-mass mergers of E galaxies with $B/T=1$ will create a remnant
  shifted to the right by only $\sim$0.3 dex, as shown by the three example arrows.}
\label{Fig_msph_rat_IP13}
\end{center}
\end{figure}

\begin{figure}
\begin{center}
\includegraphics[trim=0.0cm 0cm 0.0cm 0cm, width=\columnwidth, angle=0]{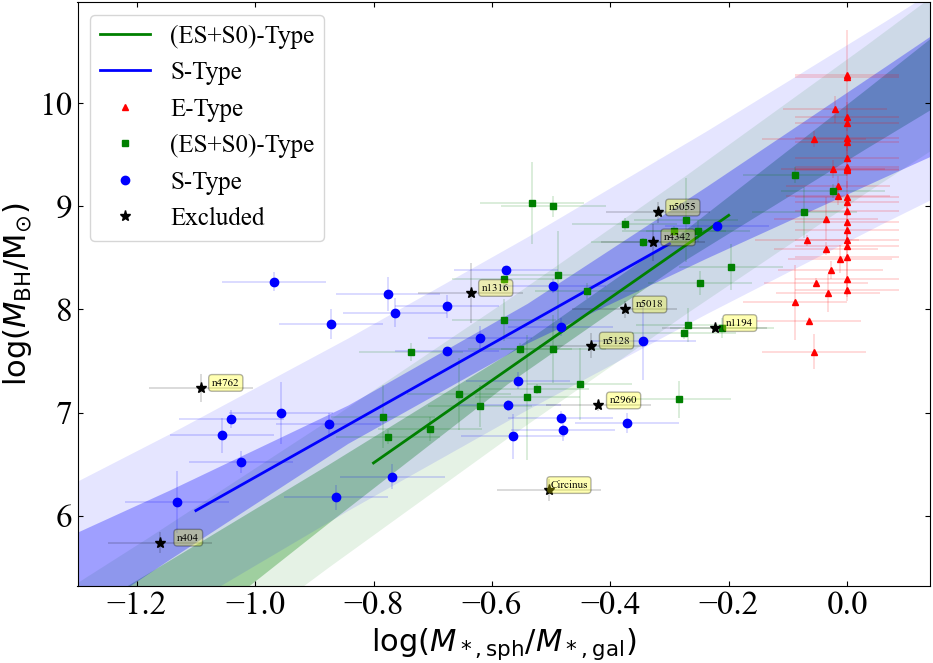} 
\caption{
Black hole mass versus the logarithm of the spheroid-to-total stellar mass
ratio for S, ES/S0 and E galaxies (for which the ratio tends to 1). 
The symbols and shading have the same meaning as in Figure~\ref{Fig_M_Msph_IP13}.  
To help understand the transition from the 
$M_{\rm bh}$--$M_{\rm *,sph}$ diagram (Figure~\ref{Fig_M_Msph_IP13}) 
to the $M_{\rm bh}$--$M_{\rm *,gal}$ diagram (Figure~\ref{Fig_MMgal_IP13}),
we have shown how the `spheroid-to-total' ratio relates to the black
hole mass.  A linear regression of $M_{\rm *,sph}/M_{\rm *,gal}$
and $M_{\rm bh}$ has been performed here (see Table~\ref{Table-IP13}). 
%
%
Note: Some E galaxies have $B/T$ ratios smaller than some
  ES galaxies due to additional undigested components and/or nuclear discs. 
%
}
\label{Fig_BonT_IP13}
\end{center}
\end{figure}

\section{Results}\label{Sec_Results}

\subsection{The $M_{\rm bh}$--$M_{\rm *,spheroid}$ and $M_{\rm bh}$--$M_{\rm
    *,galaxy}$ diagrams}\label{Sec_Diagrams}

In Figure~\ref{Fig_M_Msph_IP13}, we show the $M_{\rm bh}$--$M_{\rm *,sph}$
diagram, using Equation~\ref{Eq_MonL_IP13}, for the three morphological types:
E, ES/S0 and S.  In the upper panel we combine the E, ES and S0 galaxies,
which represent the ETGs.  In the middle panel, there is no grouping of the 
different galaxy types, while in the 
lower panel, we combine the ES, S0 and S galaxies, representing the disc galaxies.
The first point we make is that the slope 
of the $M_{\rm bh}$--$M_{\rm *,sph}$ relation for bulges in either S galaxies
or ES/S0 galaxies, and the slope of the
$M_{\rm bh}$--$M_{\rm *,sph}$ relation for elliptical galaxies, is neither equal to
1 nor close to it.  This is quantified in Table~\ref{Table-IP13} 
and described further in Section~\ref{Sec-Relations}.

The different relations for the bulges and
elliptical galaxies can also be seen in the $(M_{\rm bh}/M_{\rm *,sph})$--$M_{\rm
  *,sph}$ diagram  (Figure~\ref{Fig_msph_rat_IP13}).  
For a given spheroid stellar-mass, Figure~\ref{Fig_msph_rat_IP13} reveals 
different $(M_{\rm bh}/M_{\rm *,sph})$ ratios for elliptical
galaxies and the bulges of disc galaxies. 
This different ratio has received little attention in the
literature and has never been explained.   The arrows in this diagram trace
the expected movement due to simple, equal-mass, dry mergers of galaxies with
some illustrative bulge-to-total stellar mass ratios, $B/T$.  The merger of two E
galaxies produces a shift to the right, while a merger of two identical S0 galaxies with
a typical $B/T=0.25$ \citep[e.g.,][]{2005MNRAS.362.1319L} produces a
considerable shift to the lower right.  Considering the mean
regression lines, the elliptical galaxies appear to be
built, on average, by just one major merger.  One can also appreciate how
brightest cluster galaxies (BCGs), which are
typically E galaxies, occupy the right-hand side of the distribution in this diagram. 

\begin{figure}
\begin{center}
\includegraphics[trim=0.0cm 0cm 0.0cm 0cm, width=0.99\columnwidth, angle=0]{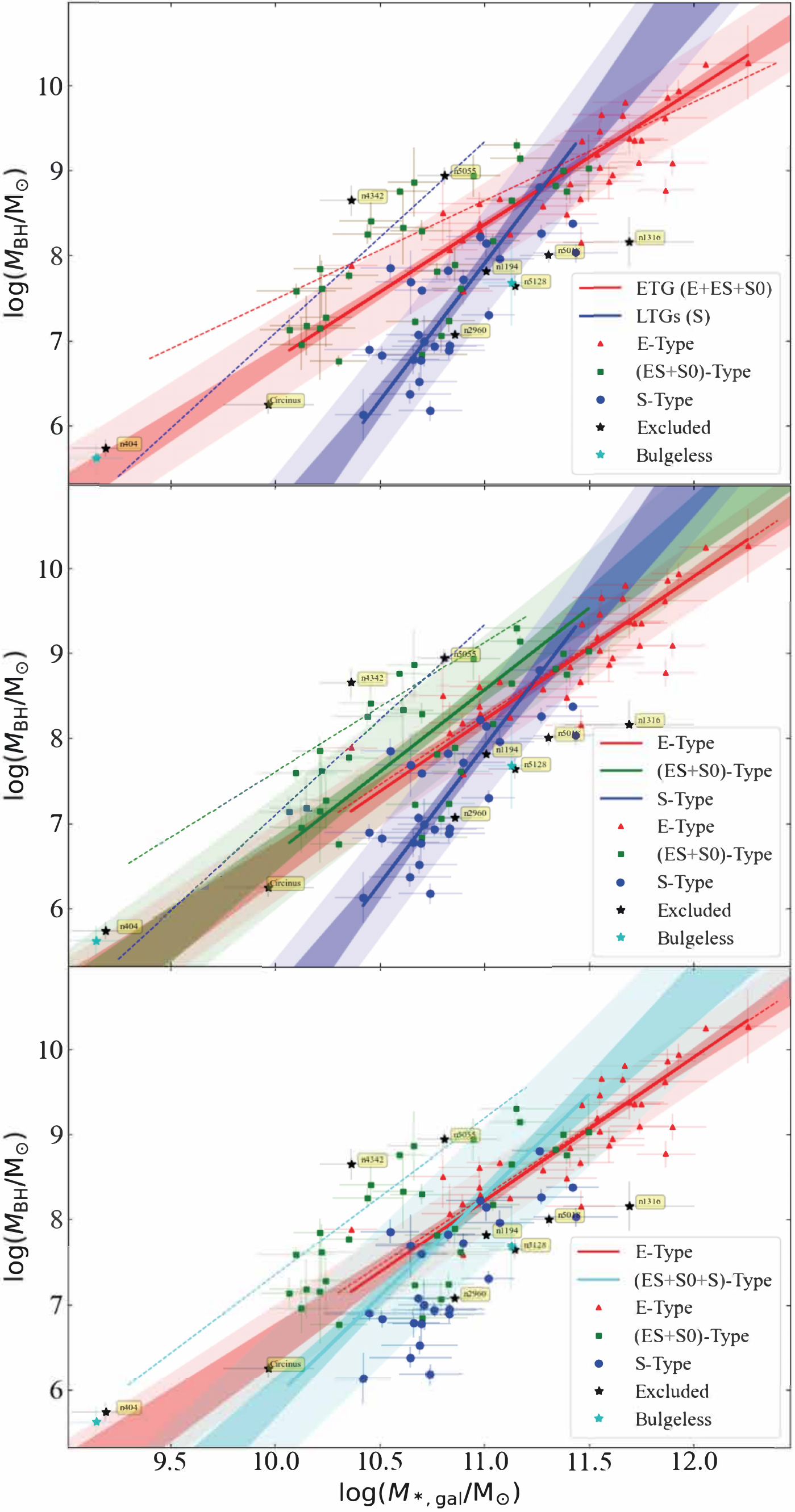} 
\caption{Similar to Figure~\ref{Fig_M_Msph_IP13} but now showing the {\em
    galaxy's} stellar mass, obtained via Equation~\ref{Eq_MonL_IP13}, rather
  than the {\em spheroid's} stellar mass.  The faint dashed lines are the
  $M_{\rm bh}$--$M_{\rm *,sph}$ relations from
  Figure~\ref{Fig_M_Msph_IP13}. The bulgeless galaxy NGC~4395, with a rather
  blue $B-V$ colour of 0.445, and thus low $M_*/L_{3.6}$ ratio, can be seen in
  the lower left next to NGC~404.  }
\label{Fig_MMgal_IP13}
\end{center}
\end{figure}

Figure~\ref{Fig_BonT_IP13} shows, for different morphological types, the trend
between black hole mass and $B/T$, or more precisely, the spheroid-to-total
stellar mass ratio. $B/T$ is not some near-constant value for
all S0 galaxies; a range of ratios is known \citep[e.g.,
][]{2008MNRAS.388.1708G}.  Aside from the exclusions mentioned in
Section~\ref{Sec_Sample}, here we exclude the ES/S0 galaxy NGC~4762 given the
excessive weight its small $B/T$ ratio has on our sample's
regression.\footnote{We do not wish to imply that S0 galaxies with low $B/T$
  ratios are in error, only that this $B/T$ data point for NGC~4762 interferes
  with the current mapping between Figures~\ref{Fig_M_Msph_IP13} and
  \ref{Fig_MMgal_IP13}.}  Figure~\ref{Fig_BonT_IP13} reveals that the S0
galaxies with the lower $B/T$ ratios have smaller black hole masses, as is
observed among the S galaxies.  These trends aid our understanding of the
transition from the $M_{\rm bh}$--$M_{\rm *,sph}$ diagram
(Figure~\ref{Fig_M_Msph_IP13}) to the $M_{\rm bh}$--$M_{\rm *,gal}$ diagram
(Figure~\ref{Fig_MMgal_IP13}).  For a given black hole mass, the smaller $B/T$
ratios in the LTGs imply that there will be a steeper $M_{\rm bh}$--$M_{\rm
  *,gal}$ relation for LTGs than for ES/S0 galaxies.  That is, given the
greater disc-to-bulge flux ratios and smaller spheroid masses when moving from
Sa to Sc galaxies,
%
the spiral galaxies transition to a steeper $M_{\rm bh}$--$M_{\rm *,gal}$
relation than the early-type disc galaxies (ES/S0).  This is seen in
Figure~\ref{Fig_MMgal_IP13}, in which the $M_{\rm bh}$--$M_{\rm *,gal}$
relation for the E galaxies is basically\footnote{Nuclear discs and
  additional, possibly undigested, galaxy components result in a slight
  difference.} the same as the $M_{\rm bh}$--$M_{\rm *,sph}$ relation for the
E galaxies.  The $M_{\rm bh}$--$M_{\rm *,gal}$ relation for the E galaxies is,
however, offset from the $M_{\rm bh}$--$M_{\rm *,gal}$ relation for the ES/S0
galaxies.  The darker shading in this diagram reveals that the relations are
not consistent with each other at the 1$\sigma$ level.  This reveals that the E and ES/S0 galaxies
are not offset from each other in the $M_{\rm bh}$--$M_{\rm *,sph}$ diagram
due to the exclusion of the non-spheroid stellar mass.

\begin{figure}
\begin{center}
\includegraphics[trim=0.0cm 0cm 0.0cm 0cm, width=0.98\columnwidth, angle=0]{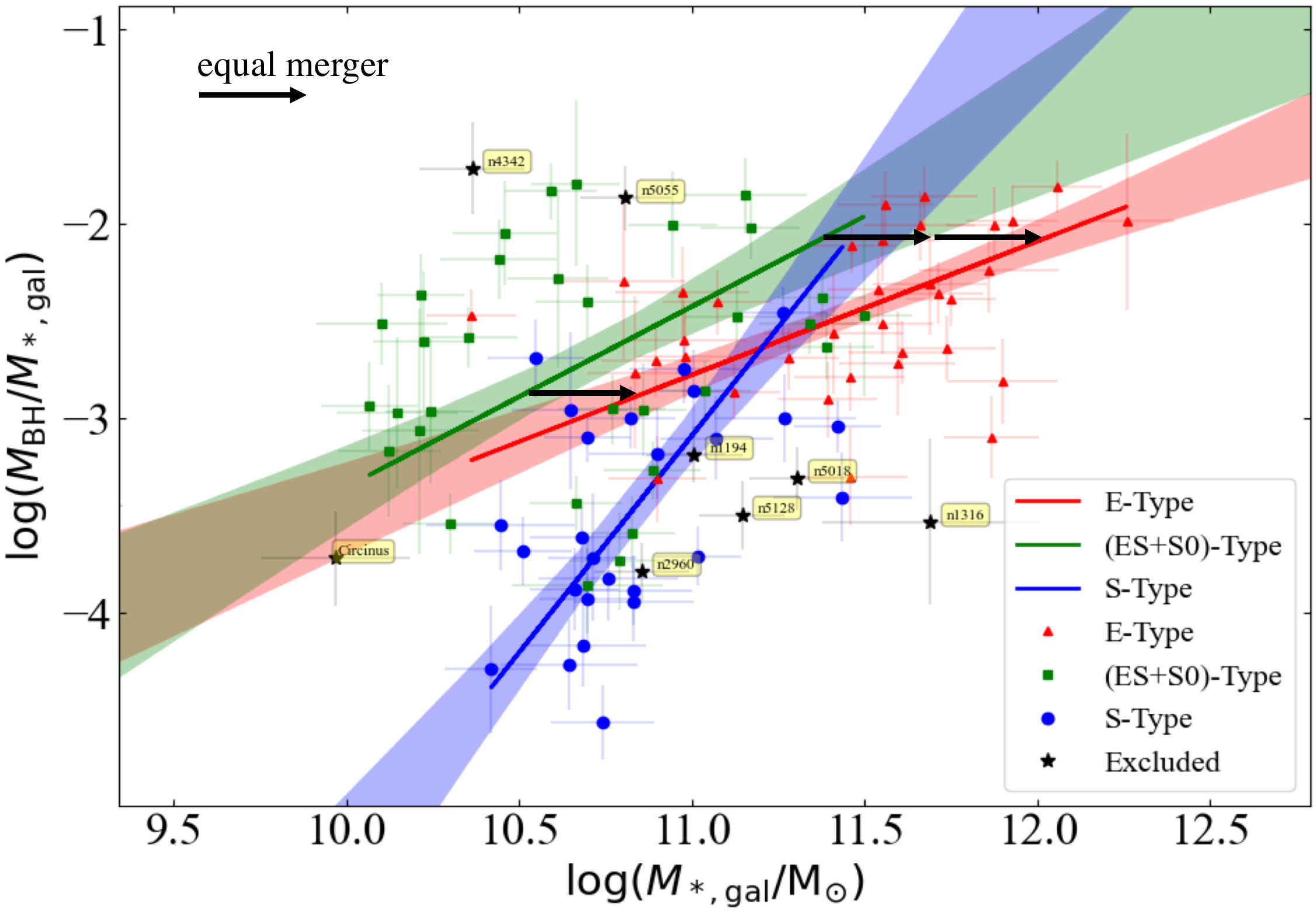}
\caption{Similar to Figure~\ref{Fig_msph_rat_IP13} but now showing the  
$M_{\rm bh}/M_{\rm *,gal}$ mass ratio versus $M_{\rm *,gal}$.  
The lines from the middle panel of Figure~\ref{Fig_MMgal_IP13} have been mapped here. 
On average, the low-mass E galaxies have 
roughly the same $M_{\rm bh}/M_{\rm *,gal}$ ratio as 
the S0 galaxies which merged to create them, but twice as much stellar mass,
in accord with a likely dry equal-mass merger. The high-mass E galaxies, including
some brightest cluster galaxies, have four times as much mass as the S0s with a similar $M_{\rm
  bh}/M_{\rm *,gal}$ ratio, reflective of two such mergers or equally one equal-mass
E galaxy merger. 
}
\label{Fig_mgal_rat_IP13}
\end{center}
\end{figure}

In Figure~\ref{Fig_mgal_rat_IP13}, we present the $(M_{\rm bh}/M_{\rm
  *,gal})$--$M_{\rm *,gal}$ diagram.  As noted in \citet{2019ApJ...876..155S},
one can {\em approximate} the ETGs (E/ES/S0) with a single relation in this
diagram if, for example, one is pursuing rough predictions for black hole
masses in other ETGs.  However, there is more detail to it than this, and this
detail is one of the keys to understanding the black hole mass scaling
diagrams.  

As with Figure~\ref{Fig_M_Msph_IP13}, one can again see that the
addition of the non-spheroid stellar mass, primarily from the disc and bar,
does not align the lenticular and elliptical
galaxies.  This reveals that the offset in the $M_{\rm bh}$--$M_{\rm *,sph}$
diagram, between the bulge component of ES/S0 galaxies and E galaxies, is not
an artifact of separating/reducing the light in some ETGs (those with discs)
but not others (those without) when we were plotting the spheroid stellar
mass.  The arrows in Figure~\ref{Fig_mgal_rat_IP13} reveal that the
distribution of elliptical galaxies is readily explained if they are built
from the dry merger of lenticular galaxies, which is widely thought to be the
case, and also the merger of elliptical galaxies.\footnote{While we have used
  simple equal-mass mergers to illustrate S0-to-E, and E-to-E,
  transformations, there are other options, such as several minor mergers.}
However, what is not well-recognised is the ensuing offset between the E and
ES/S0 galaxies in various black hole mass scaling diagrams populated with real
data. 
Here, we have: 
\begin{itemize}
\item built on \citet{2019ApJ...876..155S} which established that there is not
  a single (fundamental) $M_{\rm bh}$--$M_{\rm *,sph}$ relation for ETGs;
\item revealed that mergers have not built a near-linear $M_{\rm bh}$--$M_{\rm
    *,sph}$ relation, due to the folding-in of the disc/bar stellar mass,
  reducing the $M_{\rm bh}$--$M_{\rm *,sph}$ ratio; and
\item established that a single $M_{\rm bh}$--$M_{\rm *,gal}$ relation is not
  a fundamental relation for ETGs, with merger-built E galaxies offset from
  the S0 galaxies.
\end{itemize} 
We will return to these points with additional supportive evidence in
Section~\ref{Sec_b2e}.

\subsection{The $M_{\rm bh}$--$M_{\rm *,sph}$ and  $M_{\rm bh}$--$M_{\rm
    *,gal}$ relations and ratios}\label{Sec-Relations}

We have used hierarchical Bayesian model fitting through the
state-of-the-art platform for statistical modelling known as {\sc Stan}
\citep{2017JSS....76....1C,
  Rstan:2016}\footnote{\url{https://mc-stan.org/}}.  The
statistical model used for our linear regression considered uncertainties in
both variables and is aimed at obtaining a symmetric relation between the two
variables.  A bivariate normal density was used to represent the distribution
of latent (`true') $\log M_*$ and $\log M_{\rm bh}$ values that might occur in
our sample. As noted in \citet{2019ApJ...873...85D}, ``this is conceptually
equivalent to the generative framework sketched by
\citet{2010arXiv1008.4686H}, in which the observed data points are imagined to
be drawn from a distribution centred around a `line of best fit', except that
here we allow Bayesian `shrinkage' by estimating the underlying distribution
along the line rather than keeping this as an improper uniform prior.''
Details of the statistical model framework are described in \citet[][their
  Appendix~A]{2019ApJ...873...85D}.  

The best-fitting relations are shown in Table~\ref{Table-IP13}, along with the
slope, $A$, and intercept, $B$, at the normalisation point.  The normalisation
point of, for example, the $M_{\rm bh}$--$M_{\rm *,sph}$ relation is
$\upsilon(5\times10^{10}\,M_\odot)$.  If using equation~\ref{Eq_MonL_IP13}
to convert light into stellar mass and thus derive a value of $M_{\rm *,sph}$
for use in the $M_{\rm bh}$--$M_{\rm *,sph}$ equation (to predict $M_{\rm
  bh}$), one has that $\upsilon=1$.  If, however, a different light-to-mass
ratio prescription is preferred and used to derive one's estimate of $M_{\rm
  *,sph}$, then one needs to apply the appropriate value of $\upsilon$, as
illustrated in \citet[][see their Figure~4 and Equations~6 to
  8]{2019ApJ...876..155S} for colour-dependent light-to-mass ratio prescriptions in different
passbands.  While we could drop this $\upsilon$ term from our equations, as
typically done before \citet[][see their Equations~10 and
  11]{2019ApJ...873...85D}, its inclusion serves to remind readers that a
specific prescription for $\Upsilon_*$ has been used to derive the equation and that they need
to apply a conversion if using an alternate prescription.  The root mean
square scatter, $\Delta_{\rm rms}$, in the $\log\,M_{\rm bh}$ direction is
also tabulated for reference, although it is noted that this is not the quantity that is
minimised with a Bayesian regression. 

%
%
%

We have also applied
three additional linear regression codes to our data, and found consistent
results with our primary Bayesian analysis.  For example, the Bisector
regression from the Bivariate Correlated Errors and Intrinsic Scatter ({\sc
  BCES}) routine \citep{1996ApJ...470..706A} gave the following slopes in the
$M_{\rm bh}$--$M_{\rm *,sph}$ diagram for the E, ES/S0, and S galaxies:
1.62$\pm$0.17; 1.49$\pm$0.13; and 2.19$\pm$0.33.  Using a symmetrical
treatment\footnote{While the modified-{\sc FITEXY} routine performs a
  non-symmetrical regression of a sample of ($X$, $Y$) data pairs, a
  symmetrical treatment can be obtained by running the regression twice, the
  second time with the $Y$ and $X$ variables swapped around with each other.
  The {\em bisector} of the resulting two regression lines provides an
  expression which effectively treats the data equally \cite[e.g.,][their
    Section~3.1.1]{2006ApJ...637...96N,2007ApJ...655...77G}.}  of the
modified-{\sc FITEXY} routine from \citet{2002ApJ...574..740T} yielded slopes
equal to 1.65$\pm$0.12, 1.53$\pm$0.11 and 2.20$\pm$0.26, respectively.
Finally, the Bayesian {\sc linmix} code from \citet{2007ApJ...665.1489K}
yielded: 1.61$\pm$0.14 (E); 1.52$\pm$0.13 (ES/S0); and 2.14$\pm$0.34 (S).
From Table~\ref{Table-IP13}, we have that $M_{\rm bh} \propto M_{\rm
  *,sph}^{2.25\pm0.39}$ for the bulges of the spiral galaxies.
This has 1$\sigma$ uncertainties which overlap with those from the
steeper relation reported by \citet{2019ApJ...873...85D}, in which $M_{\rm bh}
\propto M_{\rm *,sph}^{2.44\pm0.33}$ for a larger sample of 40 spiral galaxies
observed with a range of filters and $M_*/L$ ratios.  It appears that the
bulges of spiral galaxies define a steeper $M_{\rm bh}$--$M_{\rm *,sph}$
relation than the bulges of S0 galaxies.  Three evolutionary pathways for the
spiral galaxy bulges are offered in Section~\ref{Sec_Disc_1}.  In fair 
agreement with the relations found here for the ETGs using the 
hierarchical Bayesian model fitting (see Table~\ref{Table-IP13}), 
\citet{2019ApJ...876..155S} report a slope of 1.86$\pm$0.20 for 36 ES/S0
galaxies and 1.90$\pm$0.20 for 40 E galaxies.  As seen in
\citet{2019ApJ...876..155S}, the present sample of E
galaxies trace a relation which is roughly parallel to that defined by the
bulges of S0 galaxies.

Using multicomponent decompositions, \citet{2016ApJ...817...21S} reported a
median $M_{\rm bh}/M_{\rm *,sph}$ value of $\sim$0.68 percent for 45 ETGs,
which they thought followed a near-linear $M_{\rm bh}$--$M_{\rm *,sph}$
relation.  This result was based on the use of an $M_*/L_{3.6}$ ratio of
$\sim$0.60 and a \citet{2003PASP..115..763C} IMF.  \citet{2010MNRAS.404.2087B}
suggest a reduction of just 0.05 dex to the logarithm of $\Upsilon_*$ ($\equiv
M_*/L$) to convert from the \citet{2003PASP..115..763C} IMF to the
\citet{2002Sci...295...82K} IMF.  Therefore, this $M_{\rm bh}/M_{\rm *,sph}$
mass ratio of 0.68 percent increases to 0.76 percent for the
\citet{2002Sci...295...82K} IMF. This is comparable to the median $M_{\rm
  bh}/M_{\rm *,sph}$ ratio for core-S\'ersic galaxies reported in
\citet{2013ApJ...764..151G}, which was obtained using $M_*/L_{K}=0.8$ from
\citet{2003ApJS..149..289B} and based on a diet-Salpeter IMF.  Their reported
$K$-band mass-to-light ratio of 0.49 drops by 0.15 dex, or to
$M_*/L_{K}=0.57$, when switching to the \citet{2002Sci...295...82K} IMF.
Consequently, their $M_{\rm bh}/M_{\rm *,sph}$ ratio of 0.49 percent increases
to 0.69 percent once calibrated against the \citet{2002Sci...295...82K} IMF,
and is thus in good agreement with the above value of 0.76
percent.\footnote{We thank Peter Behroozi for pointing out this issue in 2017,
  surrounding clarification of the adopted IMF before comparing $M_{\rm
    bh}/M_{\rm *,sph}$.}  However, as \citet{2019ApJ...876..155S} uncovered,
and as can be seen in Figure~\ref{Fig_msph_rat_IP13}, 
this near-constant $M_{\rm bh}/M_{\rm *,sph}$ 
mass ratio of $\sim$0.7 percent for ETGs is misleading. 
Individual ratios, at a fixed $M_{\rm *,sph}$, differ by an order of
magnitude depending on whether the system is an S0 or an E galaxy.
Furthermore, the ratio can vary by an order of magnitude within either of
these two galaxy types.
Turning to the {\em galaxy} masses, 
Figure~\ref{Fig_mgal_rat_IP13} illustrates that while the S galaxies tend to have
lower $M_{\rm bh}/M_{\rm *,gal}$ ratios than the ETGs in our sample, due in
part to the greater disc-to-bulge ratios in S galaxies, there is more to it
than that. On average, for a given $M_{\rm bh}/M_{\rm *,gal}$ ratio, the E 
galaxies have higher masses than the S0 galaxies, which is expected if S0 
galaxies merge to form E galaxies.  This observation also expresses itself as
a lower $M_{\rm bh}/M_{\rm *,gal}$ ratio in E galaxies than S0 galaxies at a
given galaxy mass, modulo the scarcity of low-mass E and high-mass S0
galaxies --- another signature of the dry merger phenomena.

In passing we note that it almost goes without saying that applying consistent
$\Upsilon_*$ ratios between different studies is vital for avoiding
artificial mismatches such as that reported in
\citet{2016MNRAS.460.3119S}.  Realised some years ago, and detailed in
\citep{SahuGrahamHon22}, this mismatch led us to introduce the mass-to-light
conversion term, $\upsilon$, in \citet{2019ApJ...873...85D}.\footnote{The lower-case
  upsilon symbol was introduced to facilitate changes to the mass-to-light
  ratio, $\Upsilon$, in a similar manner to how $h$ can enact changes to the
  adopted Hubble-Lema\^itre constant H$_0$.} 
This was developed further in \citet{2019ApJ...876..155S} and explains the
$\upsilon$ term included in Table~\ref{Table-IP13}.


\begin{table}
\centering
\caption{Black hole mass scaling relations}\label{Table-IP13}
\begin{tabular}{lccc}
\hline
Galaxy type   &   Slope (A) & Intercept (B) & $\Delta_{\rm rms}$ \\
\hline 
\multicolumn{4}{c}{$\log(M_{\rm bh}/M_\odot) = A\log[M_{\rm
      *,sph}/\upsilon(5\times10^{10}\,M_\odot)] +B$} \\
E   (35)        &  1.64$\pm$0.17  &  7.79$\pm$0.17   & 0.38 \\
ES/S0  (32)     &  1.53$\pm$0.15  &  8.67$\pm$0.15   & 0.44 \\
S      (26)     &  2.25$\pm$0.39  &  8.66$\pm$0.28   & 0.57 \\
ES/S0 \& S (58) &  1.84$\pm$0.16  &  8.63$\pm$0.14   & 0.55 \\
E \& ES/S0 (67) &  1.16$\pm$0.07  &  8.30$\pm$0.11   & 0.43 \\
\hline
\multicolumn{4}{c}{$\log(M_{\rm bh}/M_\odot) = A\log[M_{\rm *,gal}/\upsilon\,10^{11}\,M_\odot]
  +B$}  \\
E   (35)        &  1.69$\pm$0.17  &  8.22$\pm$0.15   & 0.38 \\
ES/S0  (32)     &  1.93$\pm$0.28  &  8.57$\pm$0.17   & 0.61 \\
S      (26)     &  3.23$\pm$0.57  &  7.91$\pm$0.18   & 0.60 \\
ES/S0 \& S (58) &  2.38$\pm$0.27  &  8.28$\pm$0.13   & 0.80 \\
E \& ES/S0 (67) &  1.59$\pm$0.11  &  8.37$\pm$0.11   & 0.49 \\
\hline
\multicolumn{4}{c}{$\log(M_{\rm bh}/M_\odot) = A\{\log[M_{\rm *,sph}/M_{\rm *,gal}]
  -(-0.5) \} + B$}  \\
ES/S0  (31)     &  4.00$\pm$0.58  &  7.71$\pm$0.16  & 0.64 \\ 
S      (26)     &  3.23$\pm$0.63  &  7.99$\pm$0.21  & 0.78 \\

\hline
\multicolumn{4}{c}{$\log(M_{\rm sph}/ \upsilon M_\odot) = A\log[R_{\rm e,sph,eq}/\rm kpc] +B$}
\\
All   (93)      &  1.14$\pm$0.04  & 10.48$\pm$0.08  & 0.29 \\
\hline
\multicolumn{4}{c}{$\log(M_{\rm bh}/M_\odot) = A\log[R_{\rm e,sph,eq}/\rm kpc] +B$}
\\
E   (35)        &  2.38$\pm$0.33  &  6.88$\pm$0.32  & 0.58 \\
ES/S0  (32)     &  1.98$\pm$0.24  &  8.52$\pm$0.16  & 0.54 \\
S      (26)     &  2.40$\pm$0.40  &  8.02$\pm$0.20  & 0.65 \\
\hline
\end{tabular}

The slope and intercept of the relations shown in
Figures~\ref{Fig_M_Msph_IP13}, \ref{Fig_BonT_IP13} 
\ref{Fig_MMgal_IP13}, \ref{Fig_R_Msph_IP13} and \ref{Fig_evolve_IP13} 
have been obtained 
using a Bayesian analysis that treats the data symmetrically.  
The root mean square (rms) scatter reported here, $\Delta_{\rm rms}$, 
is the vertical scatter about each relation. 
The spheroid and galaxy stellar masses have been obtained using
the $M_*/L$ prescription given in Equation~\ref{Eq_MonL_IP13}.
The $\upsilon$ term is mentioned towards the end of
Section~\ref{Sec-Relations} is equal to 1 if one uses stellar masses
consistent with those obtained via Equation~\ref{Eq_MonL_IP13}.
\end{table}

Here, we find $M_{\rm bh}/M_{\rm *,sph} = 0.0018$ (0.18 percent) for elliptical galaxies with
$M_{\rm *,sph}=10^{11}$ M$_\odot$, and 1.7 percent for bulges of the same stellar
mass.  The order of magnitude difference between these  morphological types can be
seen in Figure~\ref{Fig_msph_rat_IP13}.  This difference appears to widen slightly when
moving to lower spheroid masses.  Furthermore, one can see how
dry mergers  of S0 galaxies, building E galaxies, can explain this observation.

\subsection{From bulges to elliptical galaxies}\label{Sec_b2e}

Much of the accretion-driven growth of black holes is known to occur in
regular star-forming disc galaxies \citep[e.g.,][]{2009ApJ...691..705G,
  2011ApJ...726...57C}.  That is, the AGNs tend to reside in normal, often
isolated, spiral galaxies
\citep[e.g.,][]{2005ApJ...627L..97G, 2012ApJ...744..148K}.  AGNs are not
particularly prevalent during or after major mergers.  Given that elliptical
galaxies have little to no star formation, it is apparent that the gaseous processes driving
growth in the $M_{\rm bh}$--$M_{\rm *,sph}$ diagram occur in bulges.  As noted
by multiple studies, the black hole accretion rate relative to the star
formation rate is such that it is not expected to establish a linear $M_{\rm
  bh}$--$M_{\rm *,gal}$ relation but instead a steeper 
relation \citep[e.g.,][]{2012ApJ...755..146S, 2013ApJ...765L..33L,
  2014A&A...566A..53D, 2019ApJ...885L..36D}. 
This bodes well for the
steeper-than-linear trend seen for spiral galaxies in the $M_{\rm bh}$--$M_{\rm *,gal}$ diagram.
Should some of the star-formation be occurring in bulges, 
then the higher (black hole accretion rates)-to-(star-formation rates) 
in spiral galaxies may mesh well with the trend seen in
the $M_{\rm bh}$--$M_{\rm *,sph}$ diagram.

By considering the sizes of the spheroids, we can build a more informed
scenario for what we are witnessing in the $M_{\rm bh}$--$M_{\rm *,sph}$ and
$M_{\rm bh}$--$M_{\rm *,gal}$ diagrams.  We will see how dry mergers can account for
the steeper-than-linear $M_{\rm bh}$--$M_{\rm *,sph}$ relation observed
for the  E galaxies.

\begin{figure}
\begin{center}
\includegraphics[trim=0.0cm 0cm 0.0cm 0cm, width=\columnwidth, angle=0]{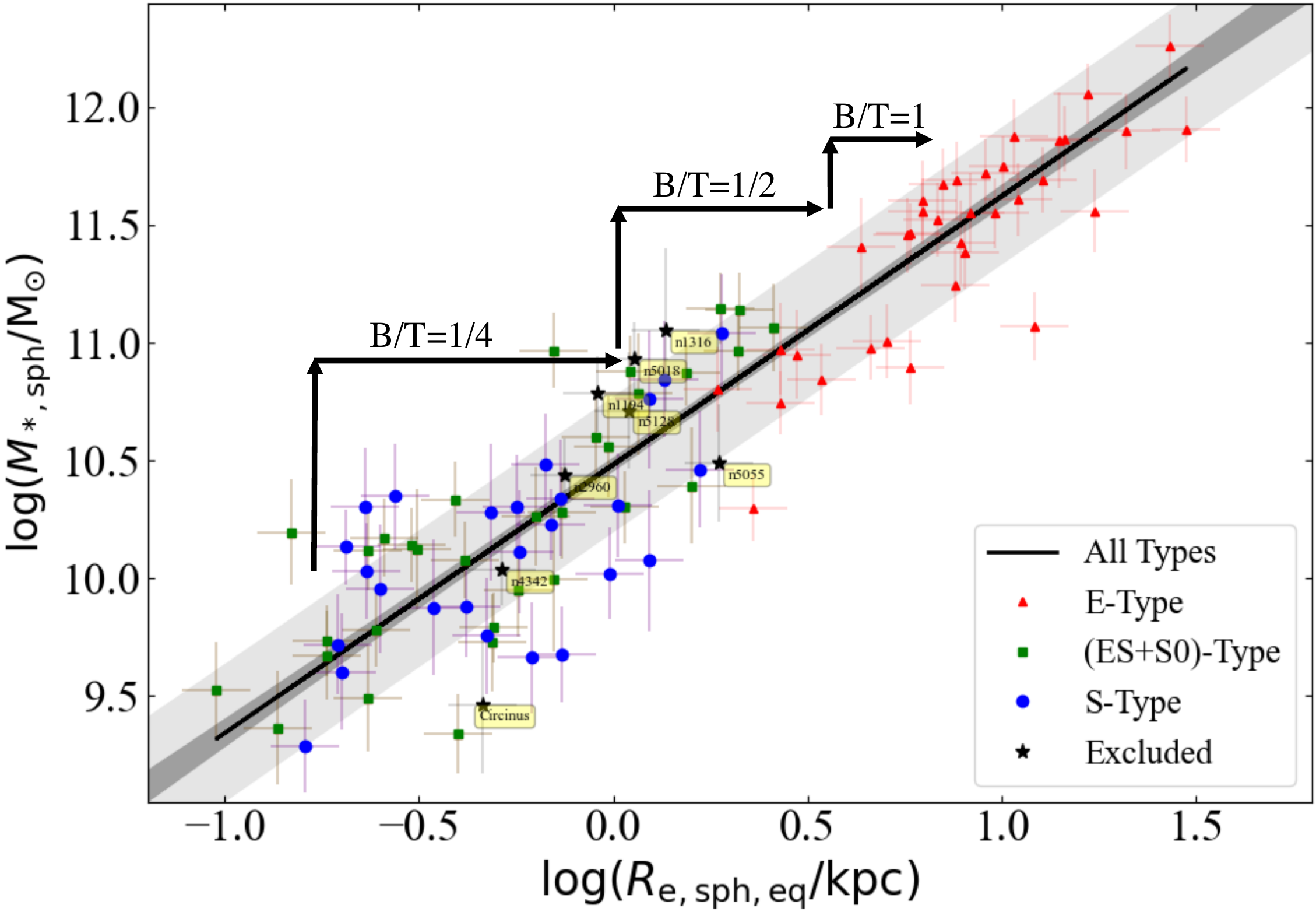}
\caption{Geometric-mean axis, aka `equivalent (circular) axis', effective
  half-light radii, $R_{\rm e,sph,eq}$, of the spheroid versus the stellar
  mass of the spheroid, $M_{\rm *,sph}$.  Modification of \citet[][their
    Figure~9]{2020ApJ...903...97S} using the stellar mass-to-light
  prescriptions in Equation~\ref{Eq_MonL_IP13} and only those galaxies with
  3.6~$\mu$m data.  The arrows show the apparent movement caused by creating
  an E galaxy from an equal-mass dry merger between two S0 galaxies with
  bulge-to-total ratios equal to one-quarter and one-half and between two
  elliptical galaxies with $B/T=1$.  The length of the horizontal arrows are
  based on maintaining the observed relation $M_{\rm *,sph} \propto R_{\rm
    e,sph,eq}^{1.14\pm0.04}$.  The E galaxies with the larger
  radii at $\log(M_{\rm *,sph}/M_\odot) \approx$ 10.25 and 11.05 
  are NGC~3377 and NGC~4697, respectively.}
\label{Fig_R_Msph_IP13}
\end{center}
\end{figure}

\begin{figure*}
\begin{center}
\includegraphics[trim=0.0cm 0cm 0.0cm 0cm, width=1.0\textwidth, angle=0]{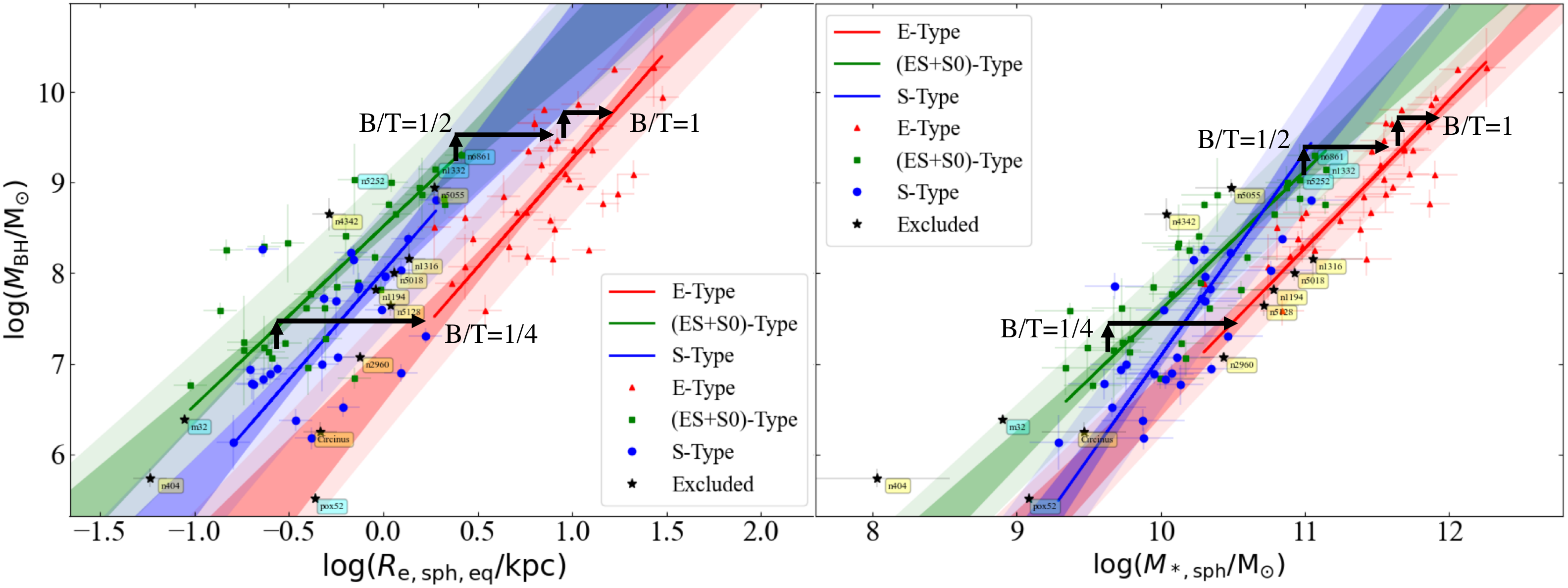}
\caption{ Left-hand panel: $M_{\rm bh}$ versus $R_{\rm e,sph,eq}$.  Adaption of Figure~13
  from \citet{2020ApJ...903...97S}, using only those galaxies with 3.6~$\mu$m
  data and updated vales reported in Table~\ref{Table-data}.  
  Right-hand panel: Evolution in the $M_{\rm bh}$--$M_{\rm *,sph}$ diagram arising from 
  dry equal-mass mergers of galaxies with bulge-to-total ($B/T$) ratios of 1.0, 0.5
  and 0.25 to produce an E galaxy.  The E galaxy sequence 
  is seen to follow a steeper-than-linear relation 
  rather than the linear relation expected from E+E galaxy mergers
  \citep{2007ApJ...671.1098P}.  The somewhat discrepant 
(low $M_{\rm *,sph}$ or high $M_{\rm bh}$) spiral and elliptical galaxy with
  $\log(M_{\rm bh}/M_\odot) \approx 7.9$ dex are NGC~1300 and NGC~3377, respectively.}
\label{Fig_evolve_IP13}
\end{center}
\end{figure*}

In Figure~\ref{Fig_R_Msph_IP13}, we show the effective half-light size of the
spheroids, $R_{\rm e,sph}$, versus their stellar mass, $M_{\rm *,sph}$.  
These radii are given in Table~\ref{Table-data}, along with the  reference
showing the modelled light profile from which the radii were measured. 
We used the geometric-mean axis, 
aka the `equivalent axis', $r=\sqrt{ab}$, along which the size of the radii are 
equivalent to a circularised version of the quasi-elliptical isophotes.  There
is no discontinuity in the $R_{\rm e,sph}$--$M_{\rm *,sph}$ diagram between
the different types of spheroids.  This continuity holds whether the spheroids
coexist with a disc that either does or does not contain a spiral pattern, or
whether they exist on their own with no disc, i.e., are an elliptical galaxy.

The $R_{\rm e,sph}$--$M_{\rm *,sph}$, or equally $M_{\rm *,sph}$--$R_{\rm
  e,sph}$, relation is seen in Figure~\ref{Fig_R_Msph_IP13} to have a slope
close to unity.  Curiously, there is little evidence for any broad curvature
in the distribution of $R_{\rm e,sph}$ and $M_{\rm *,sph}$.  This differs from
what is seen in the $M_{\rm *,gal}$--$R_{\rm e,gal}$ relation for ETGs
\citep[][their Figure~1b]{2006AJ....132.2711G} due to the presence and then
dominance of discs as one moves to lower masses.  Not surprisingly, this
near-linear slope matches that seen at the bright end of the $M_{\rm
  *,gal}$--$R_{\rm e,gal}$ relation for ETGs, which is dominated by E galaxies
\citep[e.g.,][their Figure~9]{2008MNRAS.388.1708G, 2018MNRAS.477.5327K,
  2019ApJ...886...80D, 2020ApJ...903...97S}, i.e., systems without discs.  The
simulations from \citet{2009ApJ...703.1531N}, involving elliptical galaxies
undergoing minor and major dry merger events, build a near-linear $M_{\rm
  *,gal}$--$R_{\rm e,gal}$ relation \citep{2009ApJ...706L..86N}.  This
relation will be explored further in \citet{HGS2022} with a sample twice that
used here and having multicomponent decompositions and a consistent set of
$\Upsilon_*$ ratios.


In the left-hand panel of Figure~\ref{Fig_evolve_IP13}, we see the black hole
masses versus the half-light radii of the host spheroids, as measured from the
geometric-mean axis.  Given the strong relation between the sizes and the
masses of the {\em spheroids} seen in Figure~\ref{Fig_R_Msph_IP13}, it is not
too surprising that the structure in the $M_{\rm bh}$--$R_{\rm e,sph}$ diagram
(Figure~\ref{Fig_evolve_IP13}) shows a similarity to that seen in the $M_{\rm
  bh}$--$M_{\rm *,sph}$ diagram (Figure~\ref{Fig_M_Msph_IP13}).  Using the
$M_{\rm *,sph}$--$R_{\rm e,sph}$ diagram (Figure~\ref{Fig_R_Msph_IP13}), one
can map the expected shift in the $M_{\rm bh}$--$M_{\rm *,sph}$ diagram for
equal-mass mergers of S0 galaxies that produce an E galaxy.  This is shown in
the right-hand panel of Figure~\ref{Fig_evolve_IP13}, and can be understood in
terms of the galaxies effectively folding in their disc stars to make the new
E galaxy and thereby lowering the $M_{\rm bh}/M_{\rm *,sph}$ ratio, as seen in
Figure~\ref{Fig_msph_rat_IP13}.  This scenario also readily explains the
offset between bulges and E galaxies seen in Figure~\ref{Fig_evolve_IP13}.
For example, a merger of two equal-mass S0 galaxies with B/T=0.25 \citep[e.g.,
][]{2005MNRAS.362.1319L, 2008MNRAS.388.1708G} will double $M_{\rm bh}$ and
increase the spheroid (now elliptical galaxy) mass 8-fold once the disc light
is incorporated.  In the $M_{\rm bh}$--$M_{\rm *,sph}$ diagram, this merger
moves a high stellar mass S0 up by 0.3 dex and across by 0.9 dex, placing it
on the sequence of elliptical galaxies.
%
%
From the relation $M_{\rm *,sph} \propto R_{\rm e,sph,eq}^{1.14\pm0.04}$ 
(based on the Spitzer sample used here), we have that a 0.9 dex
increase in $M_{\rm *,sph}$ is associated with a 0.79 dex 
increase in $R_{\rm e,sph,eq}$.  Such an increase from a major merger event is
plotted in Figures~\ref{Fig_R_Msph_IP13} and \ref{Fig_evolve_IP13}.
Figure~\ref{Fig_evolve_IP13} provides a previously unstated measure-of-sorts
of the average number of major mergers the E galaxies in our sample have
experienced.  There is evidence here, and in Figure~\ref{Fig_mgal_rat_IP13},
that BCGs 
have experienced a greater number of mergers, and this will be explored in more
detail in a subsequent paper.

\section{Discussion}\label{Sec_Disc_1}

\subsection{Overmassive and undermassive black holes}\label{Sec_Disc_2}

\begin{figure}
\begin{center}
\includegraphics[trim=0.0cm 0cm 0.0cm 0cm, width=1.0\columnwidth,
  angle=0]{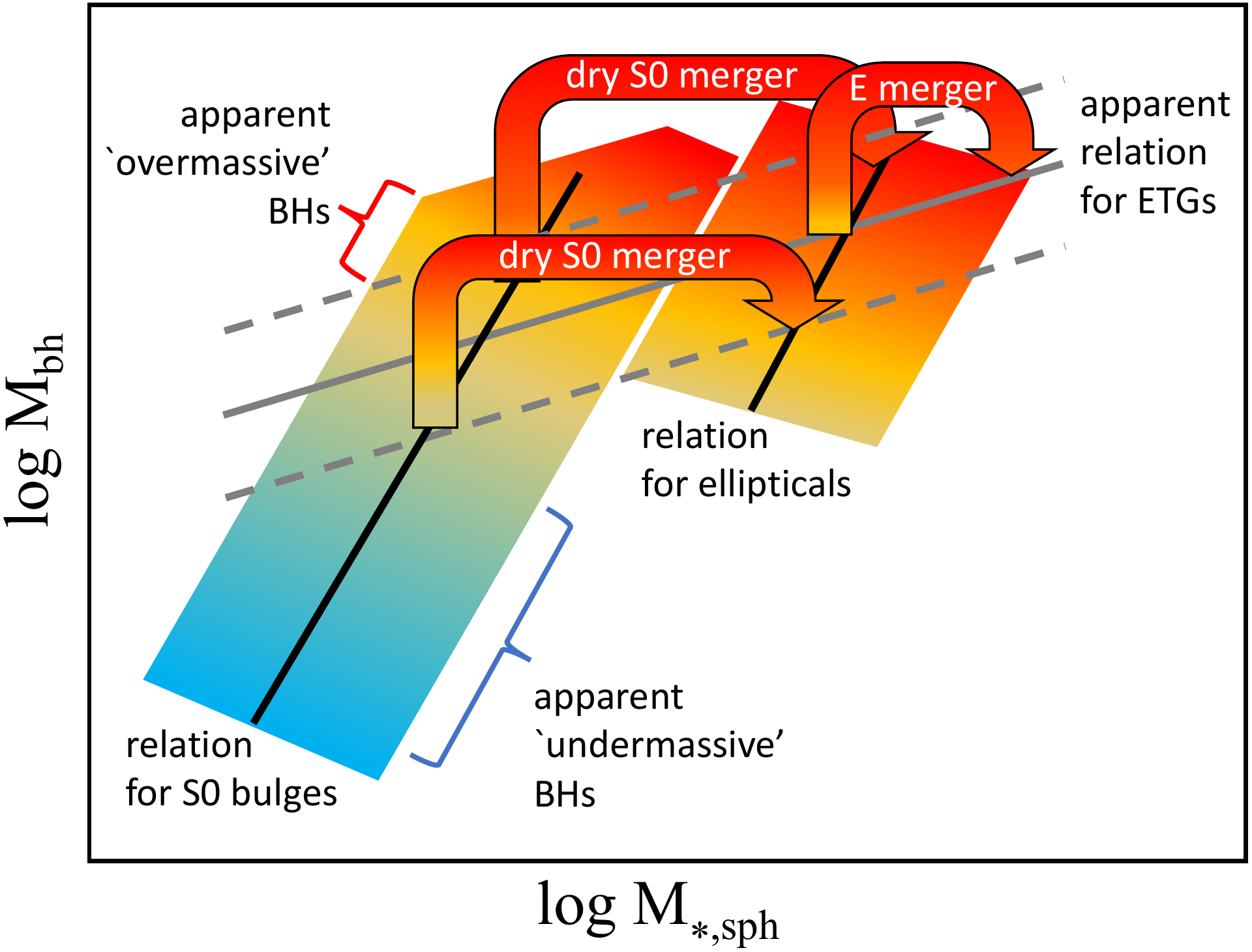} 
\caption{Schematic of the $M_{\rm bh}$--$M_{\rm *,sph}$ diagram for ETGs. 
The steep band on the left shows a relation for bulges (in S0 and ES
galaxies). 
The steep band on the right shows the relation for E galaxies. 
The shallower relation, shown in grey, tracks an apparent 
`red sequence' obtained by sampling some massive bulges and some elliptical galaxies. 
Past claims for apparently overmassive and undermassive black holes, 
relative to this near-linear `red sequence', can be understood in
terms of the host bulge belonging to the (steeper) bulge sequence.
}  
\label{Fig_schematic}
\end{center}
\end{figure}

Early observational bias favouring the detection of systems with big black
holes led to samples dominated by elliptical galaxies and lenticular galaxies
with massive bulges.  This sample selection produced an apparent near-linear
$M_{\rm bh}$--$M_{\rm *,sph}$ relation, which steepened as lower-mass black
holes were gradually included.  Figure~\ref{Fig_schematic} reveals how this
near-linear `red sequence' in the $M_{\rm bh}$--$M_{\rm *,sph}$ diagram arises
by sampling both massive bulges and elliptical galaxies.  For many such
elliptical galaxies, their spheroid mass may be dominated by the disc masses
of their progenitor galaxies.  This explains the approximately order of
magnitude lower $M_{\rm bh}/M_{\rm *,sph}$ ratio in elliptical galaxies when
compared to bulges of the same `spheroid' mass
(Figure~\ref{Fig_msph_rat_IP13}).  As noted earlier, this is not because the
galaxies' disc masses are excluded from the $M_{\rm bh}$--$M_{\rm *,sph}$
diagram; the $M_{\rm bh}/M_{\rm *,gal}$ ratio is not equal among ES/S0 and E
galaxies at a given galaxy stellar-mass (Figure~\ref{Fig_mgal_rat_IP13}).

The notion of a `red sequence' representing the underlying fundamental
connection between black holes and `classical bulges', i.e., bulges built by
mergers, introduces problems that disappear when considering the $M_{\rm bh}$--$M_{\rm
  *,sph}$ diagram in terms of a bulge sequence and an offset merger-built
population of elliptical galaxies.  Most obvious is that the E galaxies do not
follow the (near-linear) red sequence but define a steeper non-linear relation
(see Table~\ref{Table-IP13}).  In addition, the low- and high-mass bulges
appear as outliers from the near-linear `red sequence', invoking a misleading perception
as to the need for separate formation physics.  It had led to the notion that
massive bulges and relic galaxies are a disconnected population with
overmassive black holes relative to galaxies on the near-linear relation (see
Figure~\ref{Fig_schematic}).
They are, however, not overmassive relative to the bulge sequence.
Furthermore, while some black holes in BCGs appear overmassive relative to
the `red sequence', they are not overmassive relative to the elliptical galaxy
$M_{\rm bh}$--$M_{\rm *,sph}$ sequence.  By appreciating the role of mergers,
we can understand how the morphology-dependent relationships arose in the
$M_{\rm bh}$--$M_{\rm *,sph}$ and $M_{\rm bh}$--$M_{\rm *,gal}$ diagrams.  The
near-linear red-sequence also resulted in claims that low-mass bulges were yet
another disconnected population with undermassive black holes relative to
galaxies on the red-sequence.  However, they are not undermassive relative to
the bulge sequence.  We again note that while our sample of bulges does
contain members which reside below the `red sequence', they are not the
(peanut shell)-shaped structures associated with unstable bars, nor are they
nuclear or inner discs which we model with separate components.

\subsubsection{Relic red nuggets at the top of the bulge sequence}\label{Sec_Disc_3}

The gaseous processes that gave rise to some bulges may have occurred long
ago.  Indeed, many local bulges could be
the descendants of the `red nuggets' observed at $z \approx 2.5 \pm 1$
\citep{2005ApJ...626..680D, 2011ApJ...739L..44D} and potentially 
now cloaked in a large-scale disc \citep[][and references
  therein]{2015ApJ...804...32G, 2016MNRAS.457.1916D, 2022MNRAS.514.3410H}.  If any of the
high-$z$ red nuggets did not acquire a disc by today---which may be likely if they
started life in a proto-(galaxy cluster), given (i) the propensity for
ram-pressure stripping of cold gas by hot gas, and (ii) the reduction in
galaxy mergers due to high fly-by speeds---, then they will remain a compact massive galaxy
today \citep[e.g.,][]{1966ApJ...143..192Z, 1968cgcg.bookR....Z,
  1971cscg.book.....Z, 2010MNRAS.408L..21S, 2010ApJ...712..226V,
  2013ApJ...777..125P, 2014ApJ...780L..20T}.  Such local `compact galaxies' are
also referred to as `relic galaxies' if their stars are old \citep[e.g.,][]{2017MNRAS.467.1929F}.  
In separating spheroid types, the `relic galaxies' are seen not to be associated 
with the merger-built E galaxies in the $M_{\rm bh}$--$M_{\rm *,sph}$ diagram 
but rather sit on the bulge sequence,
with the most massive red relics located at the top.  
Consequently, Figure~\ref{Fig_schematic} reveals
how massive relic galaxies appear to have overmassive black holes relative to
the near-linear `red sequence' but are consistent with the bulge sequence for ETGs.

NGC~1332 is not an elliptical galaxy but a 
relic ES galaxy which has not acquired/built a large-scale disc.
The dominant spheroidal component in NGC~1332 has $\log(M_{\rm *,sph}/M_{\odot}) =
11.15\pm0.15$ dex and a geometric mean radius $R_{\rm e,sph} \approx 1.9$~kpc.
It has the second highest black hole mass of the
ES+S0 galaxies in our sample, and can be seen to reside at the top of the
bulge sequence in Figure~\ref{Fig_evolve_IP13}.
We have also labelled the ES galaxy NGC~6861 in Figure~\ref{Fig_evolve_IP13}, which has $\log(M_{\rm
  *,sph}/M_{\odot}) = 11.07\pm0.19$ and $R_{\rm e,sph} \approx 2.6$~kpc.



If a high-$z$ `red nugget' acquired a 
disc over time, then today the `red nugget' would be the compact massive spheroid of a disc galaxy.  
NGC~5252, for example, likely has such a relic bulge \citep{2019ApJ...876..155S}; 
and also a relic quasar \citep{2005A&A...431..465C}.  Given the old ages of
discs in massive lenticular galaxies, the bulk of their stars formed long
ago, no doubt acquired through direct accretion and mergers but also
possibly via star-formation in their gas discs at cosmic noon \citep[e.g.,
][]{2020MNRAS.497.3273F}.  


\subsubsection{Merger-built Brightest Cluster Galaxies}

The creation of BCGs via (multiple) mergers produces the largest elliptical
galaxies, found at the centres of galaxy groups and clusters.  The
steeper-than-linear $M_{\rm bh} \propto M_{\rm *,sph}^{1.72}$ relation seen in
\citet[][their Figure~3]{2018ApJ...852..131B} for BCGs is explained here as a
combination of mergers folding in the disc mass and a steep origin relation
for the bulges of the pre-merged progenitor galaxies in the $M_{\rm
  bh}$--$M_{\rm *,sph}$ diagram.  That is, the elliptical galaxies, which
include the BCGs\footnote{Our galaxy stellar-masses are derived by excluding
  the surrounding ICL light, either because it was fit as a component during
  the galaxy decomposition and excluded here, or because the images and light
  profiles were not deep enough for the ICL to have been an issue.}, should
not be thought of as a departure from a near-linear $M_{\rm bh}$--$M_{\rm
  *,sph}$ relation.  Instead, they represent a shift to a somewhat parallel
relation to that defined by the bulges of ETGs.  Of course, when E$+$E dry
mergers build new E galaxies, the evolutionary path in the $M_{\rm
  bh}$--$M_{\rm *,sph}$ diagram will be along a vector with a slope of 1.  One
might imagine that in the Universe's distant future, one would start to see a
linear relation for BCGs emerge from the top end of the current relation for
the E galaxies.  This is, however, something which we will leave for the
simulators.

\subsection{The stripped S0 galaxy M32}\label{Sec_Disc_4}

In Figure~\ref{Fig_R_Msph_IP13}, 
M32 appears to the left of the $M_{\rm bh}$--$M_{\rm *,sph}$ relation defined 
by bulges in ETGs. However, it resides within the 1$\sigma$ scatter about this
relation.  The slight preference to the left may reflect that the bulge, along with the disc, in
M32 has been eroded by its massive neighbour, M31.  This process can reduce
the bulge mass and inflate its half-light radius.  `Compact elliptical' (cE) 
galaxies like M32, which have lost the gravitational tug-of-war to
retain ownership of their stars \citep{Roche:1850, 1972ApJ...178..623T}, 
stand out in the galaxy colour-magnitude 
diagram due to their low luminosity for their colour 
\citep[e.g.,][their Figure~11]{2019MNRAS.484..794G}.
For M32, 
the $V-I$ (Vega) colour of $\sim$1.2--1.4 mag
\citep{1998AJ....116.2263L} implies $M_*/L_I \approx 3 M_\odot/L_\odot$
\citep{2022arXiv220202290S}, or $\sim$2.4 after converting to a \citet{2002Sci...295...82K}
IMF.  Coupled with 0.09 mag of Galactic extinction, the
absolute magnitude for the spheroidal component of M32,
$M_I = -17.0$ mag \citep[Vega: ][]{2009MNRAS.397.2148G}, corresponds to
$M_* \approx 0.8\times10^9\,M_\odot$.  
Performing a multicomponent fit to M32's 
major-axis light profile, \citet{2009MNRAS.397.2148G} measured an effective
half light-radius of 26$\arcsec$.3 for the bulge component. 
For an ellipticity of 0.3 at this radius, this translates to an equivalent-axis $R_{\rm
  e,sph,eq}=22\arcsec$.  Using a scale of 4~pc per 1$\arcsec$, this angular size is
equal to a physical size of 88~pc, as shown in Figure~\ref{Fig_evolve_IP13}. 

We add the dwarf E galaxy Pox~52 (93 Mpc distant), with 
$M_{\rm bh} = (3.2\pm1)\times10^5$ M$_\odot$ and 
$M_{\rm *,sph}=1.2\times10^9$ M$_\odot$.  We use 
$R_{\rm e,eq}=436$~pc \citep{2008ApJ...686..892T}, 
based on a minor-to-major axis ratio $b/a = 0.79$ \citep{2004ApJ...607...90B} and 
$R_{\rm e,maj}=490$~pc \citep{2008ApJ...686..892T}. 
Pox~52 follows both the $M_{\rm bh}$--$M_{\rm *,sph}$ and $M_{\rm
  bh}$--$R_{\rm e,sph}$ relations well. 

There is another spheroid in our sample, albeit not used in our regression
analyses, with a smaller mass and  size than that of M32. 
The dwarf S0 galaxy NGC~404 
can be seen in Figure~\ref{Fig_evolve_IP13} to follow the 
S0 galaxy sequence in the $M_{\rm bh}$--$R_{\rm e,sph}$ diagram but reside to
the left of the S0 galaxy sequence in the $M_{\rm bh}$--$M_{\rm *,sph}$ diagram. 
This LINER galaxy has a bright ($r \approx 2\arcsec$) nuclear spiral pattern \citep[compare
  CG~611: ][]{2017ApJ...840...68G} and 
is encircled by a much larger H\,{\footnotesize I} gas disc with knotty, irregular 
tendrils of UV hotspots and H\,{\footnotesize II} regions 
\citep{2010ApJ...716...71W, 2013ApJ...772L..23B}. 
However, this galaxy is excluded from the fitting process because 
its location at the lower extremum of our data might excessively torque the fitted
relation.  This becomes problematic using such a datum if its 
measurements are in error or if the scaling relation does not extend linearly to such
low black hole masses.  \citet{2017ApJ...836..237N} reported a
3$\sigma$ upper limit to the black hole mass of $1.5\times10^4$ M$_\odot$,
which was recently revised to  
$M_{\rm bh} = 5.5^{+4.1}_{-3.8}\times10^5$ M$_\odot$
\citep{2020MNRAS.496.4061D}.   In passing, we note how this discrepancy highlights the affect of
systematic errors not captured by the small formal/random errors typically reported for
most black hole mass measurements.  We also attach a 0.5 dex uncertainty to
our spheroid mass, which may be three times less massive than our adopted
value from \citet{2019ApJ...876..155S} if this galaxy has an anti-truncated
disc \citep{Graham:Sahu:22}, resulting in a steeper inner-disc component at
the expense of the bulge.

LEDA~87000 is a galaxy that likely harbours a central intermediate-mass black
hole \citep{2015ApJ...809L..14B}. 
Although \citet{2016ApJ...818..172G} observed it to follow the near-quadratic 
$M_{\rm bh}$--$M_{\rm *,sph}$ relation followed by LTGs, inspection of subsequent Hubble
Space Telescope images reveals that the previously poorly-resolved `barge'
component\footnote{`Barge' is an amalgamation of Bar$+$Bulge.} ---
as seen in ground-based images --- was all bar and no bulge
\citep{2017ApJ...850..196B}. 
This represents something of a growing trend in which the closer one looks,
the more `bulges' --- when simply defined as the excess of light above the inward
extrapolation of an outer exponential disc --- retreat by giving up ground to
bars or other features \citep[e.g.,][]{2003ApJ...582L..79B,
  2003ApJ...597..929E, 2005MNRAS.362.1319L, 2022MNRAS.514.3410H}.





\subsection{The primary relation}
\label{Sec_Disc_5}

The larger, merger-built elliptical galaxies are seen to define a secondary, or at
least subsequent, relation in the $M_{\rm bh}$--$M_{\rm *,sph}$ diagram.  
In terms of Darwinian evolution on a galaxy scale, they can be thought of 
as coming into existence via punctuated equilibrium rather than gradualism.
Major dry mergers between S0 galaxies, 
in which the S0 galaxies effectively fold in all their disc stellar mass to create an
elliptical galaxy, are accompanied by a substantial oversized jump in the stellar mass
(relative to the jump in the black hole mass) and a large jump in
the half-light size of the new spheroid, i.e., the elliptical
galaxy.  Such evolution explains the two prominent relations
observed in the $M_{\rm bh}$--$M_{\rm *,sph}$ diagram and the
$M_{\rm bh}$--$R_{\rm e,sph,eq}$ diagram for ETGs (Figure~\ref{Fig_evolve_IP13}).


Broadly speaking, some bulges may have arisen from a kind of rapid monolithic
collapse, or at least result from an early-formation process that 
created the observed high-$z$ `red nuggets', 
while most elliptical galaxies likely formed 
from a binary merger or hierarchical merging (in the case of the BCGs)
over the age of the Universe.  
As such, a meaningful cosmological probe into the evolution of the galaxy/black hole scaling
relations needs to be mindful of the galaxy morphology.  For example, a sample of
elliptical galaxies at $z=1$ can not be directly compared with a
sample of bulges at $z=0$; to do so would be comparing apples and oranges. 

To summarise, the notion of a single near-linear $M_{\rm bh}$--$M_{\rm
  *,sph}$ relation is inadequate and seems to offer misdirection in
understanding galaxies and black holes.  
The averaging of black hole and galaxy masses 
through mergers has established neither the expected nor an observed 
near-linear $M_{\rm bh}$--$M_{\rm *,sph}$ relation. 
While we have presented the most accurate 
$M_{\rm bh}$--$M_{\rm *,sph}$ diagram to date, and interpreted the 
broad brush stroke near--parallel relations shown in
\citet{2019ApJ...876..155S}, 
there is further information to be gleaned from this diagram. 
Mergers, both wet and dry, which do not fold in all of the disc's stellar mass will be
addressed in \citet{Graham:Sahu:22}, where we develop something of a phylogenetic
tree diagram within the bivariate space of $M_{\rm bh}$ versus $M_{\rm
  *,sph}$.  

One should expect to observe morphology-dependent substructure in other black hole scaling
diagrams. For example, as previously noted, the broad red/blue sequence for
ETGs/LTGs in the $M_{\rm bh}$--$M_{\rm *,gal}$ diagram has been observed in
the $M_{\rm bh}$--colour diagram \citep{2020ApJ...898...83D}.  This broad division 
may also appear in the $M_{\rm bh}$--metallicity, $Z$, diagram
\citep{2003ApJ...596...72W, 2008MNRAS.390..814K}. 
Depending how the number of globular clusters, $N_{\rm GC}$, traces a galaxy's
stellar mass \citep[][and references therein]{2014A&A...565L...6M}, one may
also expect the ETGs and LTGs to follow different trends in the $M_{\rm
  bh}$--$N_{\rm GC}$ diagram \citep{2009MNRAS.392L...1S, 2010ApJ...720..516B}.
The number of red and blue globular clusters 
around each galaxy may yield yet further subdivisions
\citep[see][]{1998AJ....116.2841K, 2002A&A...395..761K}, as may their
kinematics \citep[e.g.,][]{2012MNRAS.426L..51S, 2013MNRAS.433..235P}.

With our new understanding of the relevance of galaxy morphology and galaxy 
formation history, the role of mergers, and refined insight into what may be
considered the primary relations versus their modified/evolved form, one is
also better placed to tackle the question of whether or not a two-dimensional
plane within a three-dimensional space may provide an improved description
over bivariate linear relations.  For example, does a third axis, in addition
to $M_{\rm bh}$ and $M_{\rm sph}$, uncover a distribution on a more
fundamental plane?  Our analysis, considering additional parameters obtained
from physically-motivated multicomponent decomposition, such as the spheroid
S\'ersic index and stellar density \citep[e.g.,][]{2007ApJ...655...77G,
  2016ApJ...818...47S, 2022ApJ...927...67S}, along with spheroid size, mass,
and velocity dispersion, will be presented in a forthcoming paper.  Here, we
restrict ourselves to briefly explaining why the combination of $M_{\rm bh}$, $M_{\rm
  *,sph}$, and $R_{\rm e,sph,eq}$ (or equally\footnote{The term $\langle I
  \rangle _{\rm e}$ is the mean intensity within $R_{\rm e}$.
Given that $M_{\rm *,sph} \propto R_{\rm e,sph,eq}^2 \langle I \rangle _{\rm
  e}$ (by definition), modulo the (stellar mass)-to-light ratio, and given $M_{\rm
  *,sph} \propto R_{\rm e,sph,eq}^{1.14}$ (Figure~\ref{Fig_R_Msph_IP13}), 
we have that 
$\langle I \rangle _{\rm e} \propto R_{\rm e,sph,eq}^{-0.86}$
and thus
$M_{\rm *,sph} \propto \langle I \rangle _{\rm e}^{-1.33}$.}  $\langle I
\rangle _{\rm e}$) may {\em not} produce a useful plane about which the
scatter in the $\log(M_{\rm bh})$ direction is less than that seen about the
$M_{\rm bh}$--$M_{\rm *,sph}$ relation. 

For the following thought experiment, we can consider two parallel relations
in the $M_{\rm bh}$--$M_{\rm *,sph}$ diagram, one for S0 galaxy bulges and the
other for an offset population of merger-built E galaxies. We can use the
knowledge that the (logarithm of the) half-light spheroid radius scales with
the (logarithm of the) spheroid stellar mass (Figure~\ref{Fig_R_Msph_IP13}).
One way to think of the problem is that we wish to introduce an $R_{\rm
  e,sph,eq}$ term to effectively shift the E galaxies to the left in the
$M_{\rm bh}$--$M_{\rm *,sph}$ diagram, to make them overlap with the bulges
and thereby reduce the scatter seen in this diagram (see
Figure~\ref{Fig_evolve_IP13}).  However, we need to bear in mind that this
procedure will also shift the bulges to the left, given that we are assuming
no knowledge of morphology and just using the parameters $M_{\rm bh}$, $M_{\rm
  *,sph}$, and $R_{\rm e,sph,eq}$.  It turns out that to achieve overlap of
the elliptical and bulge samples, the necessary subtraction of a $\log\,R_{\rm
  e}$ term from the $\log\,M_{\rm *,sph}$ values in the $M_{\rm bh}$--$M_{\rm
  *,sph}$ diagram results in a near-vertical distribution of points with a
near-infinite slope.  Remember, $R_{\rm e,sph,eq}$ scales almost linearly with
$M_{\rm *,sph}$.  This shall be shown in a forthcoming paper but we felt it
was of sufficient interest to provide some initial insight here.
%


\subsection{Is there a role for AGN feedback in shaping the turnover and (high
  mass)-end of the galaxy mass function?}\label{Sec_Disc_6}

The observational results herein represent a considerable departure
from the connection galaxies are often claimed or thought to have with their
central black hole.  More accurate spheroid masses --- particularly from a
greater awareness that many ETGs are S0s rather than Es --- 
have revealed how the coevolution of {\em bulges} and supermassive black holes have built a
super-linear\footnote{We use the term `super-linear' to denote a power-law
   with a slope steeper than 1 but not as high as 2.} or near-quadratic\footnote{We use the term
  `near-quadratic' to describe a power-law with a slope close to a value of 2.}
$M_{\rm bh}$--$M_{\rm *,sph}$ relation.  
\citet{2012ApJ...746..113G} and \citet{2013ApJ...764..151G} highlighted this steeper 
slope and discussed how dry mergers 
might be producing an offshoot of core-S\'ersic galaxies, creating (what was
thought to be) a 
near-linear slope at high black hole masses in the $M_{\rm bh}$--$M_{\rm *,sph}$ diagram.
However, this idea did not 
account for the incoming disc mass during some mergers, or for the more recent observation
that merger-built elliptical galaxies (with and without depleted cores) do
not follow a near-linear $M_{\rm bh}$--$M_{\rm *,sph}$ relation.  
Obviously, AGN feedback has thus also not produced a near-linear $M_{\rm bh}$--$M_{\rm *,sph}$
relation for the elliptical galaxies.  Moreover, the 
location of the elliptical galaxies in the $M_{\rm bh}$--$M_{\rm *,sph}$ diagram
would appear to not be due to AGN feedback 
but rather major mergers in 
which the angular momenta of the progenitor galaxy's discs have largely
cancelled.  This observation is apparent from 
the $M_{\rm bh}$--$M_{\rm *,sph}$ diagram, 
the $M_{\rm bh}/M_{\rm *,sph}$ ratios (Figure~\ref{Fig_msph_rat_IP13}), 
the $M_{\rm bh}/M_{\rm *,gal}$ ratios (Figure~\ref{Fig_mgal_rat_IP13}) 
and the $M_{\rm bh}$--$R_{\rm e,sph}$ diagram (Figure~\ref{Fig_evolve_IP13}).

While `quasar mode' AGN feedback (discussed in Section~\ref{Sec_imp}) 
might contribute to a link between black hole mass 
and {\em bulge} mass for some lower mass systems, it is not yet well
established how much it may regulate the gas and star formation in the discs
of galaxies \citep[e.g.,][]{2014MNRAS.441.1615G, 2018ApJ...869..113D}.  Given
that most of the stellar mass in disc galaxies resides in their discs, with
$B/T < 0.5$ for most S0 and S galaxies \citep{2008MNRAS.388.1708G}, the role
of AGN in shaping the {\em galaxy} stellar-mass function appears limited.  Given that
mergers, rather than AGN feedback, have likely built the elliptical galaxies which
dominate the high-mass end of the galaxy mass function \citep[e.g.][]{2022MNRAS.513..439D}, 
the scope for AGN feedback driving and shaping coevolution in high-mass galaxies
appears quenched \citep[e.g.,][]{2003ApJ...599...38B, 2013MNRAS.436.1750R,
  2022MNRAS.511..506C}.

\begin{figure}
\begin{center}
\includegraphics[trim=0.0cm 0cm 0.0cm 0cm, width=1.0\columnwidth,
  angle=0]{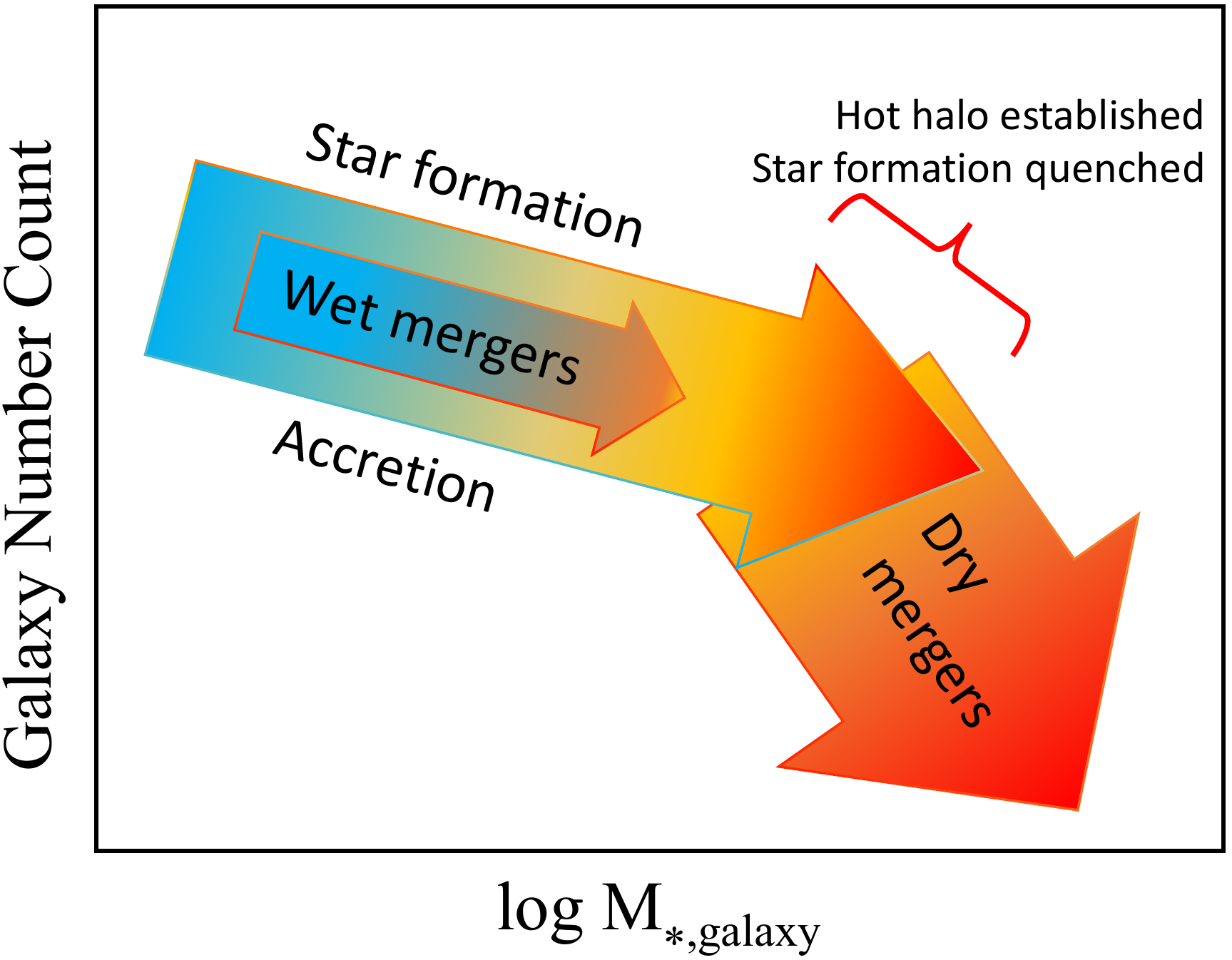} 
\caption{Schematic of the galaxy stellar-mass function.  Here we speculate, with
  reasoning in Section~\ref{Sec_Disc_6}, as to the nature of the galaxy
  stellar-mass function.  While AGN feedback might regulate bulge growth and
  help to establish the $M_{\rm bh}$--$M_{\rm *,sph}$ relation in some disc
    galaxies, it may be the hot gas halo which curtails star formation in discs
    and keeps it truncated in elliptical galaxies. Mergers, rather than AGN
    feedback, appear to have established the elliptical galaxy $M_{\rm bh}$--$M_{\rm
      *,sph}$ relation (Figures~\ref{Fig_evolve_IP13} and \ref{Fig_schematic}).
}
\label{Fig_schematic_MF}
\end{center}
\end{figure}

Of course, a galaxy does not need to blow out its gas --- via, say, supernovae
or an AGN --- in order to cause a cessation of star formation.  A galaxy could
instead prevent the cooling of gas which might form stars
\citep{2003ApJ...599...38B}.  Creating a hot gas halo in/around massive
pressure-supported spheroids may have this effect \citep[e.g.,][see our
  Figure~\ref{Fig_schematic_MF}]{2020MNRAS.491.1311M}.  While star formation
and stellar winds might not generate the escape speeds required to clear gas
from a massive galaxy (and its dark matter halo), they contribute a hot gas
source, as does gas `shock-heating' during a galaxy collision \citep[e.g.,
][]{2019ApJ...878..161J}.  X-ray sputtering from hot gas
\citep{2021A&A...649A..18G} also breaks up dust clouds and thereby removes the
shielding from ionising radiation that dust may have provided potential
stellar nurseries.  Furthermore, these winds can keep the AGN `pilot light' on
by supplying low-level fuelling \citep[e.g.,][]{1991ApJ...376..380C,
  2006ApJ...640..143S} for the AGN. We term such an energy source a `Benson
burner'.\footnote{This is a play on words combining Bunsen burner - used for
  heating and sterilisation - and a reference to the idea sparked
  by \citep{2003ApJ...599...38B}.}

Should hot gas halos efficiently suppress star formation, 
then rather than ejecting gas which might form stars, 
it is about acquiring and retaining (hot) gas.  The system needs to
be capable of maintaining, and thus also massive enough to retain, a hot gas halo rather than
have it evaporate or collapse into a disk where it may cool and form stars.  
A {\em hot 'n dry} (hot gas and dry merger) combination may help 
explain the upper-end of the galaxy mass function where 
star formation has dwindled or ceased.
Unlike energetic but directional AGN jets \citep[which can both
suppress and trigger star formation: ][]{2013ApJ...772..112S, 
  2014AN....335..531G, 2015ApJ...799...82C}, a hot gas halo can permeate the
entire galaxy, including the disc. 
The relation between black hole mass and both X-ray gas temperature and luminosity 
\citep{2018ApJ...852..131B, 2019ApJ...884..169G, 2019MNRAS.488L.134L} 
may add credibility to this picture. 
Low levels of omni-directional particle outflows and electromagnetic radiation
from the central `Benson burner' would also help counter cooling \citep{2006MNRAS.368L..67B,
  2006MNRAS.370..645B, 2006MNRAS.365...11C, 2017MNRAS.465...32B} 
--- seen as X-ray radiation coming from the hot gas halo 
\citep[e.g.,][]{1976ApJ...207..460S, 1984MNRAS.208..185N}.  This would help
hold star-formation at bay, at least in a closed-box model with no substantial
infall of cold gas \citep[e.g.,][]{2009Natur.457..451D}.  
When cooler gas is available, sporadic feeding and associated percolation
events may produce bubbles and cavities observed at various wavelengths 
\citep{1984Natur.310..568S, 1985PASJ...37..359T, 2003ApJ...582..246B, 2004ApJ...607..800B}. 
However, this so-called `radio mode' AGN feedback would only maintain
the $M_{\rm bh}$--$M_{\rm *,sph}$ relation, which we have argued is established by other means
\citep{2016MNRAS.463.3948D, 2019SSRv..215....5W}.  

For the first time, 
we have used the black hole scaling relations to confirm that 
AGN primarily have a caretaker role among elliptical galaxies, and we have
revealed how mergers rule the roost and dictate the $M_{\rm bh}/M_{\rm *,sph}$ ratio and
presumably also the $M_{\rm *,sph}/M_{\rm dark matter}$ ratio
\citep{2016MNRAS.463.3948D, 2021MNRAS.507.4274M}.   
This result is tied to the offset trend seen for elliptical galaxies in
the $M_{\rm bh}$--$M_{\rm *,sph}$, $M_{\rm bh}$--$M_{\rm *,gal}$ and $M_{\rm
  bh}$--$R_{\rm e,sph}$ diagrams.  It is not due to spheroids with 
partially depleted cores, which some E galaxies have but others do not, and
which some S0 galaxy bulges possess.  Such spheroids, whose central `phase space' is
depleted of stars, tend to occupy the $M_{\rm bh}$--$M_{\rm *,sph}$ `red
sequence', which is a `red herring' due to the
partial picture it provided. In particular, it missed the wet and damp
mergers, 
and thus the steep $M_{\rm bh}$--$M_{\rm *,sph}$ relation for the ensemble of 
elliptical galaxies.  We will pursue this further by addressing 
merger-built lenticular galaxies with depleted cores, such as NGC~5813, 
and major wet mergers, for example NGC~5128, in \citet{Graham:Sahu:22}.



\subsection{Some further thoughts}\label{Sec_imp}

It is evident that the coevolution of bulges and their central black holes
have not produced a simple near-linear $M_{\rm bh}$--$M_{\rm *,sph}$ relation.
The steep $M_{\rm bh}$--$M_{\rm *,sph}$ relation for bulges has implications
for countless simulations, semi-analytic works, theories, and papers that may
have calibrated themselves to a near-linear $M_{\rm bh}$--$M_{\rm *,sph}$
relation.  For example, as shown by \citet[][their
  Figure~6]{2018ApJ...852..131B}, while the Horizon-AGN simulation
\citep{2014MNRAS.444.1453D} produces an $M_{\rm bh}$--$M_{\rm *,sph}$ `red
sequence' with a slope around 1.1 to 1.2, it does not have the scatter to
accommodate the steeper relations defined by either the bulges or the
elliptical galaxies.  That is, it appears to have not captured the key
merger-induced jump from bulges to elliptical galaxies.  While some studies
are ahead of the pack, producing steeper relations
\citep[e.g.,][]{2006MNRAS.373.1173F, 2012MNRAS.420.2662D, 2012MNRAS.423.2397K,
  2017MNRAS.472L.109A, 2018MNRAS.479.4056W, 2019ApJ...885L..36D,
  2020MNRAS.494.2747M, 2022MNRAS.513.3768I, 2022MNRAS.511.5756T}, it has been hard for the notion
of a steep $M_{\rm bh}$--$M_{\rm *,sph}$ relation to get oxygen given the
significant paradigm shift that it implies.  It is, therefore, perhaps worth
reiterating an element from \citet[][their section~4.3]{2013ApJ...764..151G},
which introduced a related revision to the `quasar mode' (aka cold-gas mode)
of black hole growth used in some semi-analytic models
\citep{2004ApJ...600..580G, 2005MNRAS.361..776S}.

The steep $M_{\rm bh}$--$M_{\rm dyn,sph}$ relation detected by
\citet{2012ApJ...746..113G}, which supplanted the single linear relation from
\citet{2004ApJ...604L..89H}, 
challenged the past assumption of accretion-induced black 
hole growth that is linearly proportional to the inflowing mass
of cold gas. 
\citet[][their Equation~8]{2006MNRAS.365...11C} and others have popularised
this black hole feeding scenario to model how AGN outflows account for what
was thought to be  a linear $M_{\rm bh}$--$M_{\rm *,sph}$ relation.  
\citet{2013ApJ...764..151G} presented a revised prescription for the 
increase in black hole mass, $\delta M_{\rm bh}$, associated with wet mergers, such that 
\begin{equation}
\delta M_{\rm bh} \propto
\left(\frac{M_{\rm min}}{M_{\rm maj}}\right)
\left[ \frac{M^X_{\rm cold}}{1+(280\, {\rm km\, s}^{-1})/V_{\rm virial}}\right].
\label{EqQ} 
\end{equation}
The exponent $X$ represents the logarithmic slope of the $M_{\rm bh}$--$M_{\rm
  *,sph}$ relation for bulges, and they specified $X=2$. 
$M_{\rm min}$ and $M_{\rm maj}$ are the total baryonic masses from the
minor and major galaxies involved in the accretion/merger event, and $M_{\rm cold}$ is their combined cold
gas mass. The velocity $V_{\rm virial}$ is the merged system's circular or
`virial' velocity, normalised at 280 km s$^{-1}$ \citep{2000MNRAS.311..576K}. 
This modified equation may prove helpful for exploring and understanding galaxy/(black
hole) evolution through semi-analytic approaches, although it does not
encompass the cessation of star-formation due to hot X-ray halos in massive
systems, or the pivotal role of dry mergers in shaping the distribution seen in the 
$M_{\rm bh}$--$M_{\rm *,sph}$ diagram.

There are also significant ramifications for predictions of gravitational waves
from space-based interferometers \citep[e.g.,][]{2005LRR.....8....8M,
  2013CQGra..30x4009S, 2020MNRAS.492..256K, 2022arXiv220505099S} and 
pulsar timing arrays monitored with ground-based radio telescopes \citep[e.g.,
][]{2010CQGra..27h4013H, 2013Sci...342..334S, 2019MNRAS.488..401C}.
For example, the steep $M_{\rm bh}$--$M_{\rm *,sph}$ relation for 
bulges should be considered, if not favoured over the near-linear
`red-sequence' when assigning BH masses to galaxies in works attempting to estimate the
background signal from binary black hole mergers. One caveat here is that the mergers
involving a BCG may involve systems on both the bulge and the elliptical
sequence.  As noted in \citet{2019MNRAS.484..794G}, 
predictions for black hole masses will be too high if using the 
original near-linear $M_{\rm bh}$--$M_{\rm *,sph}$ relation at low spheroid
masses.  This over-prediction can result in over-looking potential populations of intermediate-mass
black holes ($10^2 < M_{\rm bh}/M_{\odot} < 10^5$).  Furthermore, application
of the steeper relation  has 
already been shown to result in an order of magnitude reduction to the 
expected detection rate of extreme mass ratio inspiral (EMRI) events from compact stellar-mass objects
around massive black holes \citep[e.g.,][]{2012A&A...542A.102M}.  That work
can be further refined based on the updated relations herein, providing better
expectations for what the European Laser
Interferometer Space Antenna ({\it LISA})
\citep{1997CQGra..14.1399D}
and TianQin \citep{2016CQGra..33c5010L} can hope to achieve based on their
current design plans. 

As noted above, the pursuit of long-wavelength gravitational waves, from the coalescence of binary
supermassive black holes \citep[e.g.,][]{2003ApJ...582L..15K,
  2006ApJ...646...49R, 2019ApJ...884...36L, 2022ApJ...926L..35O}, 
is an endeavour underway via monitoring  pulsar arrival
times using radio telescopes.   These studies will benefit from an improved knowledge of 
the varying $M_{\rm bh}/M_{\rm *,sph}$, and $M_{\rm bh}/M_{\rm *,gal}$, ratios
in pre-merged galaxies.  
This can enable revised predictions for, and possibly aid
in the tentative confirmation of, a long-wavelength gravitational wave
background \citep{2020ApJ...905L..34A, 2021MNRAS.508.4970C}. 

Related to the EMRI events are the nuclear stars clusters that coexist with
\citep[e.g.,][]{2008AJ....135..747G, 2008ApJ...678..116S, 
2009MNRAS.397.2148G} and feed \citep[e.g.,][]{2002RvMA...15...27K, 
  2002ApJ...576..753L} the central black hole in galaxies.  The
revised/steeper 
$M_{\rm bh}$--$M_{\rm *,sph}$ relations, coupled with the $M_{\rm
  nsc}$--$M_{\rm *,sph}$ relations \citep{2003ApJ...582L..79B,
  2003AJ....125.2936G}, led to the discovery of the $M_{\rm bh}$--$M_{\rm
  nsc}$ relation \citep{2016IAUS..312..269G, 2020MNRAS.492.3263G}.
This should be  
useful for modelling not only gravitational radiation events but also the
expected frequency of
tidal disruption events  \citep[TDEs:][]{2001astro.ph..6422K,
  2004ApJ...600..149W, 2015JHEAp...7..148K, 2016MNRAS.455..859S, 2017MNRAS.465.3840C,
  2020MNRAS.498..507T}, which have been observed in data dating back to 1990. 
There are currently 
around 100 such known events.\footnote{\url{https://tde.space/}}

If a non-rotating Schwarzschild-Droste \citep{1916AbhKP1916..189S,
  1917KNAB...19..197D} 
black hole is more massive than $\sim$10$^8$ M$_\odot$, and thus the 
gravitational gradient at, and beyond, the `event horizon' 
is not strong enough to pull a star apart, there will be no TDE
\citep{1999MNRAS.309..447M}.  
The star will cross the event horizon and disappear without
displaying its hot interior.  As we have seen, 
most of the systems with $M_{\rm bh} \lesssim 10^8$ M$_\odot$ 
follow the near-quadratic $M_{\rm bh}$--$M_{\rm *,sph}$ relation for bulges,
suggesting the need to use this steeper relation rather than the near-linear
`red-sequence', 
which pertains to (some) systems with $M_{\rm bh} \gtrsim 10^8$ M$_\odot$.
One should, however, be mindful that the 
spin-reduced size of the event horizon in 
a rotating \citet{1963PhRvL..11..237K} black hole can result in a 
star's tidal disruption radius being greater than the event horizon for black
hole masses up $\sim$7$\times 10^8$ M$_\odot$ for maximally spinning black
holes \citep{1992MNRAS.259..209B, 2012PhRvD..85b4037K}. 

The morphology-dependent black hole scaling relations also demand a 
re-examination of the virial $f$-factors used to convert AGN virial masses into
black hole masses \citep[e.g.,][]{2009ApJ...694L.166B, 2018ApJ...864..146B}. 
Failing to account for the different morphologies and formation history of
the spheroids hosting the AGN or inactive black holes will produce erroneous results. 

As noted in \citet{2012ApJ...746..113G}, there is a wealth of additional and
immediate  
implications and insight from the steeper-than-linear $M_{\rm bh}$--$M_{\rm *,sph}$
relations. These include black hole mass predictions in other galaxies,  
constructing the black hole mass function, and deriving the black hole mass 
density based on reliable bulge and elliptical galaxy mass functions. 
In passing, it is noted that some care with the Hubble-Lema\^itre constant, or
little $h$, is required for such calculations, as noted in \citet{2007MNRAS.380L..15G}
and \citet{2013PASA...30...52C}.


\section*{Acknowledgements}

This research was supported under the Australian Research Council's funding scheme DP17012923.
Part of this research was conducted within the Australian Research Council's
  Centre of Excellence for Gravitational Wave Discovery (OzGrav), through
  project number CE170100004.
This research has made use of the NASA/IPAC Extragalactic Database (NED) and
the NASA/IPAC Infrared Science Archive. 
We used the {\sc Rstan} package available at \url{https://mc-stan.org/}. 
We also used python packages numpy \citep{harris2020array}, 
matplotlib \citep{Hunter:2007} and SciPy \citep{2020SciPy-NMeth}.

\section{Data Availability}

The data underlying this article are available in the NASA/IPAC Infrared Science Archive.

\bibliographystyle{mnras}
\bibliography{Paper-BH-mass}{}

\appendix

\section{NIR-Optical colours and an alternative (stellar mass)-to-light ratio,
$\Upsilon$} 
\label{Apdx1}

As reported by \citet{2013MNRAS.430.2715I}, and previously noticed by
\citet{2001ApJ...550..212B}, optical-NIR colours are better indicators
of metallicity than stellar mass-to-light ratios.  They lack the sensitivity to
age which the optical colours can have.  Ironically, this may be advantageous
for our situation.  If bulges are mostly old \citep{1999MNRAS.310..703P,
  2009MNRAS.395...28M}\footnote{Old ages for (the stellar component of) bulges does not necessitate that
  they formed at high-$z$; they could be recently built bulges made out of old
  stars from the progenitor galaxies in a `dry merger' event, i.e., a gas-poor galaxy collision.}  
and the more recent star formation in the galaxies 
has occurred in their discs, then the galaxy optical colours used thus far in
this work could lead us 
astray.  Therefore, in this section, we repeat our analysis using the latest
(optical-infrared colour)-dependent $M_*/L$ ratios.  The outcome is that
our main conclusion stands, reflective of the small variation in the value of
$M_*/L$ at 3.6~$\mu$m.

\begin{figure*}
\begin{center}
\includegraphics[trim=0.0cm 0.0cm 0.0cm 0.0cm, height=0.3\textwidth, angle=0]{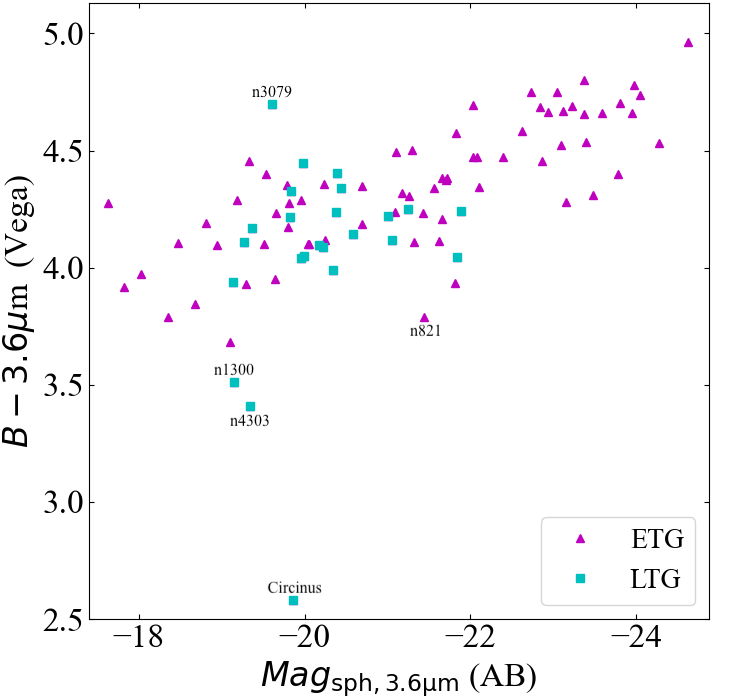} 
\includegraphics[trim=0.0cm 0.0cm 0.0cm 0.0cm, height=0.3\textwidth, angle=0]{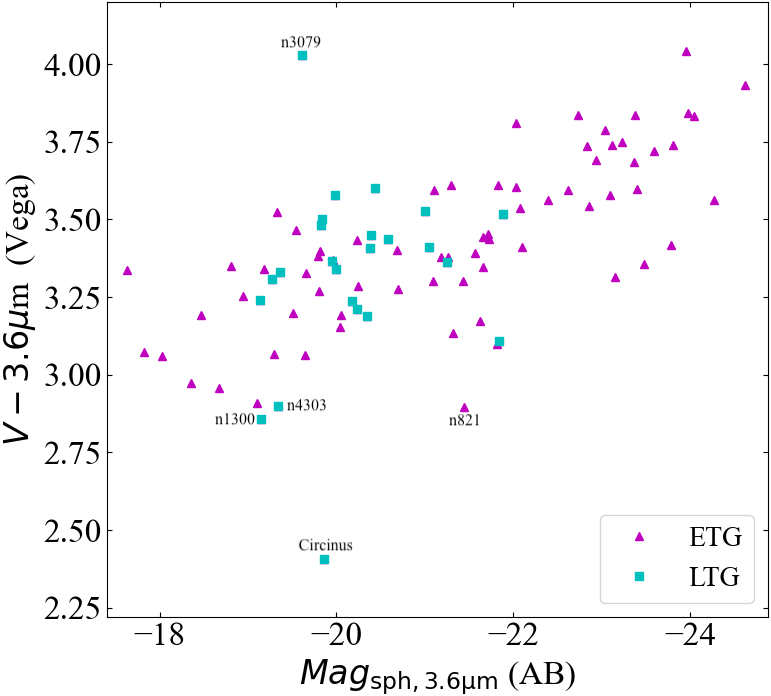} 
\includegraphics[trim=0.0cm 0.0cm 0.0cm 0.0cm, height=0.3\textwidth, angle=0]{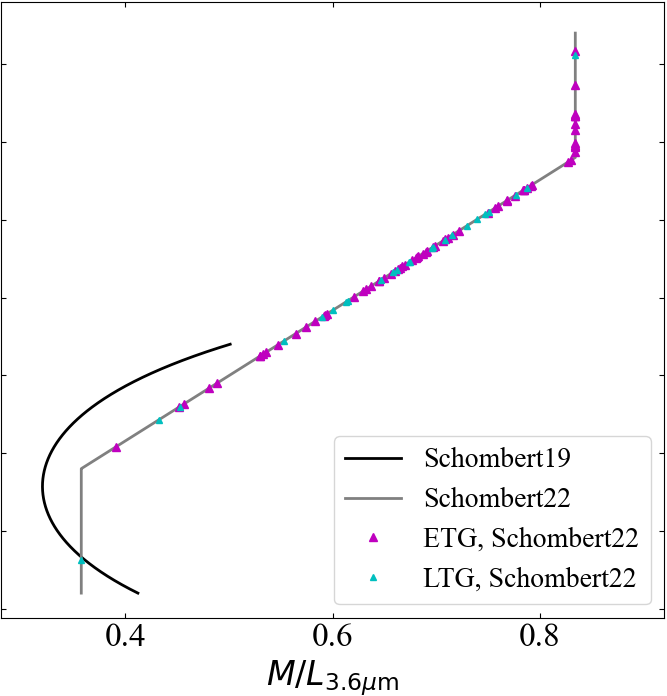} 
\caption{
Left-hand and middle panel: similar to the left-hand panel of Figure~\ref{Fig_colour_IP13}, but showing the 
  $m_B-m_{3.6}$ and $m_V-m_{3.6}$ (Vega) galaxy colour versus the spheroid 
absolute magnitude (AB).  We used $m_{\rm Vega} = m_{\rm AB} -2.76$ mag at
3.6~$\mu$m for the conversion. 
Right-hand panel: stellar 
  mass-to-light ratios at 3.6~$\mu$m from the baseline model in \citet[][their
    Table~2]{2019MNRAS.483.1496S} is shown by the black curve (Eq.~\ref{Eq-Schom-19}) and
  the new prescription from \citet{2022arXiv220202290S} is shown by the grey
  line with triangles (Eq.~\ref{Eq-Schom-21}).  Both 
  models use the $m_V-m_{3.6}$ colour and have been converted here to a \citet{2002Sci...295...82K} IMF. 
Note: The ETG with the bluest $V-$[3.6] colour is NGC~404; it is not seen in the
middle panel due to the zoomed in view. 
}
\label{Fig_colour_Sch}
\end{center}
\end{figure*}

\subsection{An (optical-NIR colour)-dependent $\Upsilon_*$}

The baseline model in \citet[][their Table~2]{2019MNRAS.483.1496S} is such
that 
\begin{eqnarray}
\log(M_*/L_{obs,3.6}) &=& 0.933(m_{V,{\rm Vega}}-m_{3.6,{\rm Vega}})^2 \nonumber \\
 &-& 4.932(m_{V,{\rm Vega}}-m_{3.6,{\rm Vega}}) +6.123, \nonumber  
\end{eqnarray}
valid for $2.3 \lesssim (m_{V,{\rm Vega}}-m_{3.6,{\rm Vega}}) \lesssim 3.1$. 
This equation is based on the \citet{2001MNRAS.322..231K} IMF, referred to as the pseudo-Kroupa IMF
by \citet{2010MNRAS.404.2087B}.  To adjust to the \citet{2002Sci...295...82K}
IMF, one needs to subtract 0.1 dex. 
To use the above equation, both the $V$-band and the 3.6~$\mu$m magnitudes need to
be calibrated on the Vega magnitude system. 
Given that our Spitzer 3.6~$\mu$m magnitudes are calibrated on the AB system, 
and the RC3 $V$-band data are on the Vega system, 
we follow
\citet{2018ApJS..236...47W}\footnote{\url{http://mips.as.arizona.edu/~cnaw/sun.html}}
and use 
\begin{equation}
m_{\rm Vega} = m_{\rm AB} -2.76 \, {\rm mag} \nonumber
\end{equation} 
at 3.6~$\mu$m  to modify the 
equation such that it can use 3.6~$\mu$m (AB) and $V$ (Vega) magnitudes.  
We also subtract 0.1 dex to bring it in line with the
\citet{2002Sci...295...82K} IMF. Doing so, we have that 
\begin{eqnarray}
\log(M_*/L_{obs,3.6}) & & =  0.933(m_{V,{\rm Vega}}-m_{3.6,{\rm AB}}+2.76)^2 - \nonumber \\
& & 4.932(m_{V,{\rm Vega}}-m_{3.6,{\rm AB}}+2.76) +6.023. 
\label{Eq-Schom-19}
\end{eqnarray} 

This relation has recently been further developed in
\citet{2022arXiv220202290S}, 
providing separate expressions for bulge and disc colours (which we do not have)
and for galaxy colours which extend to redder colours than those applicable in 
Equation~\ref{Eq-Schom-19}. 
We approximate the Bulge+Disc galaxy model from 
\citet[][see their Figure~3]{2022arXiv220202290S} --- which is based on the
\citet{2001MNRAS.322..231K} IMF --- by the equations 
\begin{eqnarray} 
M_*/L_{obs,3.6} &=& 0.45, \, {\rm for} \, (m_{V,{\rm Vega}}-m_{3.6,{\rm Vega}}) < 2.7, \nonumber \\ 
 &=& 0.45 + 0.6(m_{V,{\rm Vega}}-m_{3.6,{\rm Vega}} -2.7), \nonumber \\  
 & &  {\rm for} \, 2.7 <  (m_{V,{\rm Vega}}-m_{3.6,{\rm Vega}}) < 3.7, \nonumber \\ 
 &=& 1.05, \, {\rm for} \, (m_{V,{\rm Vega}}-m_{3.6,{\rm Vega}}) > 3.7. \nonumber  
\end{eqnarray} 
Switching to the \citet{2002Sci...295...82K} IMF, by multiplying 
by $10^{-0.1} \approx 0.794$, and switching to the AB magnitude 
system for the 3.6~$\mu$m magnitudes, we have that 
\begin{eqnarray}\label{Eq-Schom-21} 
M_*/L_{obs,3.6} &=& 0.36, \, {\rm for} \, (m_{V,{\rm Vega}}-m_{3.6,{\rm AB}}) < -0.06, \nonumber \\
 &=& 0.39 + 0.48( m_{V,{\rm Vega}}-m_{3.6,{\rm AB}} ), \nonumber \\  
 & & {\rm for} \, -0.06 < (m_{V,{\rm Vega}}-m_{3.6,{\rm AB}}) < 0.94, \nonumber \\ 
 &=& 0.83, \, {\rm for} \, (m_{V,{\rm Vega}}-m_{3.6,{\rm AB}}) > 0.94. 
\end{eqnarray}  
We assign a 15 percent uncertainty to these, reflecting the 0.05 dex
uncertainty in $\log M_*$ noted by \citet{2022arXiv220202290S} and allowing
for a reasonable error in the assigned colour.
This relation (Equation~\ref{Eq-Schom-21}), along with
Equation~\ref{Eq-Schom-19}, is shown in the right-hand panel of
Figure~\ref{Fig_colour_Sch}.  One can see that they are similar over their 
colour range of applicability.

\begin{figure*}
\begin{center}
\includegraphics[trim=0.0cm 0cm 0.0cm 0cm, height=0.3\textwidth, angle=0]{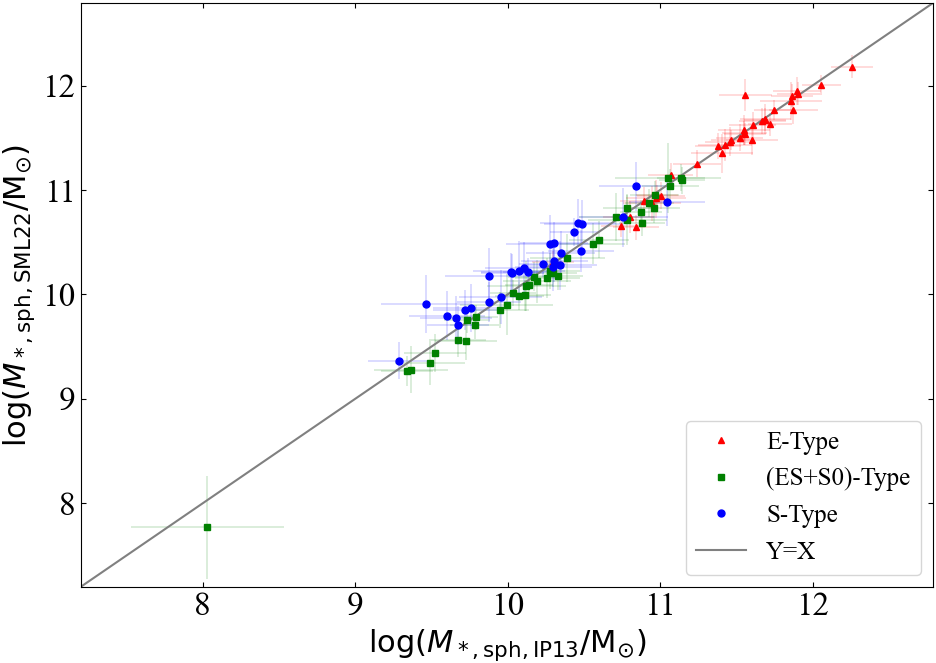}
\includegraphics[trim=0.0cm 0cm 0.0cm 0cm, height=0.3\textwidth, angle=0]{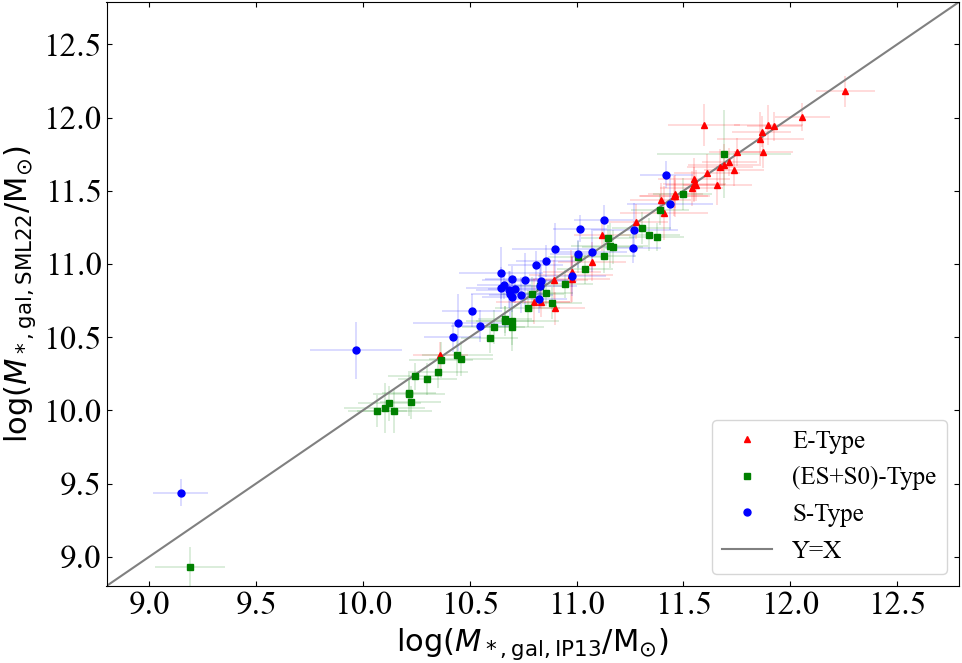}
\caption{Comparison of the spheroid and galaxy stellar masses derived using 
the (optical colour)-dependent $\Upsilon_*$ ratios from IP13, as given by our 
Equation~\ref{Eq_MonL_IP13}, and using the (optical-infrared colour)-dependent
$\Upsilon_*$ ratios from \citet{2022arXiv220202290S}, as given
by our Equation~\ref{Eq-Schom-21}. The differences seen here will result in
different black hole scaling relations to those in the main text
(Table~\ref{Table-IP13}). 
}
\label{Fig_comp}
\end{center}
\end{figure*}

Comparing Figure\ref{Fig_colour_Sch} with Figure~\ref{Fig_colour_IP13}, it is 
apparent that there are some differences.  For instances, while the ETGs
typically have red colours and $0.6 \lesssim M_*/L_{3.6} \lesssim 0.9$ in
Figure~\ref{Fig_colour_IP13}, they follow a colour-magnitude relation in
Figure\ref{Fig_colour_Sch} and have $0.6 \lesssim M_*/L_{3.6} \lesssim 0.9$.
The LTGs in Figure\ref{Fig_colour_Sch} appear to define a blue cloud, with 
$0.4 \lesssim M_*/L_{3.6} \lesssim 0.7$, while their cloud-like distribution
appears consistent with the ETG colour-magnitude relation in
Figure\ref{Fig_colour_Sch}, where they have $0.45 \lesssim M_*/L_{3.6} \lesssim 0.75$.

A few of our galaxies can also be seen to stand
out in the colour-magnitude diagram shown in the left-hand and middle panels of
Figure~\ref{Fig_colour_Sch}. 
The ETG with the faintest spheroid magnitude is NGC~2787, remodelled in
\citet{Graham:Sahu:22}. 
Circinus was already discussed in Section~\ref{Sec_Sample}.
The Seyfert galaxy NGC~3079 is a particularly dusty and rather edge-on
(80$^{\circ}$) spiral galaxy. It is notable for having blown a rather large
bubble \citep{2001ApJ...555..338C} from its central region, somewhat akin to
the Fermi Bubble in the Milky Way \citep{1984Natur.310..568S,
  2010ApJ...724.1044S}.  \citet{1991rc3..book.....D} suggested a dust
correction of 1.1 mag to the observed $B$- and $V$-band magnitudes of
NGC~3079.  The infrared glow of warm dust and the reduced optical flux may
have conspired to produce this outlier with an excessively red $V-$[3.6]
colour. Had the $V$-band magnitude been 0.5 mag brighter, this galaxy would
not stand out in Figure~\ref{Fig_colour_Sch} and its mass-to-light ratio would
be 10 percent smaller.  It does not stand out in the $M_{\rm bh}$--$M_{\rm
  *,sph}$ diagram.
NGC~821 also appears as something of an outlier; it is too blue by
$\sim$0.7~mag.  This may be due to contamination from the overlapping V$\approx$9.3 mag (Vega)
star SIMBAD BD+10~293.

For eight galaxies without a
reliable $V$-band magnitude, we used a bisector regression on the available
($V-$[3.6])--$\mathfrak{M}_{\rm sph,3.6}$ data 
(Figure~\ref{Fig_colour_Sch}) to obtain a $V-$[3.6]
colour based on their $\mathfrak{M}_{\rm sph,3.6}$ values.  
Specifically, we derived and used the expression 
$V-$[3.6] (Vega) = 1.45 - 0.09 $\mathfrak{M}_{\rm sph,3.6}$ (AB). 
These colours are marked with brackets in Table~\ref{Table-data}.
%

In passing, we note that, based on the pseudo-Kroupa IMF,
\citet{2022arXiv220202290S} suggest $M_*/L_{obs,3.6} \approx 1$ to 1.05 for elliptical galaxies, 
$\approx$0.97 for red bulges, which are likely in ETGs, and $\approx$0.86 
for blue bulges, likely in the later spiral galaxy types. For the 
\citet{2002Sci...295...82K} IMF, this equates to
0.79 to 0.83 (E), 0.77 (red bulges), and 0.68 (blue bulges).  
Such a step function differs from the continuous $M_*/L$ ratios for the 
galaxy models, while qualitatively mimics the approach in 
\citet{2019ApJ...876..155S}, which used $M_*/L_{obs,3.6}=0.6$ for ETGs and
0.453 for LTGs, based on the \citet{2003PASP..115..763C} IMF, or 0.53 and 0.40
for 
the \citet{2002Sci...295...82K} IMF. 
We are, however, keen to avoid (or only use) a prescription to convert light
into mass with such a discontinuity, given that we are checking for potential
discontinuity or 
offsets in the $M_{\rm bh}$--$M_{\rm *,sph}$ diagram. 
We proceed using Equation~\ref{Eq-Schom-21}. 

Figure~\ref{Fig_comp} compares the spheroid and galaxy stellar masses
obtained using Equations~\ref{Eq_MonL_IP13} and \ref{Eq-Schom-21}.  Here, one
can better visualise the differences between the distribution of mass-to-light
ratios seen in Figures~\ref{Fig_colour_IP13} and \ref{Fig_colour_Sch}. 
Apparently, for the full ensemble of galaxy types used in our study,
the two prescriptions we have used for $\Upsilon_*$ --- from 
\citet{2013MNRAS.430.2715I} and \citet{2022arXiv220202290S} after conversion
to the same IMF --- yield consistent results. However, taking things to the
next level by looking at dependencies on the galaxy morphology, it is
also apparent that further information can be gleaned. 
Given the slight differences in slope and zero-point (at some mass) for each
morphological type (E, ES/S0 and S), we 
have repeated the diagrams and analysis previously based on 
Equation~\ref{Eq_MonL_IP13}.

\subsection{Results}

\begin{figure*}
\begin{center}
\includegraphics[trim=0.0cm 0cm 0.0cm 0cm, width=1.0\textwidth,  angle=0]{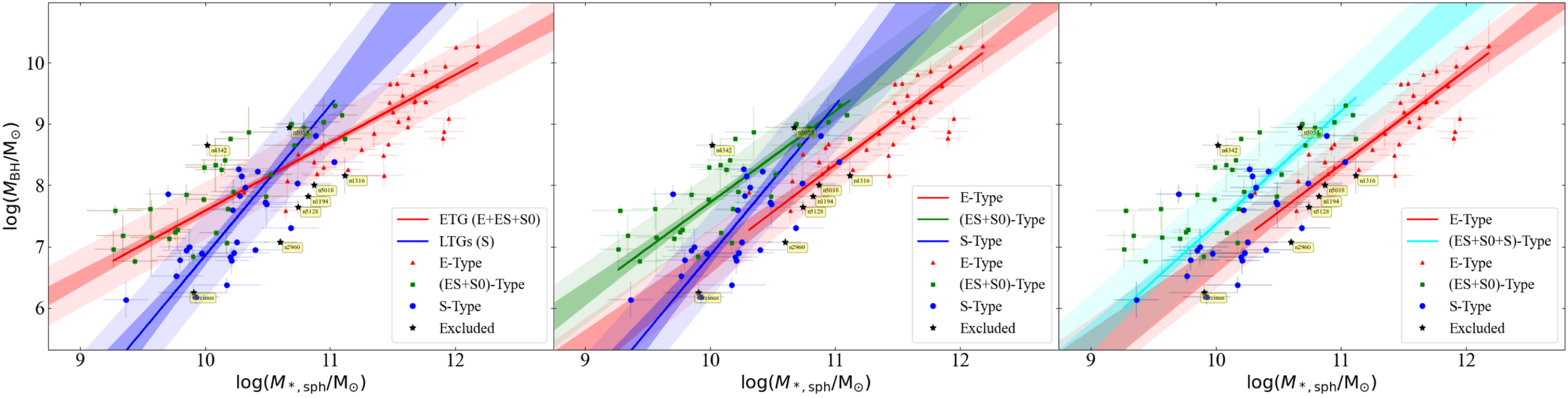}
\caption{Similar to Figure~\ref{Fig_M_Msph_IP13} but using 
  Equation~\ref{Eq-Schom-21} to derive the spheroid stellar masses. 
This yields black hole scaling relations with similar relative offsets 
  to those seen in \citet{2019ApJ...876..155S}, which used
  $M_*/L_{*,3.6}=0.6$ for all, after applying a 25 percent reduction to
  the LTGs' luminosity $L_{obs,3.6}$ due to dust glow in the LTGs \citep{2015ApJS..219....5Q}.  
Here, the middle panel appears 
  to be the optimal separation of morphological types.
The bulges of spiral
  galaxies and lenticular galaxies appear to converge at high masses. 
}
\label{Fig_M_Msph_Sch}
\end{center}
\end{figure*}

It is apparent from Figure~\ref{Fig_M_Msph_Sch} 
that, qualitatively, the
large offset between the ES/S0 and E galaxies in the $M_{\rm bh}$--$M_{\rm
  *,sph}$ diagram remains.  That is, this is 
a robust result. 
The steeper $M_{\rm bh}$--$M_{\rm *,sph}$ relation 
for the bulges of spiral galaxies is in agreement with that observed when using 
$M_*/L_{*,3.6,obs} = 0.453$ for every spiral galaxy with {\it Spitzer} data
\citep{2019ApJ...873...85D}.  This may seem counterintuitive until it is
remembered that this recent study included additional non-{\it Spitzer} data, plus
we have updated some of the photometric decompositions. 
The results suggest that the evolution of
bulges in relatively gas-poor versus gas-rich discs may proceed along
different paths, converging at high bulge masses (Figure~\ref{Fig_M_Msph_Sch},
middle panel). 
Perhaps stellar feedback has limited the black hole accretion in the spiral galaxies,
explaining their lower $M_{\rm bh}/M_{\rm *,sph}$ ratios than in many
similarly-massed ES/S0 galaxy bulges. Alternatively, 
perhaps the black holes in the spiral galaxies are yet
to grow up by exhausting their fuel supply and largely drying out, as has occurred in
many of the ES/S0 galaxies.

Using Equation~\ref{Eq-Schom-21}, the $M_{\rm bh}$--$M_{\rm *,gal}$ relations have
also been re-derived for the E, ES/S0, and S galaxies
(Figure~\ref{Fig_MMgal_Sch}), as has the associated ($M_{\rm
  bh}/M_{\rm *,gal}$)--$M_{\rm *,gal}$ diagram
(Figure~\ref{Fig_mgal_rat_Sch}). 
Table~\ref{Table-Sch} provides the mass scaling relations based on
Equation~\ref{Eq-Schom-21}.  Of note is the $\sim$0.3 dex, nearly 2$\sigma$, change to the
intercept of the $M_{\rm bh}$--$M_{\rm *,gal}$ relation for the spiral 
galaxies.  This translates to predictions for black hole masses in spiral
galaxies that are a factor of $\sim$2 different to what one obtains when using the
scaling relations in the main text.  This highlights the need to pay attention
to the $\upsilon$ term in these equations.

\begin{figure*}
\begin{center}
\includegraphics[trim=0.0cm 0cm 0.0cm 0cm, height=0.3\textwidth, angle=0]{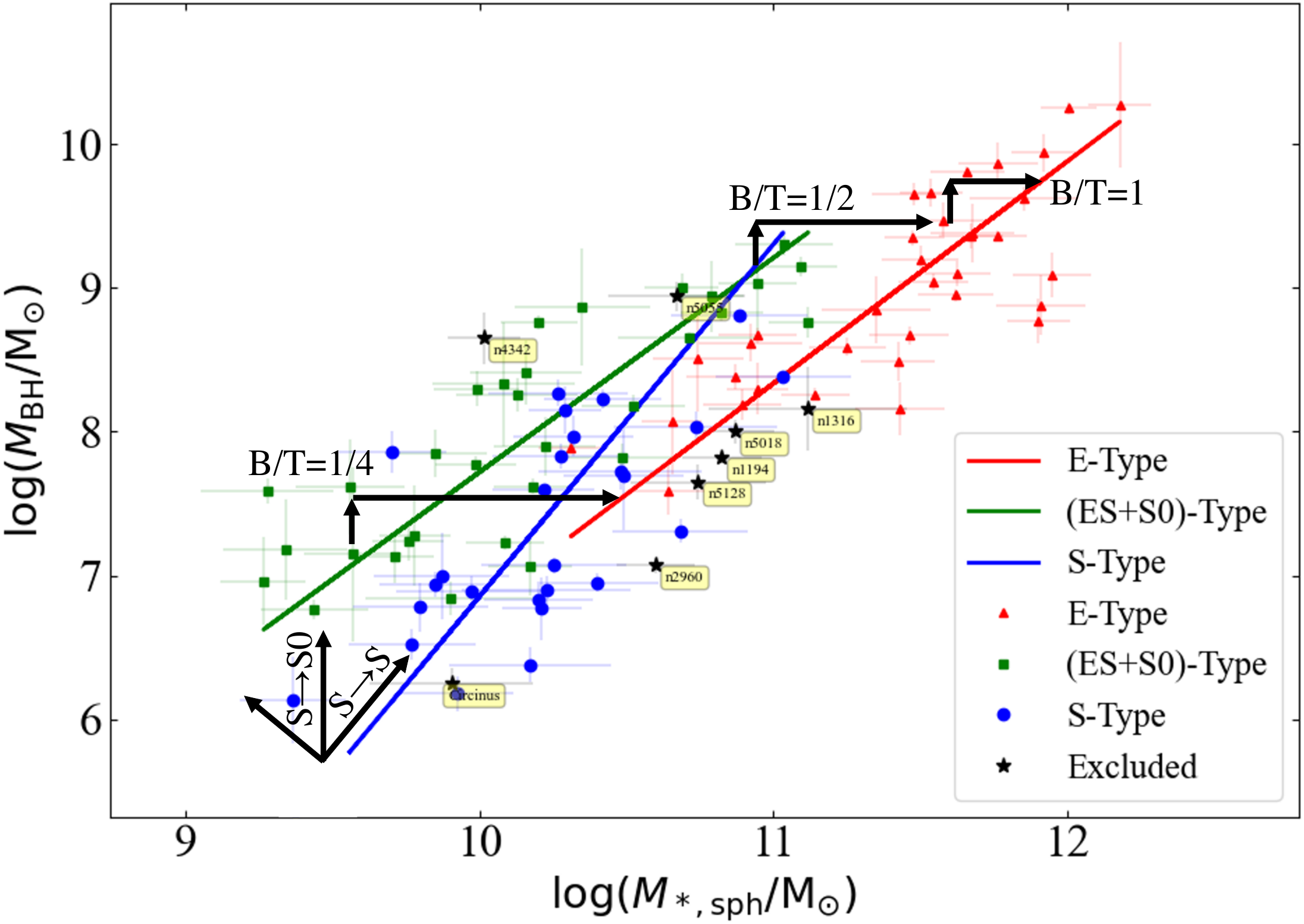}
\includegraphics[trim=0.0cm 0cm 0.0cm 0cm, height=0.3\textwidth, angle=0]{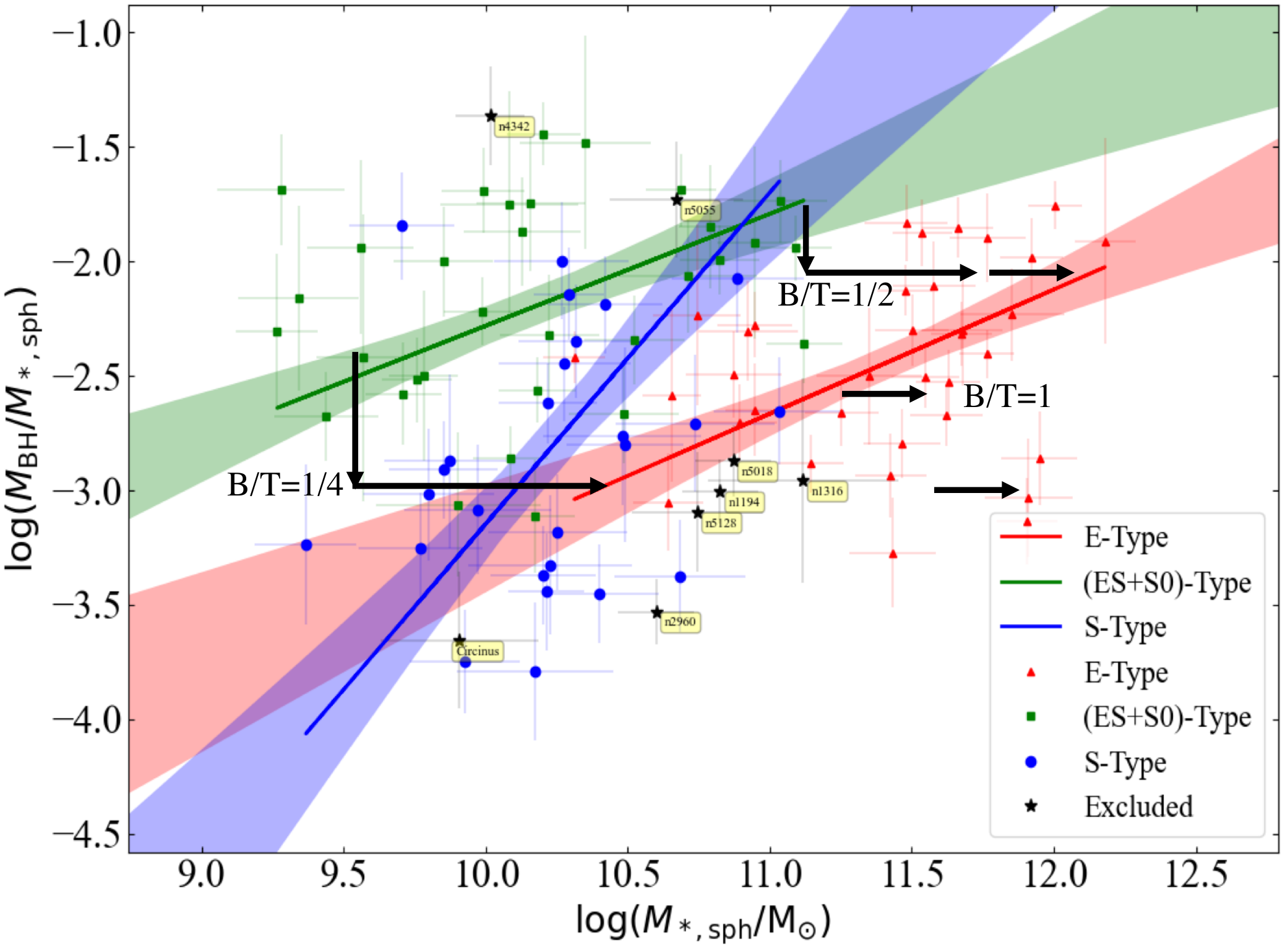} 
\caption{
Left-hand panel:  Similar to the right-hand panel of Figure~\ref{Fig_evolve_IP13} but
using Equation~\ref{Eq-Schom-21} to derive the spheroid stellar masses and
with the shading removed. 
Right-hand panel: Similar to Figure~\ref{Fig_msph_rat_IP13} but using
Equation~\ref{Eq-Schom-21}. 
The lines have been propagated from the middle panel of Figure~\ref{Fig_M_Msph_Sch}.   
For a given spheroid mass, the average $M_{\rm bh}/M_{\rm *,sph}$ ratio 
depends on the galaxy's morphological type and thus formation history. 
}
\label{Fig_msph_rat_Sch}
\end{center}
\end{figure*}

\begin{figure*}
\begin{center}
\includegraphics[trim=0.0cm 0cm 0.0cm 0cm, width=1.0\textwidth, angle=0]{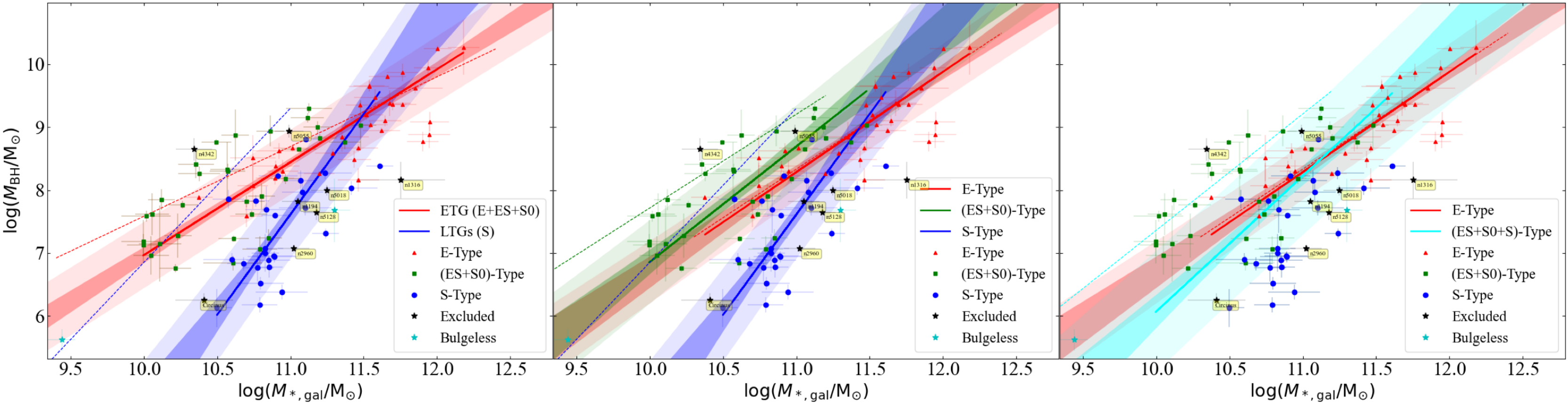} 
\caption{Similar to Figure~\ref{Fig_MMgal_IP13}, but using Equation~\ref{Eq-Schom-21}. 
The solid lines show the galaxy relations while the dashed lines show the
associated spheroid relations from Figure~\ref{Fig_M_Msph_Sch}. 
}
\label{Fig_MMgal_Sch}
\end{center}
\end{figure*}

\begin{figure*}
\begin{center}
\includegraphics[trim=0.0cm 0cm 0.0cm 0cm, height=0.3\textwidth, angle=0]{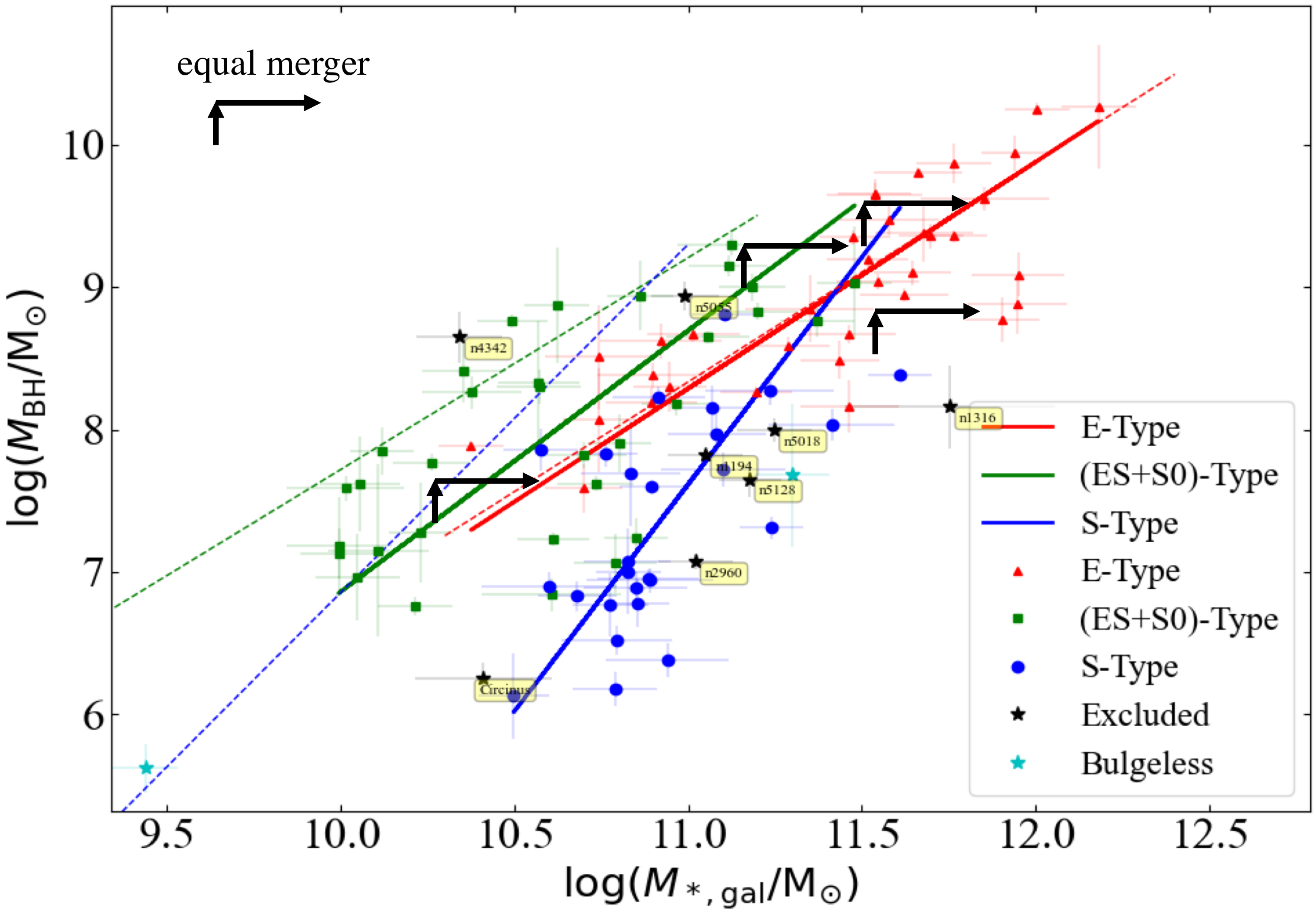}
\includegraphics[trim=0.0cm 0cm 0.0cm 0cm, height=0.3\textwidth, angle=0]{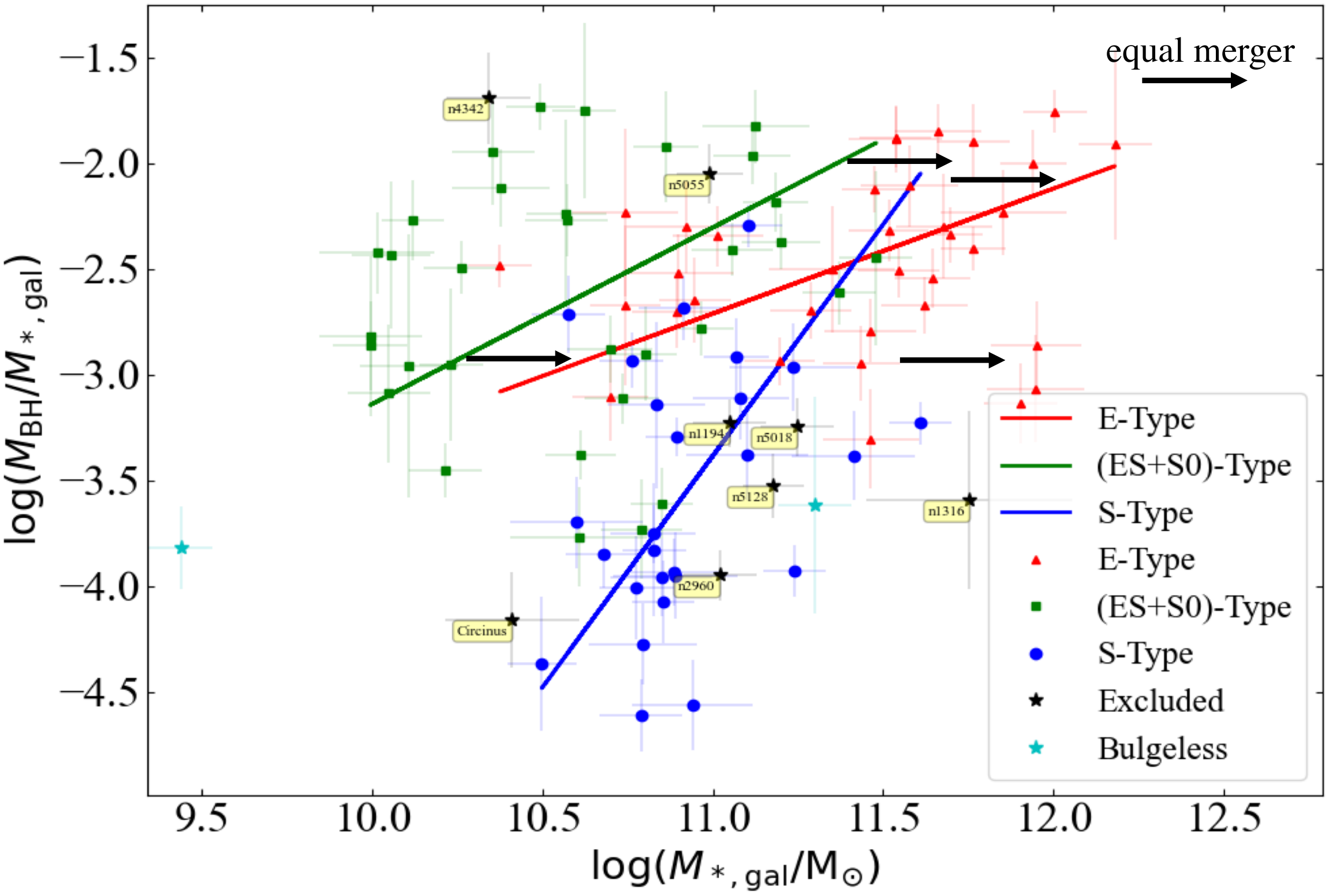}
\caption{Left-hand panel: Similar to the middle panel of Figure~\ref{Fig_MMgal_Sch} but without the shading. 
Right-hand panel: Similar to Figure~\ref{Fig_mgal_rat_IP13}, but using Equation~\ref{Eq-Schom-21}. 
The lines have been propagated from the middle panel of Figure~\ref{Fig_MMgal_Sch}. 
The arrows show the movement due to a dry equal-mass merger. 
}
\label{Fig_mgal_rat_Sch}
\end{center}
\end{figure*}

\begin{figure}
\begin{center}
\includegraphics[trim=0.0cm 0cm 0.0cm 0cm, width=1.0\columnwidth, angle=0]{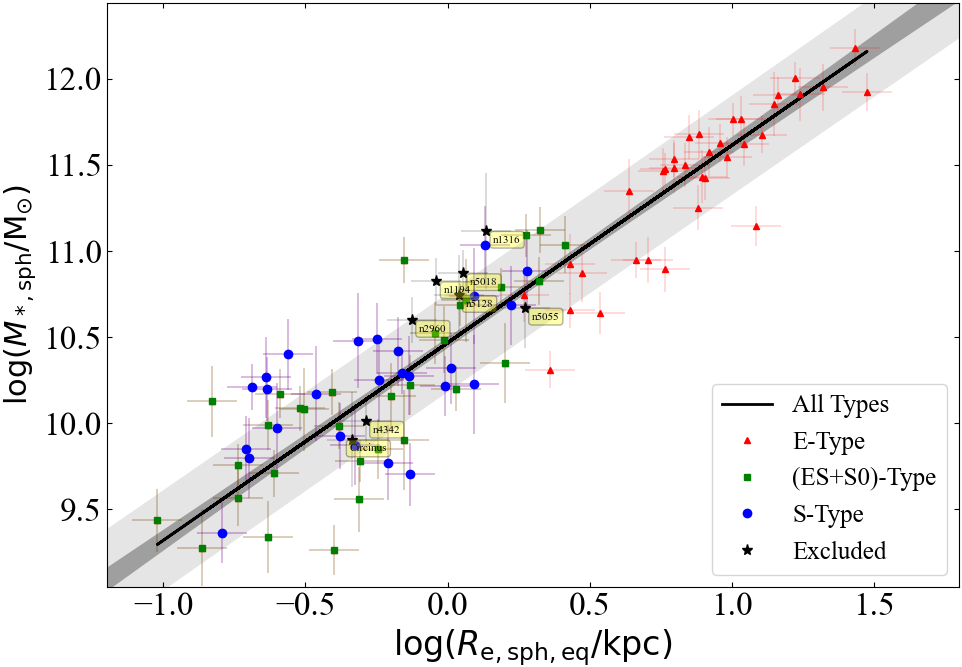}
\caption{$R_{\rm e,sph,eq}$ versus $M_{\rm *,sph}$. 
Similar to Figure~\ref{Fig_R_Msph_IP13} but using Equation~\ref{Eq-Schom-21}
for the spheroid masses. 
} 
\label{Fig_R_Msph_Sch}
\end{center}
\end{figure}

The spheroid size-mass diagram has been
reproduced in Figure~\ref{Fig_R_Msph_Sch} using the stellar mass-to-light
ratios obtained via Equation~\ref{Eq-Schom-21}.  It continues to reveal an apparent
continuity between bulges and elliptical galaxies 
that supports the merger-built evolutionary
process (shown in Figures~\ref{Fig_msph_rat_Sch} and \ref{Fig_mgal_rat_Sch}) 
for explaining the demographics of massive ETGs in the 
$M_{\rm bh}$--$M_{\rm *,sph}$ diagram. 

\begin{figure}
\begin{center}
\includegraphics[trim=0.0cm 0cm 0.0cm 0cm, width=1.0\columnwidth,
  angle=0]{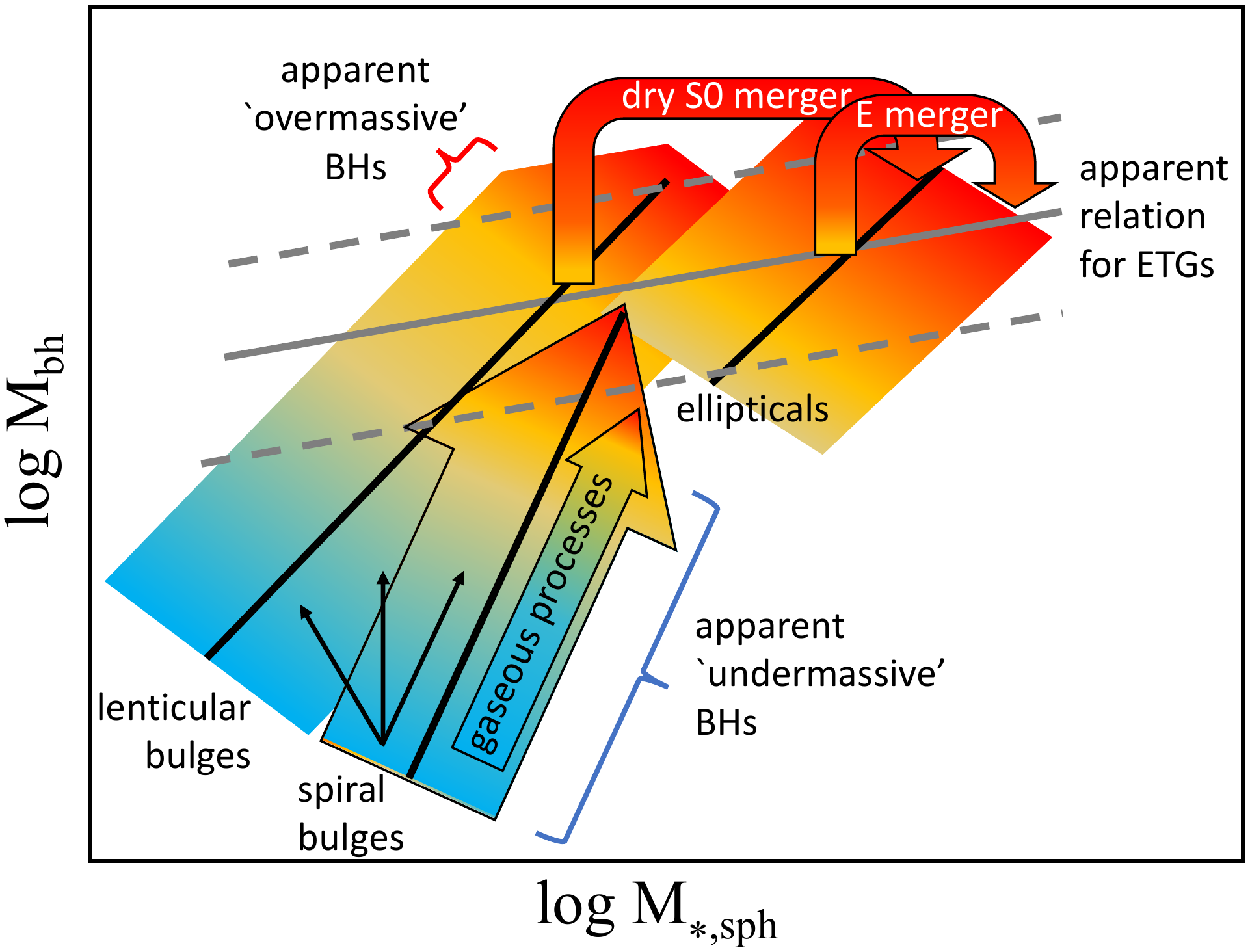} 
\caption{Modification of Figure~\ref{Fig_schematic}. 
The band on the left shows the relation for bulges in lenticular galaxies
while the steeper band in the middle is for the bulges of spiral galaxies, and the 
band on the right shows the relation for elliptical galaxies. 
The shallower relation, shown in grey, tracks an apparent 
`red sequence', or equally `red herring'. 
Regarding the three thin arrows in the lower part of the diagram: 
if the gaseous evolution of spiral galaxies predominantly grows the disc
and AGN but not the bulge, then it will evolve upwards in this diagram rather 
than along the spiral galaxy sequence. Alternatively, mass loss from stellar winds may fuel
the AGN and decrease the spheroid stellar mass. 
}  
\label{Fig_schematic_2}
\end{center}
\end{figure}

Building on Figure~\ref{Fig_schematic}, 
Figure~\ref{Fig_schematic_2} shows additional pathways for the bulges of
spiral galaxies. 
While we have introduced colour-dependent $M_*/L$ ratios and provided a
previously absent interpretation for the $M_{\rm bh}$--$M_{\rm *,sph}$ and 
$M_{\rm bh}$--$M_{\rm *,gal}$ diagrams, we identify areas where
further improvement can be pursued.  This includes the derivation of bulge,
rather than galaxy, colours.  These may come from the decomposition of optical
images or sampling regions of the bulge largely unaffected by dust
\citep[e.g.,][]{1994AJ....107..135B}. Furthermore, with integral field units
(IFUs), one can pursue star formation histories and rates, extinction,
metallicity and ages \citep[e.g.,][and references
  therein]{2019A&A...621A.120G} on a spaxel-by-spaxel basis.  Of course, some kind of
bulge/disc separation will still be desirable for many spaxels, just as it is
when dealing with a global galaxy colour.  Nonetheless, 
these measurements should enable refined
stellar $M_*/L$ ratios for the bulges, which will help to better establish the  
LTG bulge sequence in the $M_{\rm bh}$--$M_{\rm *,sph}$ diagram. 
Understanding the potential offset between the
bulges of LTGs and the bulges of ETGs offers the 
promise of yet further insight into the intriguing lives of galaxies and their
black holes. 
Colour and metallicity gradients \citep[e.g.,][]{2009ApJ...691L.138S} may 
also offer clues to how some systems formed, 
whether through wet or dry mergers or relics from an early-Universe. 
Here, we have implicitly focussed on dry mergers in which the bulk of the
orbital angular momentum (both in the progenitors and between them) cancels 
and an E galaxy is generated. 
We will explore merger-built systems that have retained substantial angular
momentum in \citet{Graham:Sahu:22}.

\begin{table}
\centering
\caption{Black hole mass scaling relations}\label{Table-Sch}
\begin{tabular}{lccc}
\hline
Galaxy type   &   Slope (A) & Intercept (B) & $\Delta_{\rm rms}$ \\
\hline 
\multicolumn{4}{c}{$\log(M_{\rm bh}/M_\odot) = A\log[M_{\rm *,sph}/\nu
    (5\times10^{10}\,M_\odot)] +B$} \\
E   (35)        &  1.54$\pm$0.16  &  7.87$\pm$0.17   & 0.41 \\
ES/S0  (32)     &  1.49$\pm$0.16  &  8.76$\pm$0.17   & 0.45 \\
S      (26)     &  2.45$\pm$0.49  &  8.56$\pm$0.29   & 0.67 \\
ES/S0 \& S (58) &  1.85$\pm$0.18  &  8.66$\pm$0.15   & 0.67 \\
E \& ES/S0 (67) &  1.11$\pm$0.07  &  8.36$\pm$0.11   & 0.45 \\
\hline
\multicolumn{4}{c}{$\log(M_{\rm bh}/M_\odot) = A\log[M_{\rm *,gal}/\nu
    10^{11}\,M_\odot] +B$}  \\
E   (35)        &  1.59$\pm$0.17  &  8.29$\pm$0.14   & 0.42 \\
ES/S0  (32)     &  1.83$\pm$0.26  &  8.70$\pm$0.19   & 0.62 \\
S      (26)     &  3.19$\pm$0.55  &  7.62$\pm$0.16   & 0.67 \\
ES/S0 \& S (58) &  2.15$\pm$0.27  &  8.23$\pm$0.13   & 0.92 \\
E \& ES/S0 (67) &  1.48$\pm$0.10  &  8.45$\pm$0.11   & 0.51 \\
\hline
\multicolumn{4}{c}{$\log(M_{\rm sph}/ \upsilon M_\odot) = A\log[R_{\rm e,sph,eq}/\rm kpc] +B$} \\
All   (93)      &  1.15$\pm$0.05  &  10.47$\pm$0.08  & 0.28 \\
\hline
\end{tabular}

Similar to Table~\ref{Table-IP13} but the stellar masses used to derive the
relations shown here came from Equation~\ref{Eq-Schom-21}.  As such, when
using these equations, $\upsilon = 1$ if one uses stellar masses
consistent with those obtained via Equation~\ref{Eq-Schom-21}. 

\end{table}

\subsection{Some notes on little $\upsilon$}

Here, we discuss the mass-to-light conversion term, $\upsilon$, introduced in
\citet{2019ApJ...873...85D}, and further developed in
\citet{2019ApJ...876..155S}, to account for switches between various
prescriptions of $\Upsilon_*$ \citep[e.g.,][]{2003ApJS..149..289B,
  2009MNRAS.400.1181Z, 2011MNRAS.418.1587T, 2013MNRAS.430.2715I,
  2015MNRAS.452.3209R, 2022arXiv220202290S}.\footnote{The lower-case upsilon
  symbol was introduced to facilitate changes to the mass-to-light ratio,
  $\Upsilon_*$, in a similar manner to how $h$ can enact changes to the
  adopted Hubble-Lema\^itre constant H$_0$.}  Such conversions are necessary if one uses 
any $M_{\rm bh}$--$M_{\rm *,sph}$ and/or $M_{\rm bh}$--$M_{\rm *,gal}$
scaling relations when an alternative $\Upsilon_*$ prescription to that used
to establish these relations is used to obtain the stellar mass of the target
for which one wishes to predict the black hole mass.  While this should sound
obvious, it can be problematic in practice and \citet{SahuGrahamHon22} show how failing to do so can
result in a bias, explaining the offset presented in
\citet{2016MNRAS.460.3119S} and the incorrect conclusion that galaxies with
directly observed black hole masses are offset from the SDSS population of
ETGs in the $M_{\rm bh}$--$\sigma$ diagram.

We raised awareness of this general issue in 
\citet[][their Eq.~10]{2019ApJ...873...85D}, through  the introduction of
the $\upsilon$ term in the relation 
 $\log(M_{\rm bh}) = A\,\log(M_*/\upsilon)+B$. 
\citet{2019ApJ...873...85D} used $M_*/L_{obs,3.6}=0.453$ for a sample of
spiral galaxies imaged with the Spitzer Space Telescope's  Infrared Array
Camera - channel 1 \citep[IRAC-1:][]{2004ApJS..154...10F},
which operated at a central wavelength of 3.6~$\mu$m. As such,  $\upsilon$
was initially introduced as 
$\upsilon_{\rm IRAC1,0.453} = \Upsilon_{\rm IRAC1}/0.453$, 
with $\Upsilon_{\rm IRAC1}$ a (potentially) different stellar 
mass-to-light ratio used by someone else to obtain the stellar masses of their
spheroid and/or galaxy sample.  
\citet[][their Figure~4 and their Equations~6--8 for $\upsilon$]{2019ApJ...876..155S} developed this
further by providing a limited set
of conversions for when the user of the scaling relation
%
%
has obtained their stellar mass in a non-IRAC1 band and thus does not simply
have an alternate $\Upsilon_{\rm IRAC1}$ value.

There are multiple approaches to deriving $\upsilon$, and we present two of them here,
using the $\Upsilon_*$ prescriptions given in this paper as an example.  

From its definition, we have that
\begin{equation}
\upsilon \equiv \Upsilon_{2} / \Upsilon_{1}, 
\end{equation}
for use in an expression like
\begin{equation}
 \log(M_{\rm bh}/M_\odot) = A_1\log[M_{\rm                                                   
      *,sph,1}/\upsilon(5\times10^{10}\,M_\odot)] +B_1
\end{equation}
in which $\Upsilon_1$ was used to define $M_{\rm *,sph,1}$. 
Now, if, for example, one uses Equation~\ref{Eq-Schom-21} to derive the stellar
mass of a spheroid, but wishes to use the equations from
Table~\ref{Table-IP13}, then one can calculate the required 
$\upsilon$ term by 
dividing the $\Upsilon_2$ ratio from Equation~\ref{Eq-Schom-21} by that given in
Equation~\ref{Eq_MonL_IP13}, to give
\begin{equation}
\upsilon = \frac{ 0.39 + 0.48( m_{V,{\rm Vega}} - m_{3.6,{\rm AB}} ) } 
{ 10^{ 1.034(m_{B,{\rm Vega}} - m_{V,{\rm Vega}} ) - 1.067 } }, 
\end{equation}
for $-0.06 < (m_{V,{\rm Vega}}-m_{3.6,{\rm AB}}) < 0.94$ and 
$m_{B,{\rm Vega}} - m_{V,{\rm Vega}} > 0.5$. 
This is fine if one has both of these colour terms, with the two magnitudes
required for each colour being measured consistently, for 
example, within the same aperture.  However, this information may not be
available, in which case an alternative method is required. 

A second approach uses the scaling 
relation obtained by comparing the masses, $M_1$ and $M_2$, of other spheroids obtained using
each of the $M_*/L$ ratios.  An example of this is seen in
Figure~\ref{Fig_comp}.  For the relation 
\begin{equation}
 \log [M_2/(5\times10^{10}\,M_\odot)] = a\log[M_1/(5\times10^{10}\,M_\odot)] +
 b, 
\end{equation}
it can be shown that 
\begin{equation}\label{Eq_up}
 \log \upsilon = \left( \frac{a-1}{a} \right) \log[M_2/(5\times10^{10}\,M_\odot)] + b/a. 
\end{equation}
Thus, knowing $a$ and $b$ provides an alternative means to calculate the value
of $\upsilon$ for your value of $M_2$ obtained using $\Upsilon_2$. 

The slightly different trends in Figure~\ref{Fig_comp} resulted in 
the somewhat different black hole scaling relations between 
Table~\ref{Table-IP13} and Table~\ref{Table-Sch}. 
With a little algebra, it can also be shown that 
\begin{equation}\label{Eq_ab}
a=\frac{2A_1}{A_1+A_2}\, {\rm and}\, b=\frac{B_1-B_2}{A_1+A_2}, 
\end{equation}
where $A_1$ and $B_1$ are the slope and intercept from the black hole scaling
relation (Table~\ref{Table-IP13}) 
derived using the $\Upsilon_1$ ratios (Equation~\ref{Eq_MonL_IP13})
and 
$A_2$ and $B_2$ are the slope and intercept from the black hole scaling
relation (Table~\ref{Table-Sch})
derived using the $\Upsilon_2$ ratios (Equation~\ref{Eq-Schom-21}. 
For our spiral galaxies, for example, we have from 
Equation~\ref{Eq_ab} that 
$a=0.96$ and $b=0.02$, while for our ES/S0
galaxies,  we have $a=1.01$ and $b=-0.03$. 
Such morphology-specific terms for the $\upsilon$ equation 
(Equation~\ref{Eq_up}) provide an additional level of sophistication to that
presented in \citet{2019ApJ...876..155S}.

\bsp    
\label{lastpage}
\end{document}